\documentclass[journal]{IEEEtran}
\usepackage{flushend,cuted}

\usepackage{cite}

\ifCLASSINFOpdf
   \usepackage[pdftex]{graphicx}
\else
\fi
\usepackage{amsmath}
\usepackage{array}

  \usepackage[caption=false,font=footnotesize]{subfig}
\usepackage{amsthm}
\usepackage{amssymb}
\usepackage{color}

\begin{document}
\onecolumn 

%
\title{Discrete Packet Management: Analysis of Age of Information of Discrete Time Status Updating Systems}
%
%
%

\author{Jixiang~Zhang, Southeast University %
\thanks{Zhang. J was with the School of Information Science and Engineering, 
Southeast University, NanJing, China. E-mail: 230179077@seu.edu.cn.}}

%
%

\markboth{Journal of \LaTeX\ Class Files,~Vol.~14, No.~8, August~2015}%
{Shell \MakeLowercase{\textit{et al.}}: Bare Demo of IEEEtran.cls for IEEE Journals}
%



\maketitle

\begin{abstract}
In this paper, we consider performing packet managements in discrete time status updating system, focusing on determining the stationary AoI-distribution of the system. Firstly, let the queue model be Ber/G/1/1, we obtain the AoI-distribution by introducing a two-dimensional AoI-stochastic process and solving its steady state, which describes the random evolutions of AoI and age of packet in system simultaneously. In this case, actually we analyze a more general queue called probabilistic preemption Ber/G/1/1, where the packet service is allowed to be preempted with certain probabilities. As a special case, stationary AoI-distribution for the system with Ber/Geo/1/1 queue is obtained either. For the system having size 2, two specific queues are considered, i.e., the Ber/Geo/1/2 and Ber/Geo/1/2* queues. The core idea to find the stationary AoI-distribution is that the random transitions of three-dimensional vector including AoI at the receiver, the packet age in service, and the age of waiting packet can be fully described, such that a three-dimensional AoI process is constituted. The stationary distribution of three-dimensional process then gives the stationary AoI distribution as one of its marginal distributions. For both cases, the explicit expressions of AoI-distribution are derived, thus giving the complete description of the steady state AoI for the system. For all the cases, since the steady state of a larger-dimensional AoI process is solved, so that except the AoI-distribution, we obtain more. For instance, the distributions of packet system time and waiting time for size-two updating system, and the so-called violation probabilities that AoI exceeds certain threshold. 
\end{abstract}


%
\IEEEpeerreviewmaketitle

\newtheorem{Definition}{Definition}  
\newtheorem{Theorem}{Theorem}  
\newtheorem{Lemma}{Lemma}
\newtheorem{Corollary}{Corollary}
\newtheorem{Proposition}{Proposition}
\newtheorem{Remark}{Remark}

\section{Introduction}
%
%
%
%

\IEEEPARstart{T}{ransmitting} information rapidly is important in many time sensitive applications. A large number of applications in IoT network require real time messages to update the state of certain nodes constantly. For example, in order to better schedule road traffics in an area, the control center needs real time changes of traffic congestion in every road. To accurately track a moving object, the changing position information is required to transmitted to monitor within a very short time.

The age of information (AoI) metric has been widely used in recent years to measure the freshness of the message obtained at the receiver and the timeliness of an information transmission system, which is defined as the time that one packet experiences since it was generated at the source \cite{1}. In the past few years, lots of articles have been published to analyze the average and peak AoI, or design optimal status updating systems that can minimize the average AoI or other AoI-related performance indices. In work \cite{2}, the authors gave a detailed survey of the recent contributions of AoI, both in theory and in practice.

A status updating system is a communication system, where the source observes the instantaneous state of a physical process and samples the process at random times. An updating packet consisting of the sampled data and timestamp is sent to receiver via a transmitter, such that the state at the receiver is updated. In the early works, AoI analysis for the system with basic queue models was carried out and many results have been obtained. The average AoI of system with M/M/1, M/D/1, and D/M/1 queues were determined in \cite{3}. Then, assume that the service discipline is last-come-first-served (LCFS), the authors of \cite{4} considered and computed the mean of the AoI for these queue models. A systematic analysis method called Stochastic Hybrid System (SHS) \cite{5} was proposed to analyze the long-term average AoI in paper \cite{6}. By this method, the author considered the updating system with multiple independent sources and determined the average AoI of packet from each source explicitly. The method has been used widely recently to find more results about analysis of AoI \cite{7,8,9,10,11,12,13}. In order to study the influence of packet multipath transmission on AoI, in work \cite{14}, information transmission system having two parallel transmitters was discussed. The authors obtained the closed-form expression of the average AoI by sophisticated random events analysis.

To enhance the timeliness of the updating system, in paper \cite{15} the authors proposed the notion of packet management, where three control policies were discussed: (i) for the system of size 1, transmitter can discard the packet if it is currently busy when the packet arrives; (ii) system keeps one packet waiting in queue for next time transmission, and discards any other packets that comes later, and (iii) similarly, a packet is allowed to wait in queue, but this waiting packet can be refreshed by the fresher ones constantly. The queue models corresponds to each case is denoted as M/M/1/1, M/M/1/2, and M/M/1/2*, respectively. For three queue models, the average AoI was characterized in paper \cite{15}. In addition, another metric called the peak AoI which denotes the maximal value of AoI before it drops, was defined and calculated assuming that the system used different queue models.

For the AoI analysis of status updating system, although many queue models have been considered and plenty of conclusions have been obtained, however, it was observed that in the majority of articles, only the average AoI is computed. From the mathematical point of view, to describe the system's AoI in the steady state, it is not sufficient to get only the first order moment. In order to obtain all the knowledge about the stationary AoI, it is necessary to find its probability distribution. Besides, within the recent AoI literatures \cite{16,17,18,19,20,21,22,23}, the optimization objective used commonly in an optimal system design problem is the average value of the AoI. But a system with low-level average AoI does not mean the system also performs well in other aspects, such as the AoI violation probability for a given threshold, the tail probability \cite{24,25,26} and its decreasing rate. If the stationary AoI distribution is known, more aspects can be taken into consideration when we try to design an excellent updating system. Finding the distribution of stationary AoI in continuous time model is very hard, even hopeless, while in this paper we will prove that for certain queues, the AoI distribution of discrete time status updating system can be determined explicitly.

We observe that the analysis of status updating system with discrete time queues has been proposed in paper \cite{27}, where the authors computed the average AoI and peak AoI for discrete time systems with Ber/G/1 and G/G/$\infty$ queues, by the methods that used in continous AoI analysis and the probability generation function method (PGF). Later, in work \cite{28}, the authors proposed a new approach to characterize a sample path of AoI stochastic process, and derived a general expression of the AoI generation function. Notice that given the generation function, by performing inverse transform the distribution of the AoI is actually determined. As the specific examples, the generation functions of AoI and peak AoI of system with G/G/1 queue were given explicitly. Recently, using the queueing theory methods, the discrete AoI and peak AoI distributions are numerically determined in series work \cite{29,30,31}. First of all, in work \cite{29} the theory of Markov Fluid Queues (MFQ) is used. The authors proposed a numerical algorithm which can be used to find the exact distributions of both the AoI and the peak AoI for the bufferless PH/PH/1/1/$P(p)$ queue with probabilistic preemption, where the preemption probability is $p$; and the single buffer M/PH/1/2/$R(r)$ queue with probabilistic replacement of the packet in queue with replacement probability $r$. The distributions of AoI and peak AoI were discussed in \cite{30}, in which the authors numerically determined those distributions in matrix-geometric form for three queueing disciplines, that is Non-Preemptive Bufferless, Preemptive Bufferless, and Non-Preemptive Single Buffer with Replacement. Finally, for an updating system with multiple independent sources, in paper \cite{31} the AoI and peak AoI distributions were calculated for each information source using matrix-analytical algorithms along with the theory of Markov fluid queues and sample path arguments. Apart from this, we note that in another paper \cite{32}, under a simple setting where the packet loss in each forward transmission is identical, the explicit probability mass function of the AoI for an $N$-hop line network was obtained.

In this paper, we solve the discrete counterpart of the packet management problem. More precisely, we determine the stationary AoI distribution for the system with Ber/G/1/1, Ber/Geo/1/2 and Ber/Geo/1/2* queues, thus obtain the complete description of steady state AoI for these cases. The essential idea is describing the random transtions of two-dimensional vector $(n,m)$ which consists of AoI $n$ and the age of packet $m$ in transmitter for the size 1 system, and as a whole characterizing the evolutions of three-dimensional stochastic vector $(n,m,l)$, including AoI $n$, age of packet $m$ which is in service, and the age of the waiting packet $l$ in queue, when the size of status updating system is 2. Then, define the corresponding two-dimensional and three-dimensional AoI stochastic process and determine their steady state by solving the so-called stationary equaitons. As long as all the stationary probabilities are obtained, the stationary AoI distribution is determined as well, which is actually one of the marginal distributions of the defined AoI process. Follow this line of thinking, eventually we obtain the explicit AoI distribution expressions for the system having all of three queue models. Notice that to find the distribution of AoI for each queue, in fact we calculate the stationary distribution of a larger-dimensional stochastic process. Therefore, apart from the AoI distribution, we obtain more. Especially for the size 2 status updating system, we show that the parameters $m$ and $l$ record the packet system time and packet waiting time, thus their distributions in steady state can be obtained as other two marginal distributions. The AoI violation probability is of interest in some articles, so we also calculate it along with its normalized negative exponent in this paper, which is defined as the the quality of service exponent (QoE) \cite{33} by some researchers.

The rest part of the paper is organized as follows. In Section II, we describe the status updating system and give the necessary definitions. In Section III, the AoI of size 1 status updating system is analyzed. Assume that the packet service can be preempted by fresher packets with certain probabilities, we derive the general expression of AoI distribution for the system with Ber/G/1/1 queue. For a set of given preemption probabilities $\tilde g(m)$, $m\geq1$, we determine the explicit formula of the stationary AoI distribution. In addition, we show that when all the preemption probabilities are cancelled and let service time is geometrically distributed, the queue model reduces to Ber/Geo/1/1, thus the AoI distribution of system with Ber/Geo/1/1 queue can be obtained easily from the general formula we have obtained before. The size 2 status updating system is considered in Section IV and Section V. Let the queue model is Ber/Geo/1/2, we calculate the AoI distribution in the first part of Section IV where a three-dimensional stochastic process is defind. As byproducts, the distributions of packet system time and packet waiting time are also obtained. Furthermore, in the last part of this Section, we discuss how the AoI distribution is determined when the queue model is generalized to Ber/Geo/1/$c$, where $c<\infty$ is an arbitrary integer. System having Ber/Geo/1/2* queue case is analyzed in Section V. The AoI distribution is given and the other two distributions are also calculated. Numerical simulations are placed in Section VI, in which we illustrate the AoI distribution curves assuming various queue models are used in the status updating system. Besides, stationary distributions of packet system time and packet waiting time are also depicted, along with the normalized negative exponent of AoI violation probabilities for system with three queue models, i.e., Ber/Geo/1/1/, Ber/Geo/1/2, and Ber/Geo/1/2*. We summarize the paper in Section VI, make some discussions and also propose the possible future work.

\section{System Model and Problem Formulation}
A status updating system consists of a source node $s$ and a receiver node $d$. The source observes one time-varying physical process continuously and samples the process at random times. An updating packet is composed of the sampled data and timestamp, and is then delivered to the destination via the system transmitter, which is modeled by a queue. In general, there may be some packets waiting in queue when the server is currently busy. After consuming a random period of time in server, a packet is sent out and obtained at the receiver immediately.

\begin{figure}[!t]
\centering
\includegraphics[width=5in]{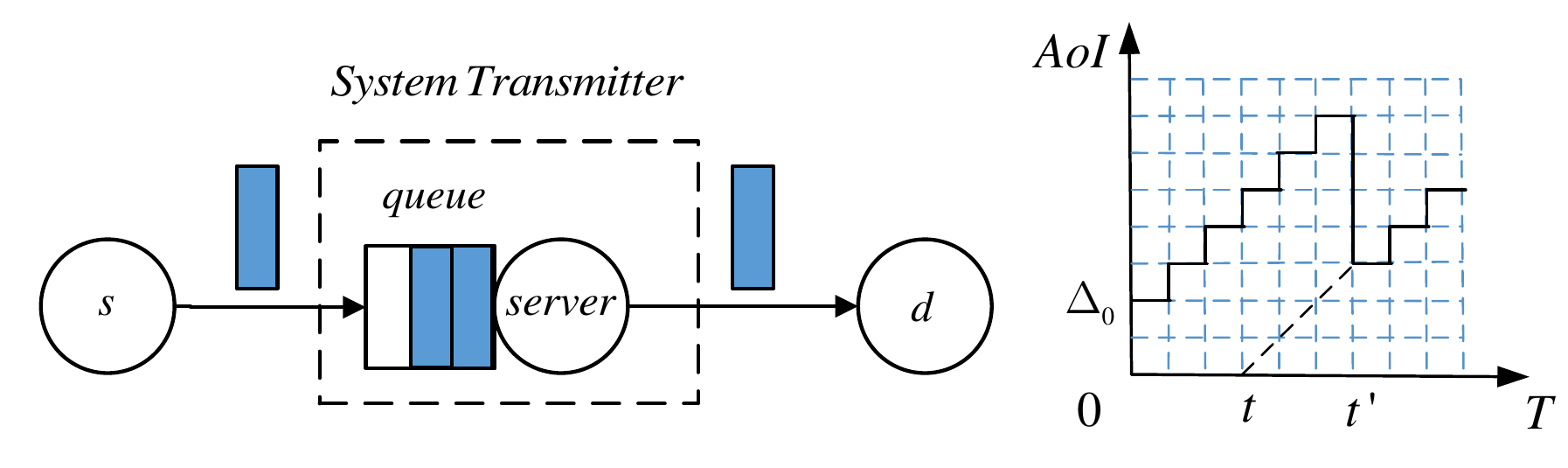}
\caption{One status updating system and an example of discrete AoI process.}
\label{fig1}
\end{figure}

The age of information (AoI) of a packet is the time a packet experiences since it was generated at the source. At the destination node, if no updating packet is obtained, the value of AoI increases constantly. Every time when a new packet arrives to receiver, then the AoI reduces to the system time of the packet, which is equal to the sum of waiting time in queue and service time in system server. We depict the model of a status updating system and offer an example sample path of the discrete AoI process in Figure \ref{fig1}.

In the discrete time model, we first give the definition of the long term average AoI. 
\begin{Definition}
Let time be discrete, the limiting time average AoI $\overline{\Delta}$ is defined as  
\begin{equation}
\overline{\Delta}=\lim_{T\to \infty}\frac1T\sum\nolimits_{k=1}^{T}a(k)   
\end{equation}
where $a(k)$ is the value of AoI at the $k$th time slot. 
\end{Definition}

We can rewrite the expression (1) as follows. 
\begin{align}
\overline{\Delta}&=\lim_{T\to \infty}\frac1T\sum\nolimits_{k=1}^{T}a(k) \notag \\
&=\lim_{T\to \infty}\frac1T\sum\nolimits_{n=1}^{M} n\cdot |\{k:a(k)=n\}|   \\
&=\sum\nolimits_{n=1}^{\infty} n\cdot \pi_{n}  \notag 
\end{align}
where 
\begin{equation}
\pi_{n}=\lim_{T\to\infty}\frac{|\{k:a(k)=n\}|}{T}   \notag 
\end{equation}
denotes the probability the AoI equals $n$ when the observation time $T$ tends to $\infty$. The number $M$ in equation (2) is defined to be $M=\max_{1\leq k\leq T}a(k)$.

Assume that the discrete AoI process is ergodic, and after sufficiently long time, the process reaches the steady state. Notice that the probabilities $\{\pi_n,n\geq 1\}$ form the stationary distribution of the discrete AoI of the system. As we mentioned before, in order to obtain the stationary AoI-distribution, not the AoI itself but a stochastic process of larger dimensional is considered, where the AoI is one dimensional of this extended process. Thus, when the steady state of the large process is solved, as the marginal distribution, the distribution of AoI is determined as well.

Following this line of thinking, in the following Sections we shall analyze and derive the stationary AoI-distribution of the status updating system with different packet management policies. Firstly, the system of size 1 is considered in Section III.

\section{Distribution of Stationary AoI: Probabilistic preemption Ber/G/1/1 queue}
In this Section, for the system of size 1, we compute the stationary distribution of the AoI. In this case, system using a more complex queue model called probabilistic preemption Ber/G/1/1 queue is discussed, where the service of a packet can be preempted by the fresher ones in a random manner. We first derive the general formula of the AoI-distribution assuming that the Ber/G/1/1 queue is used and an arbitrary group of preemption probabilities are defined. Next, in the second part of this Section, for the system with Ber/Geo/1/1 queue and the given preemption probabilities, we figure out the explicit expressions of the stationary AoI-distribution, which can be further reduced to the AoI-distribution of system having basic Ber/Geo/1/1 queue by letting all the preemption probabilities be zero.

\subsection{Distribution of the AoI: general formula}
Assume that in each time slot an updating packet is generated independently and with identical probability $p$. Equivalently, the random interarrival time $A$ between succesive packets is distributed as 
\begin{equation}
\Pr\{A=j\}=(1-p)^{j-1}p  \qquad  (j\geq1)  \notag  
\end{equation}
which is a geometric distribution with parameter $p$. On the other hand, the packet service time can be an arbitrary random variable. Observing that the settings are more general than the system with M/M/1/1 queue in continuous time model. In order to enhance the timeliness of the system further, we also assume that the service of a packet can be preempted by fresher ones with certain preemption probabilities.

Invoking a two-dimensional state vector $(n_k,m_k)$, $n_k> m_k\geq0$, where we define $n_k$ as the AoI at the receiver and let $m_k$ be the packet age at $k$th time slot. Constituting the two-dimensional AoI stochastic process $AoI_{Ber/G/1/1/g(m)}$ as 
\begin{equation}
AoI_{Ber/G/1/1/g(m)}=\{(n_k,m_k), n_k> m_k\geq0, k\in \mathbb{N}\}  \notag  
\end{equation}
where $g(m)$, $m\geq1$ is the service preemption probability and is defined as  
\begin{equation}
g(m)=\Pr\{\text{a fresher packet preempts the packet of age $m$}\}   \notag  
\end{equation}
which in general is an increasing function of packet age $m$.

Next, we analyze the random transfers of the age-states $(n,m)$, $n>m\geq0$, establish the stationary equations, and find the steady state of the process $AoI_{Ber/G/1/1/g(m)}$ in the end. Since the first component of the state vector, $n$, records the AoI along with time, then the probability distribution of the AoI can be obtained as marginal distribution of the first age-component.

Notice that the terms state vector, age-state, and simply state are used interchangeably throughout the rest of paper. We use $A$ to denote the interarrival time between successive updating packets and the packet service time is represented by $B$. To analyze the possible state transitions of $(n,m)$ from current time slot to next one, at most we have to consider three events simultaneously, that is, the random packet arrivials; whether or not the service is preempted; and if the service of the packet is complete.

First of all, assume that there is one packet in system such that the system is full, since the size of system is supposed to be 1. In this case, the state $(n,m)$ satisfies $n>m\geq1$. If no packet arrives at next time slot, then the state vector can only jump to either $(n+1,m+1)$ or $(m+1,0)$, depending on whether the packet service is complete. Since the service preemption is allowable, when a fresher packet comes but does not replace the old one, the analysis is identical to that we have discussed above. However, if a newer packet arrives and service preemption occurs, then with probability $\Pr\{B>1\}$ the state $(n,m)$ transfers to $(n+1,1)$. Otherwise, we show that the state vector $(1,0)$ is obtained at next time slot. Since if the arriving packet experiences only single time slot service in system’s server, then a packet of age 1 will arrive to receiver node. As a result, the AoI is reduced to 1 at next time slot.

Let $P_{(n,m),(i,j)}$ be the transition probability that the age-state transfers to $(i,j)$ at the next time slot from the current state $(n,m)$. Summarize above discussions, we have following transition probabilities  
\begin{align}
P_{(n,m),(n+1,m+1)}&=(1-p)\Pr\{B>m | B>m-1\}  + p\left[1-g(m)\right] \Pr\{B>m | B>m-1\}  \notag \\ 
&=\left[1-pg(m)\right] \Pr\{B>m | B>m-1\} 
\end{align}

\begin{align}
P_{(n,m),(m+1,0)}&= (1-p)\Pr\{B=m | B>m-1\} + p\left[1-g(m)\right] \Pr\{B=m | B>m-1\}  \notag \\ 
&= \left[1-pg(m)\right] \Pr\{B=m | B>m-1\}   
\end{align}

\begin{equation}
P_{(n,m),(n+1,1)} = pg(m)\Pr\{B>1\}  \notag 
\end{equation}
and
\begin{equation}
P_{(n,m),(1,0)} = pg(m)\Pr\{B=1\}  \notag 
\end{equation}

We use conditional probability in equations (3) because if the service time $B$ is greater than $m$ time slots, we must know the event $\{B>m-1\}$. The reason of applying conditional probability in equation (4) is the same.

Then, consider the case the system is empty. For this case, the state vector is represented as $(n,0)$, $n\geq1$. The analysis of state transitions is simpler, since there is no need to consider the probabilistic service preemption. It is easy to see that the age-state jumps to $(n+1,0)$ if no packet arrives to transmitter in the next time slot. Otherwise, when such a new packet comes, depending on the service time $B$ equal to or greater than 1, the state vector will transfer to $(1,0)$, or $(n+1,1)$, respectively. These transition probabilities are determined as 
\begin{equation}
P_{(n,0),(n+1,0)} = 1-p  \notag 
\end{equation}

\begin{equation}
P_{(n,0),(n+1,1)} = p\Pr\{B>1\}  \notag 
\end{equation}
and 
\begin{equation}
P_{(n,0),(1,0)} = p\Pr\{B=1\}  \notag 
\end{equation}

\begin{Remark}
Notice that in discrete time model, a state vector maintains in entire time slot. For example, for the state vector $(n,m)$ at the $k$th time slot, strictly speaking, the packet age is always $m$ in the left closed right open interval $[k-1,k)$. As a result, if we observe the state at the left-endpoint of the time slot, the packet age is $m$; while if the state is recorded near the right-endpoint, then actually the age of the packet is approximately $m+1$. This inconsistency will cause small confusion about the accurate service time when we call “the service is complete at $k$th time slot”. Therefore, we use the convention that the state vectors are all considered at the left-endpoint of the time slot, such that when the state $(n,m)$ transfers to $(m+1,0)$, we agree that the service time is $m$.  
\end{Remark}

So far, all the state transfers are analyzed and the corresponding transition probabilities are determined. Define $\pi_{(n,m)}$, $n>m\geq0$ be the stationary probability of age-state $(n,m)$. Assume that the AoI process $AoI_{Ber/G/1/1/g(m)}$ reaches the steady state, we give the stationary equations of the process in following Theorem 1. 
\begin{Theorem}
When the stochastic process $AoI_{Ber/G/1/1/g(m)}$ runs in steady state, the stationary probabilities $\pi_{(n,m)}$, $n>m\geq0$ satisfy the following equations 
\begin{equation}
\begin{cases}
\pi_{(n,m)}=\pi_{(n-1,m-1)}[1-pg(m-1)] \Pr\{B>m-1|B>m-2\}  &  (n>m\geq2)   \\
\pi_{(n,1)}=\pi_{(n-1,0)}p\Pr\{B>1\} + \left(\sum\nolimits_{j=1}^{n-2}\pi_{(n-1,j)}pg(j) \right) \Pr\{B>1\}  &  (n\geq3)  \\
\pi_{(2,1)}=\pi_{(1,0)}p\Pr\{B>1\}    \\
\pi_{(n,0)}=\pi_{(n-1,0)}(1-p) +  \left(\sum\nolimits_{k=n}^{\infty}\pi_{(k,n-1)}\right) [1-pg(n-1)]\Pr\{B=n-1|B>n-2\}  &  (n\geq2)  \\
\pi_{(1,0)}=\left(\sum\nolimits_{n=1}^{\infty}\pi_{(n,0)}\right)p\Pr\{B=1\}+ \left(\sum\nolimits_{n=2}^{\infty}\sum\nolimits_{m=1}^{n-1}\pi_{(n,m)}pg(m)\right)\Pr\{B=1\} 
\end{cases}
\end{equation}
\end{Theorem}

\begin{proof}
We explain each line of equaitons (5) as follows. First of all, observing that a packet of age $m$ in system’s server is replaced by a fresher one with probability $pg(m)$, then the probability that the age state $(n-1,m-1)$ transfers to $(n,m)$ equals $[1-pg(m-1)]$ multiplying $\Pr\{B>m-1|B>m-2\}$, i.e., the probability that the packet service is not complete. This gives the first line of (5). 

The state $(n,1)$ is obtained from $(n-1,j)$, $0\leq j\leq n-2$, if an updating packet arrives at the next time slot, replaces the old one, and the service time is greater than 1. Notice that when $j=0$, the system is empty such that there is no service preemptions. Thus, the second equation in (5) is obtained.  

To get the state vector $(n,0)$, it is easy to see that as long as no packet arrives, age-state $(n-1,0)$ changes to $(n,0)$ in next time slot. On the other hand, let the current state be $(k,n-1)$, $k\geq n$, assume that in the next time slot the service is complete, we can also obtain the state $(n,0)$. Since replacing a new packet does not cut the number of packet in system, so that what we discussed above have covered all the cases that can transfer to $(n,0)$. 

At last, beginning with any state $(n,m)$, a new packet arrives, preempts the old one, and finishes the service in single time slot will make the age state transfer to $(1,0)$. Similarly, when the system is originally empty, the preemption of packet service is nonexistent. Therefore, we prove the last line of (5), i.e., the stationary equation for state vector $(1,0)$ and complete the proof of Theorem 1.  
\end{proof}

We give the solution of the system of equations (5) in Theorem 2, the proof of which is postponed to Appendix A. 

For convenience, the distribution of the packet service time $B$ is denoted as $\Pr\{B=j\}=q_j$, $j\geq1$. 
\begin{Theorem}
Assume that the AoI process $AoI_{Ber/G/1/1/g(m)}$ reaches the steady state, the stationary probabilities $\pi_{(n,m)}$, $n>m\geq0$ are determined as follows. 

Firstly, for $n\geq1$,  
\begin{equation}
\pi_{(n,0)}=\pi_{(1,0)}\left\{(1-p)^{n-1} + \frac{1-q_1}{q_1}\bigg[\sum\nolimits_{j=0}^{n-2}(1-p)^j  \left(\prod\nolimits_{l=1}^{n-1-j}[1-pg(l)] \right)q_{n-1-j}\bigg] \right\} 
\end{equation}
where $\pi_{(1,0)}$ is determined by 
\begin{align}
\frac{1}{\pi_{(1,0)}}&=\frac{1}{p}+\frac{1-q_1}{q_1}\sum\nolimits_{n=2}^{\infty}\bigg[\sum\nolimits_{j=0}^{n-2}(1-p)^j  \left( \prod\nolimits_{l=1}^{n-1-j}[1-pg(l)]\right)q_{n-1-j} \bigg]   \notag \\
& \qquad \qquad \qquad \qquad  \quad + \frac{1-q_1}{q_1} \left[\sum\nolimits_{m=1}^{\infty}\left(\prod\nolimits_{l=1}^{m-1}[1-pg(l)]\right)\left(\sum\nolimits_{l=m}^{\infty}q_l\right)  \right] 
\end{align}

The probabilities $\pi_{(n,m)}$, $n>m\geq1$ can be obtained from the relation 
\begin{equation}
\pi_{(n,m)} = \pi_{(n-m+1,1)}\left(\prod\nolimits_{l=1}^{m-1}[1-pg(l)]\right) \left(\sum\nolimits_{l=m}^{\infty}q_l\right)  
\end{equation}

The probabilities $(n,1)$, $n\geq3$ are derived by solving the following difference equation
\begin{equation}
\pi_{(n,1)}=\pi_{(n-1,0)}p(1-q_1) + p(1-q_1) \left[\sum\nolimits_{j=1}^{n-2}\pi_{(n-j,1)} \left(\prod\nolimits_{l=1}^{j-1}[1-pg(l)]\right)\left(\sum\nolimits_{l=j}^{\infty}q_l\right) g(j) \right]
\end{equation}

In above equations, denote that $\prod\nolimits_{l=1}^{0}[1-pg(l)]=1$. 
\end{Theorem}

Since the first state component $n$ represents the AoI at the receiver, then we have  
\begin{equation}
\Pr\{\Delta=n\}=\sum\nolimits_{m=0}^{n-1}\pi_{(n,m)} \quad (n\geq1) 
\end{equation}
which form the stationary distribution of the AoI. In addition, the AoI-cumulative probabilities are calculated as 
\begin{equation}
\Pr\{\Delta \leq k\}=\sum\nolimits_{n=1}^{k}\sum\nolimits_{m=0}^{n-1}\pi_{(n,m)} \quad (k\geq1)
\end{equation}

Equations (6)-(9) give the solution of (5) for the general service time distribution and preemption probabilities $g(m)$, $m\geq1$. In next subsection, assume that the service time $B$ is also geometrically distributed, and for a specific preemption function $\tilde g(m)$, we determine all the stationary probabilities $\pi_{(n,m)}$. Then, according to equations (10) and (11), the explicit formulas of the AoI-distribution and its cumulative probability distribution can be obtained.

\subsection{Explicit expression of AoI-distribution: status updating system with Ber/Geo/1/1/$\tilde g(m)$ queue}
In this subsection, assume that $B$ is a geometric random variable with service intensity $\gamma$ where $p<\gamma$, i.e., the queue model used in status updating system is specialized to Ber/Geo/1/1. We compute the explicit expressions of the probabilities $\pi_{(n,m)}$ by solving the stationary equations of corresponding AoI stochastic process, so that the specific formula of stationary AoI-distribution can be obtained. Here, suppose that the preemption probabilities are defined as 
\begin{equation}
\tilde g(m)=\frac{1}{p}\left[1-(1-p)\frac{N_p+m+1}{N_p+m}\right]   \quad  (m\geq1)  
\end{equation}
where $N_p$ is chosen as the minimum integer that makes $p > 1/(N_p+m+1)$ for any $m\geq1$. 

Notice that $\tilde g(m)$ is an increasing function of $m$, and tends to 1 when $m\to\infty$. To ensure that $\tilde g(m)>0$ for any $m\geq1$ and any packet arriving rate $p$, it requires 
\begin{equation}
1-(1-p)\frac{N_p+m+1}{N_p+m}>0 \text{ for any } m\geq1  \notag 
\end{equation}
which is equivalent to 
\begin{equation}
p> \frac{1}{N_p+m+1} \text{ for all }m\geq1  \notag 
\end{equation}

Observing that for any $p>0$, there always exists $N_p\in \mathbb{N}$, such that $p> 1/(N_p+m+1)$, $m\geq1$ holds.

Applying these assumptions, the stationary equations of the stochastic process $AoI_{Ber/Geo/1/1/\tilde g(m)}$ are written as  
\begin{equation}
\begin{cases}
\pi_{(n,m)}=\pi_{(n-1,m-1)}(1-p) \frac{N_p+m}{N_p+m-1}(1-\gamma)   &   (n>m\geq2) \\
\pi_{(n,1)}=\pi_{(n-1,0)}p(1-\gamma) + \left(\sum\nolimits_{j=1}^{n-2}\pi_{(n-1,j)}\left[1-(1-p)\frac{N_p+j+1}{N_p+j} \right]  \right)(1-\gamma)  &  (n\geq3)  \\
\pi_{(2,1)}=\pi_{(1,0)}p(1-\gamma)  \\
\pi_{(n,0)}=\pi_{(n-1,0)}(1-p) + \left(\sum\nolimits_{k=n}^{\infty}\pi_{(k,n-1)}\right) (1-p)\frac{N_p+n}{N_p+n-1}\gamma  &   (n\geq2) \\
\pi_{(1,0)}=\left( \sum\nolimits_{n=1}^{\infty}\pi_{(n,0)} \right)p\gamma + \left(\sum\nolimits_{n=2}^{\infty}\sum\nolimits_{m=1}^{n-1}\pi_{(n,m)}\left[1-(1-p)\frac{N_p+m+1}{N_p+m}\right] \right)\gamma  
\end{cases}
\end{equation}

The probabilities $\pi_{(n,m)}$, $n>m\geq0$ can be obtained by solving the system of equations (13), or directly calculating the general expressions (6)-(9). We give the final results in Theorem 3 below, and put the long calculations in Appendix B.     
\begin{Theorem}
Let the AoI process $AoI_{Ber/Geo/1/1/\tilde g(m)}$ reach the steady state, the stationary probabilities are calculated as follows. Firstly, we show that 
\begin{equation}
\pi_{(n,0)}=\xi_1(1-p)^{n-1} -\xi_2\left[1+(N_p+n)\gamma\right]\left[(1-p)(1-\gamma)\right]^{n-1}  \quad (n\geq1)
\end{equation}
in which $\xi_1$ and $\xi_2$ are defined as 
\begin{equation}
\xi_1=\frac{(N_p\gamma+1)p(p+\gamma-p\gamma)}{N_p\gamma(p+\gamma-p\gamma)+\gamma}, 
\qquad
\xi_2=\frac{p(1-\gamma)(p+\gamma-p\gamma)}{N_p\gamma(p+\gamma-p\gamma)+\gamma} \notag 
\end{equation}

The other probabilities $\pi_{(n,m)}$, $n>m\geq1$ are obtained by using the relation 
\begin{equation}
\pi_{(n,m)}=\pi_{(n-m+1,1)}\left[(1-p)(1-\gamma)\right]^{m-1}\frac{N_p+m}{N_p+1}  
\end{equation}
where $\pi_{(n,1)}$ are given as 
\begin{equation}
\pi_{(n,1)}=\begin{cases}
\xi_1p\gamma^2(1-\gamma)   & \quad   (n=2) \\
c_1(1-p)^n+c_2(1-\gamma)^n +c_3\left(\frac{N_p(1-p)(1-\gamma)}{N_p+1}\right)^n   & \quad  (n\geq3)
\end{cases}
\end{equation}

According to three initial probabilities of $\pi_{(k,1)}$, $k=3,4,5$, the undetermined coefficients $c_1$, $c_2$, and $c_3$ can be determined by finding the solution of following system of equations 
\begin{equation}
\pi_{(k,1)}=c_1(1-p)^k+c_2(1-\gamma)^k +c_3\left(\frac{N_p(1-p)(1-\gamma)}{N_p+1}\right)^k \qquad (k=3,4,5) \notag 
\end{equation}
\end{Theorem}

\begin{Remark}
In order to obtain the probabilities $\pi_{(n,1)}$, $n\geq3$, we have to solve the difference equation (9). For the case considered in this subsection, we prove that (9) can be reduced to an order three difference equation with constant coefficients in Appendix B, whose characterization equation has three roots, i.e., $r_1=1-p$, $r_2=1-\gamma$, and $r_3=\frac{N_p(1-p)(1-\gamma)}{N_p+1}$. Thus, the general expression of probability $\pi_{(n,1)}$ is written as equation (16), where three coefficients $c_1$, $c_2$, and $c_3$ are unknown. Finding these numbers are not difficult, but the exact expressions are very complex. Therefore, we do not calculate these undetermined coefficients further, while directly determining their values when the numerical results of AoI-distribution are considered in Section VI.  
\end{Remark}

It is not hard to determine a few of initial probabilities of $\pi_{(n,1)}$, $n\geq3$. For example, from the second line of (13), we have 
\begin{align}
\pi_{(3,1)}&=\pi_{(2,0)}p(1-\gamma) + \pi_{(2,1)}\left[1-(1-p)\frac{N_p+2}{N_p+1}\right](1-\gamma)  \notag \\
&=\pi_{(2,0)}p(1-\gamma) + \pi_{(1,0)}p(1-\gamma) \left(p-\frac{1-p}{N_p+1}\right)(1-\gamma)  \notag \\
&=\pi_{(2,0)}p(1-\gamma) + \pi_{(1,0)}p(1-\gamma)^2 \left(p-\frac{1-p}{N_p+1}\right)  \notag 
\end{align}
where $\pi_{(1,0)}$ and $\pi_{(2,0)}$ are determined by equation (14).

\begin{Corollary}
For the status updating system with probabilistic preemption Ber/Geo/1/1/$\tilde g(m)$ queue, the distribution of AoI is given as 
\begin{equation}
\Pr\{\Delta=1\}=\pi_{(1,0)}=\frac{(N_p+1)p\gamma(p+\gamma-p\gamma)}{N_p(p+\gamma-p\gamma)+1} 
\end{equation}
and for $n\geq2$, the probability that the AoI takes value $n$ is equal to 
\begin{align}
\Pr\{\Delta=n\}&=\xi_1(1-p)^{n-1} - \xi_2 \bigg\{1+\gamma+(N_p+n-1)\gamma \left[1-\frac{p\gamma}{(1-p)(1-\gamma)}\right] \bigg\} \left[(1-p)(1-\gamma)\right]^{n-1} \notag \\
&\qquad + \sum\nolimits_{i=1}^{3}\frac{c_ir_i^n}{N_p+1}\bigg( \frac{N_p}{1-\delta_i} + \frac{1-[1+(1-\delta_i)(N_p+n-2)]\delta_i^{n-2}}{(1-\delta_i)^2} \bigg)  \qquad  (n\geq2)
\end{align}
where $\delta_i=(1-p)(1-\gamma)/r_i$, $i=1,2,3$. 
\end{Corollary}

\begin{proof}
Given all the probabilities $\pi_{(n,m)}$, the stationary distribution of the AoI is obtained by calculating equation (10). Firstly, when $n=1$, it shows that 
\begin{equation}
\Pr\{\Delta=1\}=\pi_{(1,0)}=\xi_1-\xi_2[1+(N_p+1)\gamma] =\frac{(N_p+1)p\gamma(p+\gamma-p\gamma)}{N_p(p+\gamma-p\gamma)+1} 
\end{equation}

For $n\geq3$, we have that
\begin{align}
\Pr\{\Delta=n\}& = \pi_{(n,0)} + \sum\nolimits_{m=1}^{n-1}\pi_{(n,m)} \notag \\
&=\pi_{(n,0)}+\sum\nolimits_{m=1}^{n-1}\pi_{(n-m+1,1)}[(1-p)(1-\gamma)]^{m-1}\frac{N_p+m}{N_p+1} \notag \\
&=\pi_{(n,0)}+\sum\nolimits_{m=1}^{n-2}\pi_{(n-m+1,1)}[(1-p)(1-\gamma)]^{m-1}\frac{N_p+m}{N_p+1} + \pi_{(2,1)}[(1-p)(1-\gamma)]^{n-2}\frac{N_p+n-1}{N_p+1}  
\end{align}
where 
\begin{align}
&\pi_{(n,0)}+\pi_{(2,1)}[(1-p)(1-\gamma)]^{n-2}\frac{N_p+n-1}{N_p+1}  \notag \\
={}& \xi_1(1-p)^{n-1} - \xi_2\left[1+(N_p+n)\gamma\right]\left[(1-p)(1-\gamma)\right]^{n-1} + \frac{\pi_{(1,0)}p(1-\gamma)}{N_p+1} (N_p+n-1)[(1-p)(1-\gamma)]^{n-2}  \notag \\
={}& \xi_1(1-p)^{n-1} - \xi_2\left[1+(N_p+n)\gamma\right]\left[(1-p)(1-\gamma)\right]^{n-1} + \xi_2p\gamma^2 (N_p+n-1)[(1-p)(1-\gamma)]^{n-2}  \notag \\
={}& \xi_1(1-p)^{n-1} - \xi_2 \bigg\{1+\gamma+(N_p+n-1) \left[\gamma-\frac{p\gamma^2}{(1-p)(1-\gamma)}\right] \bigg\} \left[(1-p)(1-\gamma)\right]^{n-1} 
\end{align}

The sum in equation (20) is calculated as 
\begin{align}
&\sum\nolimits_{m=1}^{n-2}\pi_{(n-m+1,1)}[(1-p)(1-\gamma)]^{m-1}\frac{N_p+m}{N_p+1} \notag \\
={}&\sum\nolimits_{m=1}^{n-2}\left(\sum\nolimits_{i=1}^3c_ir_i^{n-m+1}\right)[(1-p)(1-\gamma)]^{m-1}\frac{N_p+m}{N_p+1}\notag \\
={}&\sum\nolimits_{i=1}^3 \frac{c_ir_i^n}{N_p+1}\sum\nolimits_{m=1}^{n-2}\left(N_p+m\right)\delta_i^{m-1} \notag \\
={}&\sum\nolimits_{i=1}^3 \frac{c_ir_i^n}{N_p+1}\bigg(\frac{N_p}{1-\delta_i} + \frac{1-[1+(1-\delta_i)(N_p+n-1)\delta_i^{n-2}]}{(1-\delta_i)^2}\bigg)
\end{align}
in which the number $\delta_i$ is defined as $(1-p)(1-\gamma)/r_i$, $i=1,2,3$. 

Substituting equations (21) and (22) into (20), we obtain the expression (18). 

To complete the proof, we have to check that (18) is also valid for the case $n=2$. It suffices to prove that the sum (22) is zero when $n=2$, since the equation (21) reduces to $\pi_{(2,0)}+\pi_{(2,1)}$, which is exactly the probability the AoI equals 2, if we let $n=2$. It can be verified directly that (22) is zero for the case $n=2$. Thus, we completes the proof of Corollary 1. 
\end{proof}

\subsection{AoI-distribution: from system with Ber/G/1/1/g(m) queue to the case system using Ber/G/1/1 and Ber/Geo/1/1 queues}
In particular, let the preemption function be zero at all $m$, i.e., in fact there is no service preemption, the conclusions in Theorem 2 and Theorem 3 are reduced to that of AoI process where the queue model used in system is Ber/G/1/1, and is Ber/Geo/1/1 when the explicit AoI-distribution is considered. Therefore, the analysis of AoI for the system with probabilistic preemption Ber/G/1/1 queue is more general.

When the service preemption is cancelled, the stationary equations (5) and (13) are greatly simplified. As a result, it is easier for us to solve the steady state of the AoI process, and then find the stationary distribution of the AoI. 

For the system with basic Ber/G/1/1 queue, we show that the stationary equations of AoI process $AoI_{Ber/G/1/1}$ are written as  
\begin{equation}
\begin{cases}
\pi_{(n,m)}=\pi_{(n-1,m-1)}\Pr\{B>m-1|B>m-2\}   &   (n>m\geq2)   \\
\pi_{(n,1)}=\pi_{(n-1,0)}p\Pr\{B>1\}   \\
\pi_{(n,0)}=\pi_{(n-1,0)}(1-p) + \left( \sum\nolimits_{k=n}^{\infty}\pi_{(k,n-1)}\right)  \Pr\{B=n-1|B>n-2\} &  (n\geq2)  \\
\pi_{(1,0)}=\left( \sum\nolimits_{k=1}^{\infty}\pi_{(k,0)}\right)p\Pr\{B=1\}  
\end{cases}
\end{equation}

Because the packet in service cannot be preempted, the age state $(n,1)$ can only be obtained from the empty state $(n-1,0)$. The second equation in (23) dramatically simplifies the calculation of the stationary probabilities $\pi_{(n,m)}$, since we can apply the first line of (23) repeatedly until $m$ is reduced to zero, i.e., we can write that 
\begin{equation}
\pi_{(n,m)}=\pi_{(n-m,0)}p\Pr\{B>1\}\Pr\{B>m-1\}  
\end{equation}
so that to obtain all the stationary probabilities, it suffices to find $\pi_{(n,0)}$, $n\geq1$.

In the following, we directly give the final results without proofs, since the analysis methods and calculation procedures are the same as that we have used for the more general case in previous subsections.  
\begin{Theorem}
For the discrete time status updating system with Ber/G/1/1 queues, the stationary probabilities for all the state vectors $(n,m)$ are given as follows. Firstly, for $n\geq1$, it shows that 
\begin{equation}
\pi_{(n,0)}=\frac{pq_1F(p,n)}{1+p(1-q_1)\sum\nolimits_{m=1}^{\infty}\sum\nolimits_{l=m}^{\infty}q_l} 
\end{equation}
and the probabilities $\pi_{(n,m)}$, $n>m\geq1$ are determined to be  
\begin{equation}
\pi_{(n,m)}=\frac{p^2q_1(1-q_1)F(p,n-m)\left(\sum\nolimits_{l=m}^{\infty}q_l \right)}{1+p(1-q_1)\sum\nolimits_{m=1}^{\infty}\sum\nolimits_{l=m}^{\infty}q_l} 
\end{equation}
in which $F(p,n)$ is defined as 
\begin{equation}
F(p,n)=(1-p)^{n-1} +\frac{1-q_1}{q_1}\left( \sum\nolimits_{j=0}^{n-2}(1-p)^{j}q_{n-1-j}\right)  \quad (n\geq 1)
\end{equation}
\end{Theorem}

\begin{Corollary}
For the status updating system with basic Ber/G/1/1 queue, assume that the AoI process runs in steady state, the probability that the AoI takes value $n$, i.e., the stationary AoI-distribution is given as 
\begin{align}
\Pr\{\Delta=n\} &= \pi_{(n,0)}+ \sum\nolimits_{m=1}^{n-1}\pi_{(n,m)} \notag \\
&=\frac{pq_1F(p,n)+p^2q_1(1-q_1)\sum\nolimits_{m=1}^{n-1}F(p,n-m)\left(\sum\nolimits_{l=m}^{\infty}q_l\right)}{1+p(1-q_1)\sum\nolimits_{m=1}^{\infty}\sum\nolimits_{l=m}^{\infty}q_l}
\end{align}

For $k\geq1$, the cumulative probability distribution of the AoI is written as 
\begin{align}
\Pr\{\Delta \leq k \} &= \sum\nolimits_{n=1}^{k}\Pr\{\Delta=n\}  \notag   \\
&= \sum\nolimits_{n=1}^k  \frac{pq_1F(p,n) + p^2q_1(1-q_1)\sum\nolimits_{m=1}^{n-1}F(p,n-m)\left(\sum\nolimits_{l=m}^{\infty}q_l\right)}{1+p(1-q_1)\sum\nolimits_{m=1}^{\infty}\sum\nolimits_{l=m}^{\infty}q_l} 
\end{align}
\end{Corollary}

\begin{Theorem}
When the queue model of the system is specialized to Ber/Geo/1/1, the stationary probabilities of all the age-states $(n,m)$ are given as 
\begin{equation}
\begin{cases}
\pi_{(n,0)}=\frac{p\gamma ^2 }{(p+\gamma -p\gamma)(\gamma -p)} \left[(1-p)^n - (1-\gamma)^n \right]   &   (n\geq 1) \\
\pi_{(n,m)}=\frac{(p\gamma )^2}{(p+\gamma -p\gamma)(\gamma -p)}  \left[(1-p)^{n-m}(1-\gamma)^{m}-(1-\gamma)^n \right] & (n>m \geq 1)
\end{cases}
\end{equation}
\end{Theorem}

\begin{Corollary}
The distribution of AoI for the status updating system with Ber/Geo/1/1 queue is given as 
\begin{equation}
\Pr\{\Delta=n\}= \frac{p(1-p)\gamma ^3}{(p+\gamma -p\gamma)(\gamma -p)^2}\left[(1-p)^n-(1-\gamma)^n \right] - \frac{(p\gamma)^2}{(p+\gamma -p\gamma)(\gamma -p)}n(1-\gamma)^n  \qquad (n \geq 1) 
\end{equation}

In addition, for $k\geq 1$, we have 
\begin{align}
\Pr\{\Delta \leq k\} &= \frac{(1-p)\gamma^2 \left[(\gamma -p)-(1-p)^{k+1}\gamma +p(1-\gamma)^{k+1} \right] }{(p+\gamma -p\gamma)(\gamma -p)^2}  - \frac{p^2 \left[(1-\gamma)-(1+k\gamma)(1-\gamma)^{k+1} \right] }{(p+\gamma -p\gamma)(\gamma -p)} \notag \\
&=1 - \frac{ (1-p)^{k+2}\gamma^3- \Big\{ p(1-p)\gamma^2 + p^2(\gamma - p) \Big\} (1- \gamma)^{k+1} - p^2\gamma(\gamma-p)k(1-\gamma)^{k+1} }{(p+\gamma-p\gamma)(\gamma-p)^2}  
\end{align}
\end{Corollary}

Let $P_{v,k}$ be the AoI violation probability, i.e., the probability that the AoI is greater than $k$. The exponent of $P_{v,k}$ is defined as the quality of exponent (QoE) of the status updating system by some researchers and is of interest in some articles. From equation (32), this exponent can be determined exactly when the queue model of the system is Ber/Geo/1/1.  

Denote $E_{v,k}$ be the QoE of the system, then 
\begin{align}
E_{v,k}&=-\frac{1}{k}\log \Pr\{\Delta>k\}  \notag \\
&=-\frac{1}{k}\log \frac{ (1-p)^{k+2}\gamma^3- \Big\{ p(1-p)\gamma^2 + p^2(\gamma - p) \Big\} (1- \gamma)^{k+1} - p^2\gamma(\gamma-p)k(1-\gamma)^{k+1} }{(p+\gamma-p\gamma)(\gamma-p)^2}  \\
&>-\frac{1}{k}\log \frac{(1-p)^2\gamma^3}{(p+\gamma-p\gamma)(\gamma-p)^2}(1-p)^k  \\
&=\log\frac{1}{1-p} - \frac{1}{k}\log \frac{(1-p)^2\gamma^3}{(p+\gamma-p\gamma)(\gamma-p)^2}
\end{align}

\begin{figure}[!t]
\centering
\includegraphics[width=4in]{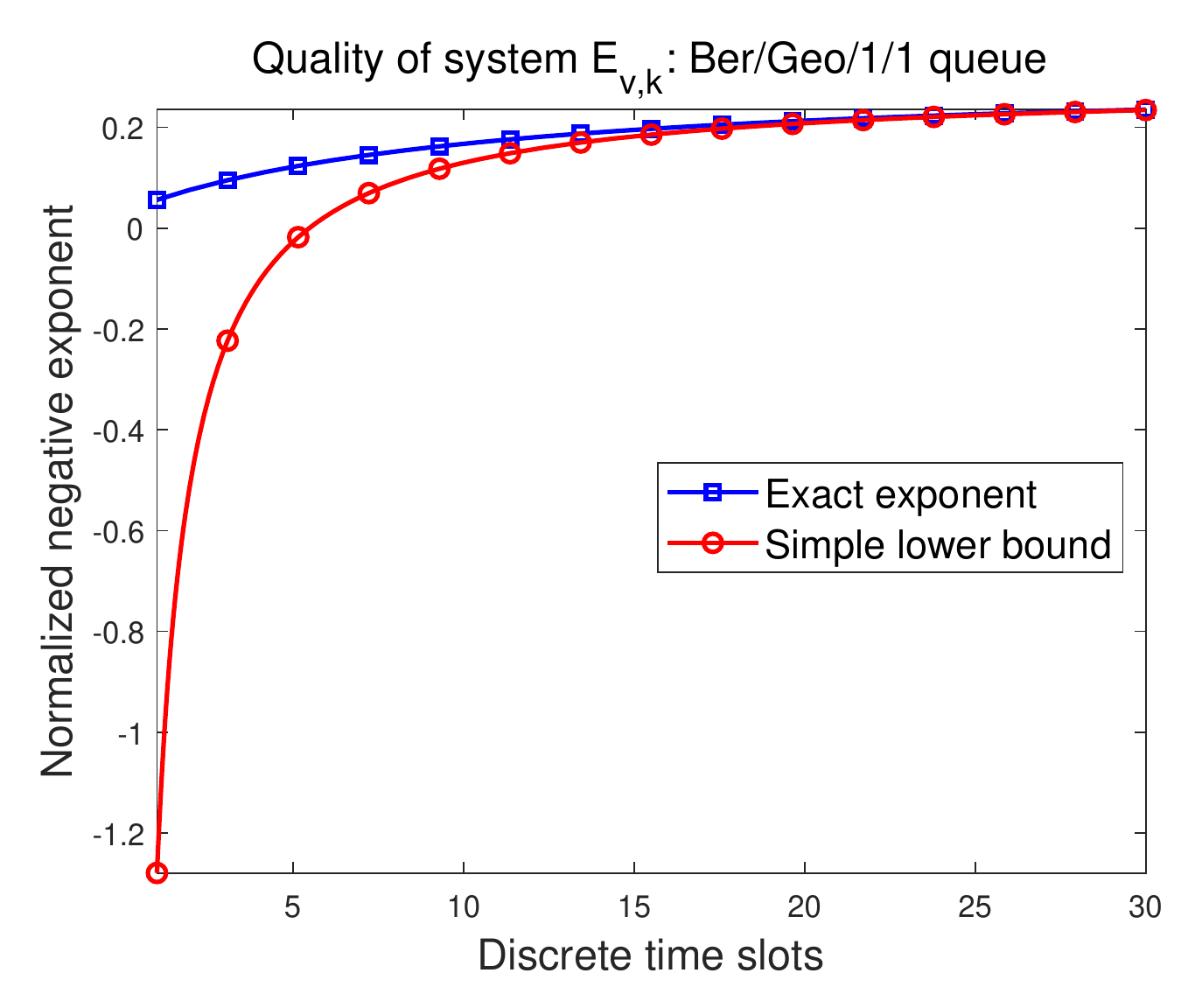}
\caption{Exponent of AoI-violation probability and the simple lower bound.}
\label{fig2}
\end{figure}

Equation (33) is directly obtained from (32). In equation (34), we give a simple lower bound of $E_{v,k}$ by omitting the negative terms in numerator part. Since $1-p>1-\gamma$, the dominant part of the exponent is dependent on the term consists of $1-p$. We depict these exponents in Figure \ref{fig2}, where the system parameters are selected as $p=0.18$ and $\gamma=0.3$.

For the system with Ber/Geo/1/1 queue, notice that in order to obtain the stationary distribution of AoI, we invoke a two-dimensional state vector $(n,m)$ where the first component denotes the instantaneous AoI at the receiver, while the latter parameter represents the age of the packet in system. Since the system size is 1 and the packet waiting time is zero, then the second component $m$ is exactly the service time of the packet, which should be geometrically distributed in this case. We next prove this claim by computing the marginal distribution of $M$, which is defined as the random variable of the second component of the state vector. 

For $m\geq 1$, we have 
\begin{align}
\Pr\{M=m\}&=\sum\nolimits_{n=m+1}^{\infty}\pi_{(n,m)} \notag \\  
&=\sum\nolimits_{n=m+1}^{\infty}\frac{(p\gamma)^2}{(p+\gamma-p\gamma)(\gamma-p)}\left[(1-p)^{n-m}(1-\gamma)^m-(1-\gamma)^n \right]   \notag \\  
&=\frac{(p\gamma)^2}{(p+\gamma-p\gamma)(\gamma-p)}\left[\frac{1-p}{p}(1-\gamma)^m - \frac{1-\gamma}{\gamma}(1-\gamma)^m \right]  \notag \\
&=\frac{p\gamma}{p+\gamma-p\gamma}(1-\gamma)^m  
\end{align}

The probability (36) should be normalized by the probability that the transmitter is not idle, since in that case we have $m=0$. It shows that  
\begin{equation}
\Pr\{\text{system is not idle}\}= 1-\sum\nolimits_{n=1}^{\infty}\pi_{(n,0)} =1-\frac{\pi_{(1,0)}}{p\gamma}=\frac{p(1-\gamma)}{p+\gamma-p\gamma} 
\end{equation}
where we have used the relation $\sum\nolimits_{n=1}^{\infty}\pi_{(n,0)}=\frac{\pi_{(1,0)}}{p\gamma}$, which is obtained from the last line of equations (23). 

Therefore, for $m\geq1$, the probability that the service time $B$ equals $m$ is given as 
\begin{equation}
\Pr\{B=m\} =\frac{\Pr\{M=m\}}{\Pr\{\text{system is not idle}\}} =(1-\gamma)^{m-1}\gamma 
\end{equation}
which is indeed the geometric distribution with service intensity $\gamma$.

\section{The distribution of AoI for status updating systems of larger sizes: Ber/Geo/1/$c$ queues}
In Section III, the stationary AoI-distribution of size 1 status updating system was considered, in which we invoked a two-dimensional state vector and constituted a two-dimensional AoI stochastic process. In this Section, we generalize this idea and the methods used before to analyze the steady state AoI for the system of larger sizes.

First of all, suppose that the queue model in system is Ber/Geo/1/2, we shall prove that as a whole, the random transfers of the three-dimensional state vector, which consists of the AoI at the receiver, the age of the packet in service, and the age of the packet in queue, can be fully described, such that the stationary distribution of the three-dimensional AoI process gives the AoI-distribution as its marginal distribution corresponding to the first age-component. We also discuss how the AoI-distribution of system with Ber/Geo/1/$c$ queue can be obtained, where system size $c$ can be an arbitrary finite integer.

\subsection{The distribution of AoI: Ber/Geo/1/2 queue}
Now, the status updating system has size $c=2$. We directly give the idea that using a three-dimensional state vector $s_k=(n_k,m_k,l_k)$, where the first component $n_{k}$ denotes the AoI at the receiver in the $k$th time slot. The other two parameters, $m_{k}$ and $l_{k}$, represent the age of the packet which is under service currently, and the age of the packet waiting in queue. 

Because the transmitter can store an extra packet, in general the waiting time of a packet is not zero. Observing that the nonzero waiting time will make the actual packet service time unknown, since we only record the instantaneous age of the packet, which is in fact equal to the packet system time. A way to overcome this difficulty is letting the packet service be history-independent, that is, whether or not the service is complete is independent in each time slot. This is the reason why we assume that the service time $B$ is also a geometric random variable. To keep the queueing system stable, it is necessary to ensure that $p$, the packet arrival rate is less than service intensity $\gamma$ strictly.  

It is known that the exponential distribution is the unique continuous time distribution having memoryless property, while the geometric distribution is the only discrete distribution with the same property. In this sense, the Ber/Geo/1/$c$ queue can be regarded as the discrete counterpart of the M/M/1/$c$ queue in continuous time model, in which the arrival-service setting is called the Poisson-Exponential pattern.

Now, we constitute the three-dimensional AoI stochastic process 
\begin{equation}
AoI_{Ber/Geo/1/2}=\{(n_k,m_k,l_k), n_k>m_k>l_k\geq0, k\in\mathbb{N}\}  \notag  
\end{equation} 
and describe the random transitions of all the state vectors $(n,m,l)$.

First of all, assume that there are two packets in system, so that the updating system is full. For this circumstance, the state vector $(n,m,l)$ satisfies $n>m>l \geq 1$. We only need to consider the state transfers caused by the random service of the packet, since at this time no other packets can enter the system. As long as the packet service is not complete at next time slot, all the components of age-state $(n,m,l)$ becomes larger by 1. Thus, we have
\begin{equation}
P_{(n,m,l),(n+1,m+1,l+1)} = \Pr\{\text{packet service is not complete} \}=1-\gamma   \notag 
\end{equation}

Otherwise, if the packet finishes the service and departs from the transmitter in next time slot, the state vector will change to $(m+1,l+1,0)$. The transition probability is equal to
\begin{equation}
P_{(n,m,l),(m+1,l+1,0)} = \Pr\{\text{packet service finishes at next time slot}\}=\gamma   \notag  
\end{equation}

Remember that every time the source node generates a new packet at the beginning of a time slot. If the updating system at that time is full, the arriving packet cannot enter the system and be discarded directly. Therefore, when there is a packet leaving from a full system, the last vector component $l$, which denotes the age of the waiting packet, must jump to zero. Actually, we show that a packet’s service finishes always cuts the number of packet in system by one, and this is the only way that the packet number in system can decrease.

Next, the case $n>m \geq 1$ and $l=0$ is considered. Now the system contains only one packet, which is being served in transmitter. At this time, other packets can enter the system and wait in queue, since the updating system is not full. Therefore, when considering the random jumps of age-states, both the random packet arrivals and the packet departures need to be discussed. It shows that 
\begin{equation}
P_{(n,m,0),(n+1,m+1,0)} = \Pr\{ \text{service does not end, no other packet comes} \} = (1-\gamma)(1-p)   \notag 
\end{equation}

In addition, we also have 
\begin{equation}
P_{(n,m,0),(n+1,m+1,1)} = \Pr\{ \text{service does not end, and a fresher packet arrives} \} = (1-\gamma)p   \notag 
\end{equation}

\begin{equation}
P_{(n,m,0),(m+1,0,0)} = \Pr\{ \text{service is complete, but no extra packet comes} \}= \gamma (1-p)   \notag  
\end{equation}

\begin{equation}
P_{(n,m,0),(m+1,1,0)} = \Pr\{ \text{service ends and a new packet arrives} \} = \gamma p   \notag  
\end{equation}

Finally, suppose that the system is empty, we analyze the possible state transitions. Let $n \geq 1$ and $m=l=0$, Observing that in this case the state transitions are totally governed by the random packet arrivals. If at next time slot no packet arrives, it is easy to see that 
\begin{equation}
P_{(n,0,0),(n+1,0,0)} = \Pr\{\text{no packet comes in next time slot} \} = 1-p  \notag  
\end{equation}

On the contrary, if a fresher packet indeed arrives and enters the system’s transmitter, we have to consider further whether or not the packet is still in system at the end of the current time slot, which depends on the service time $B$ larger than or equal to 1. Specifically, we show that    
\begin{equation}
P_{(n,0,0),(n+1,1,0)}  = \Pr\{\text{a packet comes, the service is not end in one time slot} \} = p\Pr\{B>1\}=p(1-\gamma)   \notag  
\end{equation}

If the coming packet experiences only single time slot service and leaves the transmitter, the age state $(n,0,0)$ transfers to $(1,0,0)$ at next time slot, because in this case the receiver will obtain an updating packet of age 1. The correspoding transition probability is  
\begin{equation}
P_{(n,0,0),(1,0,0)} = \Pr\{\text{a packet comes but leaves system after single time slot} \}= p\Pr\{B=1\}=p\gamma   \notag  
\end{equation}

So far, we have analyzed the state transfers among all the age-states $(n,m,l)$ and determined their transition probabilities. Let $\pi_{(n,m,l)}$ be the stationary probability of the state vector $(n,m,l)$, in the following we establish the stationary equations of the AoI process $AoI_{Ber/Geo/1/2}$. 
\begin{Theorem}
Assume that the queue model used in a discrete time status updating system is Ber/Geo/1/2. Let the stochastic process $AoI_{Ber/Geo/1/2}$ runs in steady state, then the stationary probabilities $\pi_{(n,m,l)}$, $n>m>l\geq0$ satisfy the following equations  
\begin{equation}
\begin{cases}
\pi_{(n,m,l)}=\pi_{(n-1,m-1,l-1)}(1-\gamma)  &   (n>m>l \geq 2)  \\
\pi_{(n,m,1)}=\pi_{(n-1,m-1,0)}p(1-\gamma)    &   (n>m \geq 2)  \\
\pi_{(n,m,0)}=\pi_{(n-1,m-1,0)}(1-p)(1-\gamma) +\left( \sum\nolimits_{k=n}^{\infty}\pi_{(k,n-1,m-1)} \right) \gamma & (n>m \geq 2)  \\
\pi_{(n,1,0)}=\pi_{(n-1,0,0)}p(1-\gamma) + \left( \sum\nolimits_{k=n}^{\infty}\pi_{(k,n-1,0)} \right) p\gamma &  (n \geq 2)  \\
\pi_{(n,0,0)}=\pi_{(n-1,0,0)}(1-p) + \left( \sum\nolimits_{k=n}^{\infty}\pi_{(k,n-1,0)} \right) (1-p)\gamma  & (n \geq 2)  \\
\pi_{(1,0,0)}=\left(\sum\nolimits_{k=1}^{\infty}\pi_{(k,0,0)} \right) p\gamma 
\end{cases}
\end{equation}
\end{Theorem}

\begin{proof}
We explain each line of (39) as follows. First of all, suppose that there are two packets in system, so that the updating system is full. The state transfers in this case is easy, because the randomness that affects the state transitions only comes from the service time the packet consumes in system’s server. All the generated packets trying to enter the system will be discarded, since the system is full. The state $(n,m,1)$ is special, because the system becomes full just now. The only way to obtain the state $(n,m,1)$ is from $(n-1,m-1,0)$, and satisfying following two conditions: (i) the packet service continues so that the first two vector components turn to $(n,m)$ at the next time slot; (ii) a new packet arrives which makes the $l$ change to 1. As for the general state $(n,m,l)$ where $n>m>l\geq 2$, begin with $(n-1,m-1,l-1)$, let the packet in service does not leave, then the age-state jumps to $(n,m,l)$ at next time slot. Notice that this is the only way that the state vector $(n,m,l)$ can be obtained. This explains first two lines of equations (39).

The next two equations corresponds to the scenario where there is one packet in system. Firstly, the state $(n,1,0)$ implies that before one time slot, the system is empty. The packet in system is generated at the current time slot. It shows that the empty state $(n-1,0,0)$ will jump to $(n,1,0)$ at next time slot, if a new packet comes and still stays in system after one time slot. Otherwise, when the packet arrives but leaves the system after single time slot service, we have mentioned before that the age-state will reduce to $(1,0,0)$. In addition, a state vector of form $(k,n-1,0)$, $k\geq n$ can also transfer to $(n,1,0)$, as long as the packet in service is sent out after its service ends, and one new packet arrives. Notice that in this case the newly coming packet have to wait in queue, since when the packet arrives, the server is busy. For the stationary equation of age-state $(n,m,0)$ where $m\geq 2$, similarly we have to consider two cases. Age-state $(n,m,0)$ can be obtained from either $(n-1,m-1,0)$ or any state having form $(k,n-1,m-1)$ with different probabilities. Apparently, getting $(n,m,0)$ from $(n-1,m-1,0)$ requires that the service of the packet does not finish and no packet comes at the next time slot. To make $(k,n-1,m-1)$ transfer to $(n,m,0)$, we just assume that the packet service is over and then the packet is sent out, regardless of the possible arriving packet because any such packet cannot get in at the beginning of the current time slot. Thus, we obtain the third and the fourth equation of (39).

Finally, assume that the system is empty. In this case, the state is denoted by $(n,0,0)$, $n\geq1$. Let an updating packet arrives at the beginning of a time slot, and departs from the transmitter after one-time-slot service. Then, a packet of age 1 is sent to the receiver and the AoI is reduced to 1. Therefore, beginning with an age-state of form $(k,0,0)$, $k\geq1$, with probability $p\gamma$ the state transfers to $(1,0,0)$ in the next time slot. Next, we consider the state transitions that can lead to $(n,0,0)$, $n\geq 2$. The system is also empty, but the instantaneous AoI at the receiver is greater than 1. From an empty state $(n-1,0,0)$, as long as no packet arrives, obviously the state changes to $(n,0,0)$ in the next time slot. On the other hand, let the only packet in system leaves and a packet of age $n$ arrives to the receiver, the empty-system state $(n,0,0)$ is obtained as well. Observing that from the state $(k,n-1,0)$, if no packet comes in next time slot, and the service of the packet is over, then we shall obtain the age-state $(n,0,0)$. This explains last two lines of (39), i.e., the stationary equations corresponding to the empty state $(n,0,0)$.

Summarizing above discussions, we complete the proof of the Theorem 6 in the end. 
\end{proof}

Quite a lot of calculations have to do in order to find the exact solutions of system of equations (39). In the following, we show that how the stationary probabilities $\pi_{(n,m,l)}$ are determined step by step. 

Firstly, we give two general formulas by using the recursive relations in (39). 
\begin{Lemma}
For the stationary probabilities $\pi_{(n,m,l)}$, we have   
\begin{equation}
\pi_{(n,m,l)}=\pi_{(n-l,m-l,0)}p(1-\gamma)^{l}  \qquad  (n>m>l \geq 1)
\end{equation}
and when $n>m\geq 1$, it shows that 
\begin{align}
\pi_{(n,m,0)}&=\pi_{(n-m,0,0)}p(1-\gamma) \left[(1-p)(1-\gamma)\right]^{m-1} \notag \\
& \qquad \qquad   +\frac{\gamma}{1-\gamma} \left(\sum\nolimits_{k=n-m+1}^{\infty}\pi_{(k,n-m,0)} \right) \left\{ (1-\gamma)^m-[(1-p)(1-\gamma)]^m \right\} 
\end{align}
\end{Lemma}

\begin{proof}
Obtaining the first formula (40) is not difficult. Applying the first line of (39) repeatedly yields that  
\begin{equation}
\pi_{(n,m,l)}=\pi_{(n-1,m-1,l-1)}(1-\gamma)  = \pi_{(n-2,m-2,l-2)}(1-\gamma)^2 =  \cdots   = \pi_{(n-l+1,m-l+1,1)}(1-\gamma)^{l-1}  \notag 
\end{equation}

To complete the last iteration, we use the second equation in (39) which gives that 
\begin{equation}
\pi_{(n,m,l)}=\pi_{(n-l,m-l,0)}p(1-\gamma)^{l}  \quad (n>m>l \geq 1)
\end{equation}

Next, the other formula (41) for the stationary probabilities $\pi_{(n,m,0)}$ is proved. The third line of (39) says 
\begin{equation}
\pi_{(n,m,0)}=\pi_{(n-1,m-1,0)}(1-p)(1-\gamma)  +  \left(\sum\nolimits_{k=n}^{\infty}\pi_{(k,n-1,m-1)} \right)\gamma  
\end{equation}

Using expression (42), the sum in equation (43) is dealt with as  
\begin{align}
\left(\sum\nolimits_{k=n}^{\infty}\pi_{(k,n-1,m-1)} \right)\gamma &= \left(\sum\nolimits_{k=n}^{\infty}\pi_{(k-m+1,n-m,0)}p(1-\gamma)^{m-1} \right)\gamma  \notag \\
&= p\gamma \left(\sum\nolimits_{k=n-m+1}^{\infty}\pi_{(k,n-m,0)} \right) (1-\gamma)^{m-1} 
\end{align}

Therefore, we obtain 
\begin{equation}
\pi_{(n,m,0)}=\pi_{(n-1,m-1,0)}(1-p)(1-\gamma)  + p\gamma \left(\sum\nolimits_{k=n-m+1}^{\infty}\pi_{(k,n-m,0)}\right)(1-\gamma)^{m-1}
\end{equation}

Iteratively applying (45) gives following equations
\begin{align}
\pi_{(n,m,0)}&=\pi_{(n-1,m-1,0)}(1-p)(1-\gamma) + p\gamma \left( \sum\nolimits_{k=n-m+1}^{\infty} \pi_{(k,n-m,0)} \right)  (1-\gamma)^{m-1}  \notag  \\
&= \bigg[ \pi_{(n-2,m-2,0)}(1-p)(1-\gamma) +p\gamma \left( \sum\nolimits_{k=n-m+1}^{\infty} \pi_{(k,n-m,0)}  \right)  (1-\gamma)^{m-2} \bigg] (1-p)(1-\gamma) \notag \\
&\qquad +p\gamma \left( \sum\nolimits_{k=n-m+1}^{\infty}\pi_{(k,n-m,0)}  \right)  (1-\gamma)^{m-1} \notag \\
&=\pi_{(n-2,m-2,0)}\left[(1-p)(1-\gamma)\right]^2  +p\gamma \left(\sum\nolimits_{k=n-m+1}^{\infty}\pi_{(k,n-m,0)} \right) \notag \\
&\qquad \times \left\{\left[(1-p)(1-\gamma)\right](1-\gamma)^{m-2}+(1-\gamma)^{m-1}\right\}  \notag \\
&\qquad \qquad \qquad \qquad \quad\quad\qquad \qquad \quad    \vdots  \notag \\
&=\pi_{(n-m+1,1,0)}\left[(1-p)(1-\gamma)\right]^{m-1} +p\gamma \left(\sum\nolimits_{k=n-m+1}^{\infty}\pi_{(k,n-m,0)} \right) \notag \\
& \qquad \times \left(\sum\nolimits_{j=0}^{m-2}\left[(1-p)(1-\gamma)\right] ^j (1-\gamma)^{m-1-j} \right)  
\end{align}

From the fourth line of (39), we have 
\begin{align}
&\pi_{(n-m+1,1,0)}\left[(1-p)(1-\gamma)\right]^{m-1}  \notag \\
={}&\left[\pi_{(n-m,0,0)}p(1-\gamma) + \left( \sum\nolimits_{k=n-m+1}^{\infty}\pi_{(k,n-m,0)} \right) p\gamma \right] [(1-p)(1-\gamma)]^{m-1}   \notag \\
={}& \pi_{(n-m,0,0)}p(1-\gamma)[(1-p)(1-\gamma)]^{m-1} + p\gamma \left(\sum\nolimits_{k=n-m+1}^{\infty}\pi_{(k,n-m,0)} \right) [(1-p)(1-\gamma)]^{m-1} 
\end{align}

Combining (46) and (47) gives final result 
\begin{align}
\pi_{(n,m,0)}&=\pi_{(n-m,0,0)}p(1-\gamma)[(1-p)(1-\gamma)]^{m-1}+p\gamma \left(\sum\nolimits_{k=n-m+1}^{\infty}\pi_{(k,n-m,0)} \right)  \left(\sum\nolimits_{j=0}^{m-1}(1-p)^j (1-\gamma)^{m-1} \right)  \notag \\
&=\pi_{(n-m,0,0)}p(1-\gamma)[(1-p)(1-\gamma)]^{m-1} + \frac{\gamma}{1-\gamma}\left( \sum\nolimits_{k=n-m+1}^{\infty}\pi_{(k,n-m,0)}\right) \left\{(1-\gamma)^m-[(1-p)(1-\gamma)]^m\right\}  
\end{align}

This finishes the proof of Lemma 1. 
\end{proof}

Lemma 1 shows that the stationary probabilities $\pi_{(n,m,l)}$, $l\geq 1$ are known if we find all the stationary probabilities $\pi_{(n,m,0)}$. Furthermore, notice that the second line of (39) shows 
\begin{equation}
\sum\nolimits_{k=n-m+1}^{\infty}\pi_{(k,n-m,0)} = \frac{\pi_{(n-m+1,0,0)}-\pi_{(n-m,0,0)}(1-p)}{(1-p)\gamma}
\end{equation}

Combining (48) and (49), it was observed that the stationary probabilities $\pi_{(n,m,0)}$ can be represented by $\{\pi_{(n,0,0)}$, $n\geq 1\}$. Therefore, in order to solve all the probabilities $\pi_{(n,m,l)}$, we only need to determine those probabilities for the subset of state vectors $\{(n,0,0),n\geq 1\}$.

In the last two lines of (39), we see that the probability $\pi_{(1,0,0)}$ is related to an infinite sum $\sum\nolimits_{k=1}^{\infty}\pi_{(k,0,0)}$, and the other probabilities $\pi_{(n,0,0)}$, $n\geq 2$ are also dependent of different infinite sums. At next step, we compute these infinite sums $\sum\nolimits_{k=n}^{\infty}\pi_{(k,n-1,0)}$, $n\geq1$, so that an explicit recursive relation for the probabilities $\pi_{(n,0,0)}$ is obtained.

Firstly, we need another two sums $M_1$ and $M_2$ which are given below.  
\begin{Lemma}
Two sums $M_1$ and $M_2$ are defined and calculated as 
\begin{equation}
M_1=\sum\nolimits_{n=1}^{\infty}\pi_{(n,0,0)}= \frac{(1-p)\gamma^2}{N(p,\gamma)},
\qquad  
M_2=\sum\nolimits_{n=2}^{\infty}\sum\nolimits_{m=1}^{n-1}\pi_{(n,m,0)}= \frac{p\gamma(1-\gamma)}{N(p,\gamma)} 
\end{equation}
where $N(p,\gamma)$ represents 
\begin{equation}
N(p,\gamma)=(p+\gamma -2p\gamma)\gamma +p^2(1-\gamma)^2 \notag
\end{equation}
\end{Lemma}

The proof of Lemma 2 is deferred to Appendix C. 

\begin{Lemma}
The infinite sums $\sum\nolimits_{k=n}^{\infty}\pi_{(k,n-1,0)}$, $n\geq 2$ are determined as 
\begin{equation}
\sum\nolimits_{k=n}^{\infty}\pi_{(k,n-1,0)}=\gamma M_2 (1-\gamma)^{n-2}
\end{equation}
in which $M_2$ is defined and given in Lemma 2. 
\end{Lemma}

For the case $n=1$, the sum $\sum\nolimits_{k=1}^{\infty}\pi_{(k,0,0)}$ is exactly equal to $M_1$, which has been obtained in Lemma 2. We will prove Lemma 3 in Appendix D. 

Provided above results, a specific recursive relation for the stationary probabilities $\pi_{(n,0,0)}$, $n\geq 1$ can be derived, which enables us to find the general expressions of probabilities $\pi_{(n,0,0)}$.  
\begin{Theorem}
The explicit expression of stationary probabilities $\pi_{(n,0,0)}$ are given as 
\begin{equation}
\pi_{(n,0,0)}=\frac{p(1-p)\gamma^3}{N(p,\gamma)(\gamma-p)}\left[(1-p)^n-(1-\gamma)^n\right]  \qquad  (n\geq1)
\end{equation}
\end{Theorem}

\begin{proof}
The basic relation in (39) shows that 
\begin{equation}
\pi_{(n,0,0)}=\pi_{(n-1,0,0)}(1-p) + \left( \sum\nolimits_{k=n}^{\infty}\pi_{(k,n-1,0)} \right)(1-p)\gamma \qquad (n\geq2) \notag
\end{equation}

According to Lemma 3, substituting the infinite sums, we obtain that 
\begin{equation}
\pi_{(n,0,0)}=\pi_{(n-1,0,0)}(1-p) + (1-p)\gamma^2 M_2 (1-\gamma)^{n-2}  
\end{equation}

By repeatedly using equation (53), eventually we can derive the general expressions of probabilities $\pi_{(n,0,0)}$, $n\geq1$. 
\begin{align}
\pi_{(n,0,0)}&=\pi_{(n-1,0,0)}(1-p) + (1-p)\gamma^2 M_2 (1-\gamma)^{n-2}  \notag \\
&=\left[ \pi_{(n-2,0,0)}(1-p)  + (1-p)\gamma^2 M_2 (1-\gamma)^{n-3} \right] (1-p)+ (1-p)\gamma^2 M_2 (1-\gamma)^{n-2} \notag \\
&=\pi_{(n-2,0,0)}(1-p)^2 + (1-p)\gamma^2 M_2 \left[ (1-p)(1-\gamma)^{n-3}+ (1-\gamma)^{n-2}\right]  \notag \\
& \qquad \qquad \qquad \qquad  \qquad \quad \qquad \qquad   \vdots  \notag \\
&=\pi_{(1,0,0)}(1-p)^{n-1}+(1-p)\gamma^2 M_2  \left( \sum\nolimits_{j=0}^{n-2}(1-p)^j (1-\gamma)^{n-2-j}  \right)  \notag \\
&=\pi_{(1,0,0)}(1-p)^{n-1}+(1-p)\gamma^2 M_2  \frac{(1-p)^{n-1}-(1-\gamma)^{n-1}}{\gamma-p} 
\end{align}

Notice that the last line of (39) shows that $\pi_{(1,0,0)}=p\gamma M_1$, where $M_1$ is given in Lemma 2. Substituting this relation into (54) and rearranging the terms, finally we have 
\begin{align}
\pi_{(n,0,0)}&=\left[p\gamma M_1 +\frac{(1-p)\gamma^2M_2}{\gamma-p} \right](1-p)^{n-1} -\frac{(1-p)\gamma^2M_2}{\gamma-p}(1-\gamma)^{n-1} \notag \\
&=\frac{p(1-p)\gamma^3}{N(p,\gamma)(\gamma-p)}\left[(1-p)^n-(1-\gamma)^n\right]  \qquad (n\geq2)
\end{align}

For the case $n=1$, it is easy to verify that expression (55) determines $\pi_{(1,0,0)}$. Thus, we obtain the general expression for probabilities $\pi_{(n,0,0)}$, $n\geq 1$ and finish the proof of Theorem 7. 
\end{proof}

Now, the other probabilities $\pi_{(n,m,0)}$, $n>m\geq 1$, and $\pi_{(n,m,l)}$ where $n>m>l\geq1$ can be calculated.  
\begin{Corollary}
The stationary probabilities $\pi_{(n,m,0)}$, $n>m\geq 1$ are determined by 
\begin{equation}
\pi_{(n,m,0)}= \frac{p^2\gamma^3}{N(p,\gamma)(\gamma-p)}\left\{(1-p)^n(1-\gamma)^m - (1-p)^m(1-\gamma)^n  \right\} + \frac{p\gamma^3}{N(p,\gamma)}\left\{(1-\gamma)^{n-1} - (1-p)^m(1-\gamma)^{n-1}  \right\}
\end{equation}

For the probabilities $\pi_{(n,m,l)}$, $n>m>l\geq 1$, we have 
\begin{equation}
\pi_{(n,m,l)} = \frac{(p\gamma)^3}{N(p,\gamma)(\gamma-p)}\bigg\{(1-p)^{n-l}(1-\gamma)^m  - (1-p)^{m-l}(1-\gamma)^n \bigg\} + \frac{p^2\gamma^3}{N(p,\gamma)} \left\{(1-\gamma)^{n-1} - (1-p)^{m-l}(1-\gamma)^{n-1}  \right\}
\end{equation}
\end{Corollary}

\begin{proof}
In order to obtain probability $\pi_{(n,m,0)}$, we substitute following equation into (41). From Theorem 7, we have
\begin{equation}
\pi_{(n-m,0,0)}=\frac{p(1-p)\gamma^3}{N(p,\gamma)(\gamma-p)}\left[(1-p)^{n-m}-(1-\gamma)^{n-m}\right] \notag  
\end{equation}
and Lemma 3 shows that  
\begin{equation}
\sum\nolimits_{k=n-m+1}^{\infty}\pi_{(k,n-m,0)}=\gamma M_2 (1-\gamma)^{n-m-1} \notag
\end{equation}

Therefore, equation (41) gives that 
\begin{align}
\pi_{(n,m,0)}&= \frac{p(1-p)\gamma^3}{N(p,\gamma)(\gamma-p)}\left[(1-p)^{n-m}-(1-\gamma)^{n-m}\right] p(1-\gamma)\left[(1-p)(1-\gamma)\right]^{m-1}  \notag \\
& \qquad \qquad \qquad + \frac{\gamma}{1-\gamma} \gamma M_2 (1-\gamma)^{n-m-1} \left\{ (1-\gamma)^m-[(1-p)(1-\gamma)]^m \right\} \notag \\
&= \frac{p^2\gamma^3\left[(1-p)^{n-m}-(1-\gamma)^{n-m}\right][(1-p)(1-\gamma)]^m}{N(p,\gamma)(\gamma-p)} + \frac{p\gamma^3\left\{(1-\gamma)^{n-1}-(1-p)^m(1-\gamma)^{n-1} \right\}}{N(p,\gamma)} \notag \\
&= \frac{p^2\gamma^3\left\{(1-p)^{n}(1-\gamma)^{m}- (1-p)^{m}(1-\gamma)^{n}  \right\}}{N(p,\gamma)(\gamma-p)} + \frac{p\gamma^3\left\{(1-\gamma)^{n-1}-(1-p)^m(1-\gamma)^{n-1} \right\}}{N(p,\gamma)}
\end{align}

This gives equation (56). For the stationary probability $\pi_{(n,m,l)}$, the relation 
\begin{equation}
\pi_{(n,m,l)}=\pi_{(n-l,m-l,0)}p(1-\gamma)^l  \notag
\end{equation}
is used and we have that  
\begin{align}
\pi_{(n,m,l)} &= \bigg\{ \frac{p^2\gamma^3 \left[ (1-p)^{n-l} (1-\gamma)^{m-l} - (1-p)^{m-l} (1-\gamma)^{n-l} \right]}{N(p,\gamma)(\gamma-p)}  \notag \\
&\qquad \qquad \qquad  + \frac{p\gamma^3 \left[(1-\gamma)^{n-l-1}-(1-p)^{m-l}(1-\gamma)^{n-l-1} \right] }{N(p,\gamma)} \bigg\} p(1-\gamma)^l  \notag\\
&= \frac{(p\gamma)^3 \left[ (1-p)^{n-l} (1-\gamma)^{m} - (1-p)^{m-l} (1-\gamma)^{n} \right]}{N(p,\gamma)(\gamma-p)} +\frac{p^2\gamma^3 \left[(1-\gamma)^{n-1}-(1-p)^{m-l}(1-\gamma)^{n-1} \right] }{N(p,\gamma)}  \notag
\end{align}

This completes the proof of Corollary 4. 
\end{proof}

So far, we have completely solved the stationary equations (39) and determine all the stationary probabilities $\pi_{(n,m,l)}$. Then, when the AoI process $AoI_{Ber/Geo/1/2}$ runs in steady state, the probability that AoI takes value $n$ can be obtained by merging all the stationary probabilities having $n$ as its first age-component. 

\begin{Theorem}
For the discrete time status updating system with Ber/Geo/1/2 queue, the stationary distribution of the AoI is calculated as  
\begin{align}
\Pr\{\Delta=n\}&= \pi_{(n,0,0)}+\sum\nolimits_{m=1}^{n-1}\pi_{(n,m,0)}+\sum\nolimits_{m=2}^{n-1}\sum\nolimits_{l=1}^{m-1}\pi_{(n,m,l)}
\notag \\ 
&=\frac{p(1-p)^2\gamma^4}{N(p,\gamma)(\gamma-p)^2}[(1-p)^n-(1-\gamma)^n]  \notag \\
&\qquad - \frac{p^2\gamma^3}{N(p,\gamma)}\left[\frac{1}{2(1-\gamma)} + \frac{1-p}{\gamma-p}\right]n(1-\gamma)^n + \frac{p^2\gamma^3}{2N(p,\gamma)}n^2(1-\gamma)^{n-1}  \qquad  (n\geq1)
\end{align}

For $k\geq1$, the AoI-cumulative probabilities are equal to 
\begin{align}
\Pr\{\Delta \leq k\}&= \sum\nolimits_{n=1}^k \Pr\{\Delta=n\}  \notag \\
&=1- \frac{(1-p)^3\gamma^4}{N(p,\gamma)(\gamma-p)^2}(1-p)^k + \frac{p(1-\gamma)}{N(p,\gamma)(\gamma-p)}\left[ \frac{(1-p)^2\gamma^3}{\gamma-p}+ p^2(1-\gamma) \right] (1-\gamma)^k \notag \\
&\qquad + \frac{p^2\gamma}{N(p,\gamma)}\left[ \frac{(1-p)(1-\gamma)\gamma}{\gamma-p}-\frac{2-\gamma}{2} \right] k(1-\gamma)^k - \frac{(p\gamma)^2}{2N(p,\gamma)}k^2(1-\gamma)^k   
\end{align}
\end{Theorem}

With these complex expressions, we obtain the complete description of the stationary AoI for the status updating system with Ber/Geo/1/2 queue. The calculation details of equations (59) and (60) are placed in Appendix E.

\subsection{Stationary distribution of packet system time and packet waiting time}
In the state vector $(n,m,l)$, the second age-component $m$ records the age of the packet in service, which is equal to the packet system time. The last parameter $l$ represents the waiting time of the packet in queue. Since we have obtained all the stationary probabilities $\pi_{(n,m,l)}$, thus the distributions of both packet system time and waiting time can be determined as well by computing the marginal distributions of the second and third age-components.   

\begin{Corollary}
In a status updating system with Ber/Geo/1/2 queue, the packet system time $T$  is distributed as 
\begin{equation}
\Pr\{T=m\}=\begin{cases}
\frac{(1-p)\gamma^2}{N(p,\gamma)}  & \quad   (m=0)  \\
\frac{p\gamma^2}{N(p,\gamma)}[1+(m-1)p](1-\gamma)^m  & \quad  (m\geq1) 
\end{cases}
\end{equation}
and the distribution of waiting time $W$ of a packet in queue is given as 
\begin{equation}
\Pr\{W=l\}=\begin{cases}
\frac{(p+\gamma-2p\gamma)\gamma}{N(p,\gamma)} & \quad  (l=0)  \\
\frac{p^2\gamma}{N(p,\gamma)}(1-\gamma)^{l+1} & \quad  (l\geq1) 
\end{cases}
\end{equation}
\end{Corollary}

\begin{proof}
Firstly, let $m=0$, we have 
\begin{equation}
\Pr\{T=0\}=\sum\nolimits_{n=1}^{\infty}\pi_{(n,0,0)}=M_1=\frac{(1-p)\gamma^2}{N(p,\gamma)}
\end{equation}

To determine the probability that $T=m$, we compute 
\begin{equation}
\Pr\{T=m\}=\sum\nolimits_{n=m+1}^{\infty}\sum\nolimits_{l=0}^{m-1}\pi_{(n,m,l)} = \sum\nolimits_{n=m+1}^{\infty}\pi_{(n,m,0)}+\sum\nolimits_{n=m+1}^{\infty}\sum\nolimits_{l=1}^{m-1}\pi_{(n,m,l)} 
\end{equation}
where the first sum is equal to 
\begin{align}
\sum\nolimits_{n=m+1}^{\infty}\pi_{(n,m,0)}&= \sum\nolimits_{n=m+1}^{\infty} \bigg\{\frac{p^2\gamma^3}{N(p,\gamma)(\gamma-p)}\bigg[(1-p)^n(1-\gamma)^m  - (1-p)^m(1-\gamma)^n \bigg]  \notag \\
&\qquad \qquad \qquad + \frac{p\gamma^3}{N(p,\gamma)} \bigg[(1-\gamma)^{n-1} - (1-p)^m(1-\gamma)^{n-1}\bigg] \bigg\}  \notag \\
&= \frac{p^2\gamma^3}{N(p,\gamma)(\gamma-p)}\bigg[\frac{(1-p)^{m+1}(1-\gamma)^m}{p} - \frac{(1-p)^m(1-\gamma)^{m+1}}{\gamma}\bigg]  \notag \\
&\qquad  \qquad \qquad + \frac{p\gamma^3}{N(p,\gamma)}\left[\frac{(1-\gamma)^m}{\gamma} - \frac{[(1-p)(1-\gamma)]^m}{\gamma}\right] \notag \\
&= \frac{p^2\gamma^3}{N(p,\gamma)(\gamma-p)} \frac{\gamma-p}{p\gamma}[(1-p)(1-\gamma)]^m + \frac{p\gamma^2}{N(p,\gamma)} (1-\gamma)^m -  \frac{p\gamma^2}{N(p,\gamma)} [(1-p)(1-\gamma)]^m \notag \\
&= \frac{p\gamma^2}{N(p,\gamma)} [(1-p)(1-\gamma)]^m   + \frac{p\gamma^2}{N(p,\gamma)} (1-\gamma)^m -  \frac{p\gamma^2}{N(p,\gamma)} [(1-p)(1-\gamma)]^m \notag \\
&= \frac{p\gamma^2}{N(p,\gamma)} (1-\gamma)^m 
\end{align}

The other sum is calculated as  
\begin{align}
&\sum\nolimits_{n=m+1}^{\infty}\sum\nolimits_{l=1}^{m-1}\pi_{(n,m,l)} \notag \\ 
={}& \sum\nolimits_{n=m+1}^{\infty}\sum\nolimits_{l=1}^{m-1} \bigg\{ \frac{(p\gamma)^3}{N(p,\gamma)(\gamma-p)}\bigg[ (1-p)^{n-l}(1-\gamma)^m  - (1-p)^{m-l}(1-\gamma)^n \bigg] \notag \\
  {}& \qquad \qquad \qquad \qquad + \frac{p^2\gamma^3}{N(p,\gamma)} \left[(1-\gamma)^{n-1} - (1-p)^{m-l}(1-\gamma)^{n-1} \right]  \bigg\}  \notag \\
={}& \sum\nolimits_{n=m+1}^{\infty} \bigg\{ \frac{(p\gamma)^3}{N(p,\gamma)(\gamma-p)}\bigg[ \frac{(1-p)^{n-m+1}-(1-p)^n}{p} (1-\gamma)^m - \frac{(1-p) - (1-p)^m}{p}(1-\gamma)^n \bigg]  \notag \\
  {}& \qquad \qquad \qquad \qquad  + \frac{p^2\gamma^3}{N(p,\gamma)} \left[(m-1)(1-\gamma)^{n-1} - \frac{(1-p) - (1-p)^m}{p}(1-\gamma)^{n-1} \right]  \bigg\} \notag \\
={}& \frac{(p\gamma)^3}{N(p,\gamma)(\gamma-p)}\bigg[ \frac{(1-\gamma)^m}{p} \left(\frac{(1-p)^2}{p}-\frac{(1-p)^{m+1}}{p}\right) - \frac{ (1-p)-(1-p)^m}{p} \frac{(1-\gamma)^{m+1}}{\gamma} \bigg]  \notag \\ 
  {}& \qquad \qquad \qquad \qquad + \frac{p^2\gamma^3}{N(p,\gamma)} \left[(m-1)\frac{(1-\gamma)^m}{\gamma} - \frac{(1-p) - (1-p)^m}{p}\frac{(1-\gamma)^m}{\gamma} \right] \bigg\} \notag \\
={}& \frac{p\gamma^2(1-\gamma)^m}{N(p,\gamma)} \Big( (1-p) - (1-p)^m \Big) + \frac{(p\gamma)^2(m-1)(1-\gamma)^m}{N(p,\gamma)} - \frac{p\gamma^2(1-\gamma)^m}{N(p,\gamma)}\Big( (1-p) - (1-p)^m \Big)  \notag \\
={}& \frac{(p\gamma)^2}{N(p,\gamma)}(m-1)(1-\gamma)^m 
\end{align}

Combining (65) and (66), we obtain that  
\begin{equation}
\Pr\{T=m\} = \frac{p\gamma^2}{N(p,\gamma)}(1-\gamma)^m + \frac{(p\gamma)^2}{N(p,\gamma)}(m-1)(1-\gamma)^m =\frac{p\gamma^2}{N(p,\gamma)}[1+(m-1)p](1-\gamma)^m   \quad  (m\geq1) 
\end{equation}

Here, we also check that all the probabilities sum up to 1, so that (67) and (63) form a proper probability distribution. It shows that 
\begin{align}
&\Pr\{T=0\} + \sum\nolimits_{m=1}^{\infty}\Pr\{T=m\} \notag \\
={}& M_1 + \sum\nolimits_{m=1}^{\infty}\frac{p\gamma^2}{N(p,\gamma)}[1+(m-1)p](1-\gamma)^m \notag \\
={}& M_1 + \frac{p\gamma^2}{N(p,\gamma)}\frac{1-\gamma}{\gamma} + \frac{(p\gamma)^2}{N(p,\gamma)}\frac{(1-\gamma)^2}{\gamma^2}  \notag \\
={}& \frac{(1-p)\gamma^2}{N(p,\gamma)}+\frac{p\gamma(1-\gamma)}{N(p,\gamma)} + \frac{p^2(1-\gamma)^2}{N(p,\gamma)} \notag \\
={}& \frac{(p+\gamma - 2p\gamma)\gamma+p^2(1-\gamma)^2}{N(p,\gamma)} \notag \\
={}&1 \notag   
\end{align}
where $N(p,\gamma)$ is defined in Lemma 2.

Now, we calculate the stationary distribution of the packet waiting time $W$. First of all, we have 
\begin{equation}
\Pr\{W=0\}=\sum\nolimits_{n=1}^{\infty}\pi_{(n,0,0)} + \sum\nolimits_{n=2}^{\infty}\sum\nolimits_{m=1}^{n-1}\pi_{(n,m,0)}=M_1 + M_2 =\frac{(p+\gamma - 2p\gamma)\gamma}{N(p,\gamma)} 
\end{equation}

For the cases $l\geq2$, notice that 
\begin{align}
\Pr\{W=l\}&=\sum\nolimits_{n=l+2}^{\infty} \sum\nolimits_{m=l+1}^{n-1}\pi_{(n,m,l)} \notag \\
&= \sum\nolimits_{n=l+2}^{\infty} \sum\nolimits_{m=l+1}^{n-1}\pi_{(n-1,m-1,l-1)}(1-\gamma)   \\
&= \left(\sum\nolimits_{\tilde n=l+1}^{\infty} \sum\nolimits_{\tilde m=l}^{\tilde n-1}\pi_{(\tilde n,\tilde m,l-1)}\right)(1-\gamma)   \\
&= \Pr\{W=l-1\}(1-\gamma)  
\end{align}
where in (69) the first relation of (39) is used, that is $\pi_{(n,m,l)}=\pi_{(n-1,m-1,l-1)}(1-\gamma)$ when $l\geq2$. To obtain equation (70), do the substitutions $n-1=\tilde n$ and $m-1=\tilde m$.

Applying (71) repeatedly gives 
\begin{equation}
\Pr\{W=l\}=\Pr\{W=1\}(1-\gamma)^{l-1}  \qquad  (l\geq2)
\end{equation}

Notice that 
\begin{align}
1&=\Pr\{W=0\} + \sum\nolimits_{l=1}^{\infty}\Pr\{W=l\} =\frac{(p+\gamma-2p\gamma)\gamma}{N(p,\gamma)} + \sum\nolimits_{l=1}^{\infty}\Pr\{W=1\}(1-\gamma)^{l-1} \notag \\
&=\frac{(p+\gamma-2p\gamma)\gamma}{N(p,\gamma)} + \frac{\Pr\{W=1\}}{\gamma}  \notag 
\end{align}
from which we can solve that 
\begin{equation}
\Pr\{W=1\} = \frac{p^2\gamma(1-\gamma)^2}{N(p,\gamma)}
\end{equation}

Substituting (73) into (72), eventually we obtain  
\begin{equation}
\Pr\{W=l\}=\frac{p^2\gamma}{N(p,\gamma)}(1-\gamma)^{l+1} \qquad  (l\geq1)
\end{equation}

Combining equations (68) and (74), we fully determine the stationary distribution of the packet waiting time $W$. This completes the proof of Corollary 5. 
\end{proof}

\subsection{Finding the stationary AoI-distribution of status updating system with Ber/Geo/1/$c$ queue}
The idea that using a multiple-dimensional state vector to describe the instantaneous age-state of a status updating system can be generalized to determine the AoI-distribution for the more general system which uses Ber/Geo/1/$c$ queue. We can introduce a $(c+1)$-dimensional state vector to record the value of the AoI at the receiver, the age of the packet in service and the ages of those packets waiting in queue, respectively. Accordingly, a $(c+1)$-dimensional discrete stochastic process, we called $AoI_{Ber/Geo/1/c}$ can be constituted. Then, the random transitions of the $(c+1)$-dimensional age-states are described so that the corresponding stationary equations of the resulting AoI process are determined. As long as the stationary probabilities of all the age-states are found, the distribution of the AoI can be determined by collecting all the probabilities whose first age-component equal to $n$, from $n=1$ to $\infty$. However, deriving the exact expression of AoI-distribution when the system's size is greater than 3 may be very hard, since as $c$ becomes larger, the amount of calculations to solve all the stationary probabilities increases as well. In such cases, in order to determine the probability that AoI equals $n$, we have to merge more stationary probabilities, whose first age-component is $n$.  

As an example, in the following we show how the AoI-distribution can be found when the queue model used for the system is Ber/Geo/1/3. We give the stationary equations of the established stochastic process $AoI_{Ber/Geo/1/3}$, and explain these equations briefly without further calculations. 

In this case, except for the packet in service, at most two updating packets are allowed to wait in queue. Define a four-dimensional state vector $s=(v_1,v_2,v_3,v_4)$, $v_1>v_2>v_3>v_4 \geq 0$, where $v_1$ denotes the AoI at the receiver, $v_2$ records the age of the packet that is currently in service, and $v_3$, $v_4$ are the ages of packets in queue. Accordingly, constituting the four-dimensional discrete stochastic process as  
\begin{equation}
AoI_{Ber/Geo/1/3} = \{(v_{1t},v_{2t},v_{3t},v_{4t}),v_{1t}>v_{2t}>v_{3t}>v_{4t}\geq 0,t\in \mathbb{N}\} \notag 
\end{equation}

\begin{Theorem}
Let $\pi_{(v_1,v_2,v_3,v_4)}$ be the stationary probability of the age-state $(v_1,v_2,v_3,v_4)$. Then, the stationary equations describing the steady state of process $AoI_{Ber/Geo/1/3}$ are written as  
\begin{align}
\begin{cases}
\pi_{(v_1,v_2,v_3,v_4)}=\pi_{(v_1-1,v_2-1,v_3-1,v_4-1)}(1-\gamma)  &   (v_1>v_2>v_3>v_4 \geq 2)     \\
\pi_{(v_1,v_2,v_3,1)}=\pi_{(v_1-1,v_2-1,v_3-1,0)}p(1-\gamma)  &  (v_1>v_2>v_3 \geq 2)    \\
\pi_{(v_1,v_2,v_3,0)}=\pi_{(v_1-1,v_2-1,v_3-1,0)}(1-p)(1-\gamma)  + \left( \sum\nolimits_{k=v_1}^{\infty} \pi_{(k,v_1-1,v_2-1,v_3-1)} \right) \gamma  &    (v_1>v_2>v_3 \geq 2)       \\
\pi_{(v_1,v_2,1,0)}=\pi_{(v_1-1,v_2-1,0,0)}p(1-\gamma)  +\left(  \sum\nolimits_{k=v_1}^{\infty}\pi_{(k,v_1-1,v_2-1,0)}\right) p\gamma  & (v_1>v_2 \geq 2)     \\
\pi_{(v_1,v_2,0,0)}=\pi_{(v_1-1,v_2-1,0,0)}(1-p)(1-\gamma) +\left( \sum\nolimits_{k=v_1}^{\infty}\pi_{(k,v_1-1,v_2-1,0)}\right) (1-p)\gamma   &   (v_{1}>v_{2} \geq 2)     \\
\pi_{(v_1,1,0,0)}=\pi_{(v_1-1,0,0,0)}p(1-\gamma)  +\left(\sum\nolimits_{k=v_1}^{\infty}\pi_{(k,v_1-1,0,0)} \right)p\gamma & (v_{1} \geq 2)   \\
\pi_{(v_1,0,0,0)}=\pi_{(v_1-1,0,0,0)} (1-p)  +\left( \sum\nolimits_{k=v_1}^{\infty}\pi_{(k,v_1-1,0,0)}\right) (1-p)\gamma  & (v_{1} \geq 2)    \\
\pi_{(1,0,0,0)}= \left( \sum\nolimits_{k=1}^{\infty}\pi_{(k,0,0,0)} \right)p\gamma  
\end{cases}
\end{align}
\end{Theorem}

Notice that equations (75) are listed in the order of the number of packets in system, from the updating system is full to the system is finally empty. All the equations above is obtained and explained by considering both the random packet arrivals and packet service in transmitter simultaneously in each time slot.

We will not solve the system of equations (75) and calculate the AoI-distribution of size 3 system in this paper. It is interesting to consider the stationary distribution of the AoI for the system with a general Ber/Geo/1/$c$ queue, and discuss the influence of queue size $c$ on the AoI-distributions when the $(c+1)$-dimensional AoI stochastic process runs in steady state, which is one of our future works.

\section{AoI-Distribution: Status Updating System with Ber/Geo/1/2* queue}  
Assume that one extra packet can wait in queue, it has been proved that replacing the packet in queue constantly can further decrease the average AoI of the system. In this Section, we shall calculate the AoI-distribution for this case, that is, the system having Ber/Geo/1/2* queue, also by constituting a three-dimensional AoI stochastic process.

The idea and analysis to characterize the state transfers of three-dimensional age-states are similar to that we have discussed in previous Section IV. Therefore, here we directly give the stationary equations of the stochastic process $AoI_{Ber/Geo/1/2^*}$, which is defined for this case.

\subsection{Describing and determining stationary AoI-distribution}
\begin{Theorem}
For the status updating system with Ber/Geo/1/2* queue, constituting the following AoI stochastic process
\begin{equation}
AoI_{Ber/Geo/1/2^*} = \{(n_k,m_k,l_k),n_k>m_k>l_k\geq0,k\in\mathbb{N}\}  \notag
\end{equation}
where $n_k$, $m_k$, and $l_k$ represent the AoI, the packet age in system's server, and the age of packet in queue in the $k$th time slot, respectively. Let the process reaches the steady state, the stationary probabilities $\pi_{(n,m,l)}$, $n>m>l\geq0$ satisfy 
\begin{equation}
\begin{cases}
\pi_{(n,m,l)}=\pi_{(n-1,m-1,l-1)}(1-p)(1-\gamma)  &  (n>m>l \geq 2)  \\
\pi_{(n,m,1)}=\pi_{(n-1,m-1,0)}p(1-\gamma) + \left(\sum\nolimits_{j=1}^{m-2}\pi_{(n-1,m-1,j)}\right) p(1-\gamma)  &  (n>m \geq 3)  \\
\pi_{(n,2,1)}=\pi_{(n-1,1,0)}p(1-\gamma)    &  (n\geq3)   \\
\pi_{(n,m,0)}=\pi_{(n-1,m-1,0)}(1-p)(1-\gamma)  + \left(\sum\nolimits_{k=n}^{\infty}\pi_{(k,n-1,m-1)}\right) (1-p)\gamma & (n>m \geq 2)  \\
\pi_{(n,1,0)}=\pi_{(n-1,0,0)}p(1-\gamma) + \left( \sum\nolimits_{k=n}^{\infty}\sum\nolimits_{j=0}^{n-2}\pi_{(k,n-1,j)} \right) p\gamma  & (n \geq 2)  \\
\pi_{(n,0,0)}=\pi_{(n-1,0,0)}(1-p)  + \left( \sum\nolimits_{k=n}^{\infty}\pi_{(k,n-1,0)} \right) (1-p)\gamma  &  (n \geq 2)  \\
\pi_{(1,0,0)}=\left(\sum\nolimits_{k=1}^{\infty}\pi_{(k,0,0)} \right) p\gamma 
\end{cases}
\end{equation}
\end{Theorem}

\begin{proof}
Since the packet in queue can be substituted, then a new packet can always enter the transmitter as if the queue is empty. As a result, to make $(n-1,m-1,l-1)$ transfer to $(n,m,l)$, it requires that no packet arrives and the packet service is not over. This gives the first line of (76).

To obtain state vector $(n,m,1)$, we consider two cases. Firstly, assume that before one time slot the age state is $(n-1,m-1,0)$, that is the queue is empty. Then, let a new packet comes and the service of the packet in server continues, we will get the state vector $(n,m,1)$. On the other hand, begin with $(n-1,m-1,j)$, $1\leq j\leq m-2$, an arriving packet will replace the packet in queue, meanwhile, suppose that the packet service is not complete, observing that the state will also jump to $(n,m,1)$. Collecting all of these cases, we obtain the second equation in (76).

The stationary equation for the age state $(n,m,0)$ is similar to that of (39), but when consider the state transfers from $(k,n-1,m-1)$, we have to ensure that the packet in queue is not replaced at that time, i.e., it requires that no packet comes.

For the state $(n-1,0,0)$, it is easy to see that let a packet arrive and stay in server after one time slot, it will jump to $(n,1,0)$. In addition, consider the state vector $(k,n-1,j)$ where $k\geq n$ and $0\leq j\leq n-2$, when the packet in queue is refreshed and the packet in service departs system's transmitter, after one time slot we get the state $(n,1,0)$ as well. Summarize all these cases yields the fourth line of equations (76). Observing that when $j=0$, the queue is empty. In this case, the packet in queue is "refreshed" means a new packet arrives. 

The explanations to last two lines are the same as that of (39), just notice that when the packet's service ends, let the packet in queue stay in system, not be substituted. With above discussions, we complete the proof of Theorem 10.   
\end{proof}

Solving the system of equations (76) is more difficult than (39). Compared with the second equation in (39), observing that the second line of (76) says that probability $\pi_{(n,m,1)}$ is related with more other stationary probabilities. In the following, we will prove that all the stationary probabilities $\pi_{(n,m,l)}$ can also be solved, thus the distribution of AoI is obtained.

We first give some intermediate results.  
\begin{Lemma}
Let the stochastic process $AoI_{Ber/Geo/1/2^*}$ reaches the steady state, the stationary probabilities $\pi_{(n,m,l)}$ satisfy 
\begin{equation}
\pi_{(n,m,l)}=\left( \sum\nolimits_{j=0}^{m-l-1}\pi_{(n-l,m-l,j)} \right)p(1-\gamma)[(1-p)(1-\gamma)]^{l-1} \qquad (n>m>l\geq1)  
\end{equation}
and 
\begin{equation}
\pi_{(n,m,0)}=\pi_{(n-m,0,0)}p(1-\gamma)[(1-p)(1-\gamma)]^{m-1}+p\gamma mt_{n-m}[(1-p)(1-\gamma)]^{m-1} \qquad (n>m\geq1)
\end{equation}
where 
\begin{equation}
t_n=\sum\nolimits_{k=n+1}^{\infty}\sum\nolimits_{j=0}^{n-1}\pi_{(k,n,j)}  \qquad  (n\geq1)   \notag  
\end{equation}
is the probability the middle age-component equals $n$.  
\end{Lemma}

\begin{proof}
According to the first two lines of (76), we have 
\begin{align}
\pi_{(n,m,l)}&=\pi_{(n-1,m-1,l-1)}(1-p)(1-\gamma)=\dots=\pi_{(n-l+1,m-l+1,1)}[(1-p)(1-\gamma)]^{l-1} \notag  \\
&=\left(\sum\nolimits_{j=0}^{m-l-1}\pi_{(n-l,m-l,j)} \right)p(1-\gamma)[(1-p)(1-\gamma)]^{l-1}
\end{align}
which gives equations (77).  

Using (77), the third equation in (76) can be rewritten as 
\begin{align}
\pi_{(n,m,0)}&=\pi_{(n-1,m-1,0)}(1-p)(1-\gamma) + \left( \sum\nolimits_{k=n}^{\infty}\pi_{(k,n-1,m-1)} \right)(1-p)\gamma  \notag \\
&=\pi_{(n-1,m-1,0)}(1-p)(1-\gamma) + \left( \sum\nolimits_{k=n}^{\infty}\left\{\left( \sum\nolimits_{j=0}^{n-m-1}\pi_{(k-m+1,n-m,j)}\right) p(1-\gamma)[(1-p)(1-\gamma)]^{m-2}\right\} \right)(1-p)\gamma  \notag \\
&=\pi_{(n-1,m-1,0)}(1-p)(1-\gamma) + p\gamma\left( \sum\nolimits_{k=n}^{\infty} \sum\nolimits_{j=0}^{n-m-1}\pi_{(k-m+1,n-m,j)}\right) [(1-p)(1-\gamma)]^{m-1}  \notag \\
&=\pi_{(n-1,m-1,0)}(1-p)(1-\gamma) + p\gamma t_{n-m}[(1-p)(1-\gamma)]^{m-1} 
\end{align}
where we define 
\begin{equation}
t_n=\sum\nolimits_{k=n+1}^{\infty} \sum\nolimits_{j=0}^{n-1}\pi_{(k,n,j)}  \qquad  (n\geq1)
\end{equation}
which is the probability that the second state component equal to $n$.

Applying equation (80) iteratively yields that 
\begin{align}
\pi_{(n,m,0)}&=\pi_{(n-1,m-1,0)}(1-p)(1-\gamma) + p\gamma t_{n-m}[(1-p)(1-\gamma)]^{m-1} \notag \\
&=\left\{ \pi_{(n-2,m-2,0)}(1-p)(1-\gamma) + p\gamma t_{n-m}[(1-p)(1-\gamma)]^{m-2} \right\}(1-p)(1-\gamma) +  p\gamma t_{n-m}[(1-p)(1-\gamma)]^{m-1} \notag \\
&=\pi_{(n-2,m-2,0)}[(1-p)(1-\gamma)]^2 + p\gamma t_{n-m} 2[(1-p)(1-\gamma)]^{m-1}   \notag \\
& \qquad \qquad \qquad \qquad \qquad \qquad \qquad \qquad \qquad \qquad   \vdots  \notag \\
&=\pi_{(n-m+1,1,0)}[(1-p)(1-\gamma)]^{m-1} + p\gamma t_{n-m} (m-1)[(1-p)(1-\gamma)]^{m-1}   \notag  \\
&=\left\{ \pi_{(n-m,0,0)}p(1-\gamma) + \left(\sum\nolimits_{k=n-m+1}^{\infty}\sum\nolimits_{j=0}^{n-m-1}\pi_{(k,n-m,j)} \right)p\gamma \right\}[(1-p)(1-\gamma)]^{m-1}  \notag \\
& \qquad  + p\gamma t_{n-m} (m-1)[(1-p)(1-\gamma)]^{m-1}   \\
&=\pi_{(n-m,0,0)}p(1-\gamma) [(1-p)(1-\gamma)]^{m-1} + p\gamma mt_{n-m} [(1-p)(1-\gamma)]^{m-1}  
\end{align}

Thus, we prove the formula (78). To obtain (82), we have used the fourth line of (76). This completes the proof of Lemma 4.  
\end{proof}

\begin{Theorem}
The stationary probabilities $\pi_{(n,0,0)}$ are derived to be 
\begin{equation}
\pi_{(n,0,0)}=p(1-p)^{n-1}-p(1-\gamma) \left[1 + \frac{p\gamma(p+\gamma-p\gamma)}{(p+\gamma-p\gamma)^2-p\gamma} n\right] [(1-p)(1-\gamma)]^{n-1} \quad  (n\geq1) 
\end{equation}
\end{Theorem}

In Theorem 11, the probabilities of state vectors $(n,0,0)$, $n\geq1$ are determined. We will prove Theorem 11 in Appendix F.

Notice that if $t_n$ is known, then according to (78), the probabilities $\pi_{(n,m,0)}$ can be obtained. We solve $t_n$, $n\geq0$ in following Lemma 5. When $n=0$, $t_0$ denotes the sum $\sum\nolimits_{k=1}^{\infty}\pi_{(k,0,0)}$, which is the probability that the middle vector component is zero. 
\begin{Lemma}
It shows that 
\begin{equation}t_n=
\begin{cases}
\frac{(1-p)\gamma^2}{(p+\gamma-p\gamma)^2-p\gamma}   & \quad   (n=0)  \\
\tilde{\delta}_1(1-\gamma)^{n-1} - \tilde{\delta}_2[(1-p)(1-\gamma)]^{n-1}   & \quad   (n\geq1) 
\end{cases}
\end{equation}
in which 
\begin{equation}
\tilde{\delta}_1=\frac{p\gamma(1-\gamma)(p+2\gamma-2p\gamma)}{(p+\gamma-p\gamma)^2-p\gamma},  \qquad  
\tilde{\delta}_2=\frac{p\gamma(1-p)(1-\gamma)(p+\gamma-p\gamma)}{(p+\gamma-p\gamma)^2-p\gamma}  \notag 
\end{equation}
\end{Lemma}

Lemma 5 is proved in Appendix G. With these results, we then give the probabilities $\pi_{(n,m,0)}$, $n>m\geq1$.   
\begin{Theorem}
The probabilities $\pi_{(n,m,0)}$ are equal to 
\begin{align}
\pi_{(n,m,0)}&=p^2(1-p)^{n-2}(1-\gamma)^m[1-(1-\gamma)^{n-m}] + \frac{(p\gamma)^2(p+2\gamma-2p\gamma)}{(p+\gamma-p\gamma)^2-p\gamma}m(1-p)^{m-1}(1-\gamma)^{n-1} \notag \\
& \qquad - \frac{p^2\gamma(1-\gamma)(p+\gamma-p\gamma)}{(p+\gamma-p\gamma)^2-p\gamma}\left[p(1-\gamma)n + (\gamma-p)m \right] [(1-p)(1-\gamma)]^{n-2}  \qquad  (n>m\geq1)
\end{align}
\end{Theorem}

\begin{proof}
According to equation (78), we have 
\begin{align}
\pi_{(n,m,0)}&=\pi_{(n-m,0,0)}p(1-\gamma)[(1-p)(1-\gamma)]^{m-1} + p\gamma mt_{n-m}[(1-p)(1-\gamma)]^{m-1}  \notag \\
&=\left\{ p(1-p)^{n-m-1}- p(1-\gamma)\left[1+\frac{p\gamma(p+\gamma-p\gamma)}{(p+\gamma-p\gamma)^2-p\gamma}(n-m)\right] [(1-p)(1-\gamma)]^{n-m-1}\right\} p(1-\gamma)[(1-p)(1-\gamma)]^{m-1}   \notag \\
& \qquad + p\gamma m \left(\tilde{\delta}_1(1-\gamma)^{n-m-1} - \tilde{\delta}_2[(1-p)(1-\gamma)]^{n-m-1} \right) [(1-p)(1-\gamma)]^{m-1}  \notag \\
&=p^2(1-p)^{n-2}(1-\gamma)^m - p^2(1-\gamma)^2[(1-p)(1-\gamma)]^{n-2} - \frac{p^3\gamma(1-\gamma)^2(p+\gamma-p\gamma)}{(p+\gamma-p\gamma)^2-p\gamma}(n-m)[(1-p)(1-\gamma)]^{n-2}  \notag \\
& \qquad +  \frac{(p\gamma)^2(1-\gamma)(p+2\gamma-2p\gamma)}{(p+\gamma-p\gamma)^2-p\gamma}m(1-p)^{m-1}(1-\gamma)^{n-2} - \frac{(p\gamma)^2(1-p)(1-\gamma)(p+\gamma-p\gamma)}{(p+\gamma-p\gamma)^2-p\gamma}m[(1-p)(1-\gamma)]^{n-2}  \notag \\
&=p^2(1-p)^{n-2}(1-\gamma)^m[1-(1-\gamma)^{n-m}] + \frac{(p\gamma)^2(p+2\gamma-2p\gamma)}{(p+\gamma-p\gamma)^2-p\gamma}m(1-p)^{m-1}(1-\gamma)^{n-1} \notag \\
& \qquad - \frac{p^2\gamma(1-\gamma)(p+\gamma-p\gamma)}{(p+\gamma-p\gamma)^2-p\gamma}\left[p(1-\gamma)n + (\gamma-p)m \right] [(1-p)(1-\gamma)]^{n-2}
\end{align}

During above calculations, we have substituted probability $\pi_{(n-m,0,0)}$ and the sum $t_{n-m}$ using equations (84) and (85). Therefore, the probabilities $\pi_{(n,m,0)}$ are obtained and the proof of Theorem 12 completes.  
\end{proof}

At last, the probabilities $\pi_{(n,m,l)}$, $n>m>l\geq1$ are calculated, which is given in following Theorem 13. 
\begin{Theorem}
The stationary probabilities $\pi_{(n,m,l)}$, $n>m>l\geq1$ are given as 
\begin{align}
\pi_{(n,m,l)}&=p^3(1-p)^{n-m+l-2}(1-\gamma)^m \left[ 1 - (1-\gamma)^{n-m} \right]  \notag \\
&\quad - \frac{p^2\gamma(p+\gamma-p\gamma)\left[p^2(1-\gamma)(n-m) + (1-p)\gamma\right]}{(p+\gamma-p\gamma)^2-p\gamma} (1-p)^{n-m+l-2}(1-\gamma)^{n-1}  \notag \\
& \quad  +  \frac{(p\gamma)^2(p+2\gamma-2p\gamma)}{(p+\gamma-p\gamma)^2-p\gamma}(1-p)^{l-1}(1-\gamma)^{n-1}\left[1 - (1-p)^{m-l} \right] + \frac{(p\gamma)^2(p+\gamma-p\gamma)}{(p+\gamma-p\gamma)^2-p\gamma}[(1-p)(1-\gamma)]^{n-1} 
\end{align}
\end{Theorem}

\begin{proof}
In equations (79), we have shown that 
\begin{equation}
\pi_{(n,m,l)}=\pi_{(n-l+1,m-l+1,1)}[(1-p)(1-\gamma)]^{l-1}  \notag 
\end{equation}
thus, in order to find $\pi_{(n,m,l)}$, we only need to calculate probabilities $\pi_{(n,m,1)}$ where $n>m\geq2$.

From the second line of equations (76), it shows that 
\begin{align}
\pi_{(n,m,1)}&=\pi_{(n-1,m-1,0)}p(1-\gamma) + \left( \sum\nolimits_{j=1}^{m-2}\pi_{(n-1,m-1,j)}\right)p(1-\gamma)   \notag \\
&=\pi_{(n-1,m-1,0)}p(1-\gamma) + \left( \sum\nolimits_{j=1}^{m-2}\pi_{(n-j,m-j,1)}[(1-p)(1-\gamma)]^{j-1}\right)p(1-\gamma)    
\end{align}

Similarly, we can write 
\begin{align}
\pi_{(n-1,m-1,1)}&=\pi_{(n-2,m-2,0)}p(1-\gamma) + \left( \sum\nolimits_{j=1}^{m-3}\pi_{(n-1-j,m-1-j,1)}[(1-p)(1-\gamma)]^{j-1}\right)p(1-\gamma)   \notag \\
&=\pi_{(n-2,m-2,0)}p(1-\gamma) + \left( \sum\nolimits_{\tilde j=2}^{m-2}\pi_{(n-\tilde j,m-\tilde j,1)}[(1-p)(1-\gamma)]^{\tilde j-2}\right)p(1-\gamma) 
\end{align}
where in (90), do the substitution $\tilde j=1+j$.

Compute the difference  
\begin{align}
&\pi_{(n,m,1)}-\pi_{(n-1,m-1,1)}(1-p)(1-\gamma) \notag \\
={}& \left[\pi_{(n-1,m-1,0)} - \pi_{(n-2,m-2,0)}(1-p)(1-\gamma) \right] p(1-\gamma) + \pi_{(n-1,m-1,1)}p(1-\gamma)  \notag \\
={}&p^2\gamma(1-\gamma) t_{n-m}[(1-p)(1-\gamma)]^{m-2} + \pi_{(n-1,m-1,1)}p(1-\gamma)  
\end{align}

In (91), we have used expression (78). Equation (91) shows that 
\begin{equation}
\pi_{(n,m,1)}=\pi_{(n-1,m-1,1)}(1-\gamma) + p^2\gamma(1-\gamma) t_{n-m}[(1-p)(1-\gamma)]^{m-2} 
\end{equation}
which gives a recursive relation of probabilities $\pi_{(n,m,1)}$. Repeatedly using (92), we have  
\begin{align}
\pi_{(n,m,1)}&=\pi_{(n-1,m-1,1)}(1-\gamma)  +  p^2\gamma(1-\gamma) t_{n-m}[(1-p)(1-\gamma)]^{m-2}  \notag \\
&=\left[\pi_{(n-2,m-2,1)}(1-\gamma)  +  p^2\gamma(1-\gamma) t_{n-m}[(1-p)(1-\gamma)]^{m-3}\right](1-\gamma) + p^2\gamma(1-\gamma) t_{n-m}[(1-p)(1-\gamma)]^{m-2}  \notag \\
&=\pi_{(n-2,m-2,1)}(1-\gamma)^2  +  p^2\gamma(1-\gamma) t_{n-m}\left\{ (1-\gamma)[(1-p)(1-\gamma)]^{m-3} + [(1-p)(1-\gamma)]^{m-2} \right\} \notag \\
& \qquad  \qquad  \qquad  \qquad  \qquad  \qquad  \qquad  \qquad \qquad  \qquad  \vdots  \notag \\
&=\pi_{(n-m+2,2,1)}(1-\gamma)^{m-2}  +  p^2\gamma(1-\gamma) t_{n-m}\left( \sum\nolimits_{j=0}^{m-3}(1-\gamma)^j [(1-p)(1-\gamma)]^{m-2-j} \right)  \notag \\
&=\pi_{(n-m+2,2,1)}(1-\gamma)^{m-2} + p\gamma(1-p)(1-\gamma) t_{n-m}\left\{ (1-\gamma)^{m-2} - [(1-p)(1-\gamma)]^{m-2}\right\}
\end{align}

The first term in (93) is calculated as 
\begin{align}
\pi_{(n-m+2,2,1)}(1-\gamma)^{m-2}&=\pi_{(n-m+1,1,0)}p(1-\gamma)^{m-1}  \notag \\
&=\left[ \pi_{(n-m,0,0)}p(1-\gamma) +\left( \sum\nolimits_{k=n-m+1}^{\infty}\sum\nolimits_{j=0}^{n-m-1}\pi_{(k,n-m,j)}\right) p\gamma  \right] p(1-\gamma)^{m-1}   \notag \\
&=\pi_{(n-m,0,0)}p^2(1-\gamma)^m + p^2\gamma t_{n-m} (1-\gamma)^{m-1} 
\end{align}

Therefore, we obtain that 
\begin{align}
\pi_{(n,m,1)}&=\pi_{(n-m,0,0)}p^2(1-\gamma)^m + p^2\gamma t_{n-m} (1-\gamma)^{m-1} + p\gamma(1-p) t_{n-m} (1-\gamma)^{m-1} - p\gamma t_{n-m}[(1-p)(1-\gamma)]^{m-1}  \notag \\
&=\pi_{(n-m,0,0)}p^2(1-\gamma)^m + p\gamma t_{n-m} (1-\gamma)^{m-1} - p\gamma t_{n-m}[(1-p)(1-\gamma)]^{m-1} \notag \\
&=\pi_{(n-m,0,0)}p^2(1-\gamma)^m + p\gamma t_{n-m}\left\{ (1-\gamma)^{m-1} - [(1-p)(1-\gamma)]^{m-1} \right\}
\end{align}

Finally, substituting expressions (84) and (85) yields that 
\begin{align}
\pi_{(n,m,1)}&=p^3(1-p)^{n-m-1}\left[ (1-\gamma)^m - (1-\gamma)^n \right] - \frac{p^2\gamma(p+\gamma-p\gamma)\left[p^2(1-\gamma)(n-m) + (1-p)\gamma\right]}{(p+\gamma-p\gamma)^2-p\gamma} (1-p)^{n-m-1}(1-\gamma)^{n-1}  \notag \\
& \quad  +  \frac{(p\gamma)^2(p+2\gamma-2p\gamma)}{(p+\gamma-p\gamma)^2-p\gamma}\left[(1-\gamma)^{n-1} - (1-p)^{m-1}(1-\gamma)^{n-1} \right] + \frac{(p\gamma)^2(p+\gamma-p\gamma)}{(p+\gamma-p\gamma)^2-p\gamma}[(1-p)(1-\gamma)]^{n-1} \notag 
\end{align}
and then we have   
\begin{align}
\pi_{(n,m,l)}&=\pi_{(n-l+1,m-l+1,1)}[(1-p)(1-\gamma)]^{l-1}  \notag \\
&=p^3(1-p)^{n-m+l-2}(1-\gamma)^m \left[ 1 - (1-\gamma)^{n-m} \right]  \notag \\
&\quad - \frac{p^2\gamma(p+\gamma-p\gamma)}{(p+\gamma-p\gamma)^2-p\gamma}\left[p^2(1-\gamma)(n-m) + (1-p)\gamma\right] (1-p)^{n-m+l-2}(1-\gamma)^{n-1}  \notag \\
& \quad  +  \frac{(p\gamma)^2(p+2\gamma-2p\gamma)}{(p+\gamma-p\gamma)^2-p\gamma}(1-p)^{l-1}(1-\gamma)^{n-1}\left[1 - (1-p)^{m-l} \right] + \frac{(p\gamma)^2(p+\gamma-p\gamma)}{(p+\gamma-p\gamma)^2-p\gamma}[(1-p)(1-\gamma)]^{n-1} 
\end{align}

Therefore, the probability $\pi_{(n,m,l)}$ is eventually obtained and the proof of Theorem 13 completes.  
\end{proof}

So far, all the stationary probabilities $\pi_{(n,m,l)}$, $n>m>l\geq0$ have been determined. Thus, by collecting all the probabilities having identical first age-component, we can obtain the stationary distribution of the AoI.   
\begin{Theorem}
For the status updating system having Ber/Geo/1/2* queue, the stationary AoI-distribution is determined as 
\begin{equation}
\Pr\{\Delta=n\}=
\begin{cases}
\frac{p(1-p)\gamma^3}{(p+\gamma-p\gamma)^2-p\gamma}  &  (n=1) \\
\pi_{(n,0,0)} + \sum\nolimits_{m=1}^{n-1}\pi_{(n,m,0)} + \sum\nolimits_{m=2}^{n-1} \sum\nolimits_{l=1}^{m-1}\pi_{(n,m,l)} & (n\geq2)
\end{cases}
\end{equation}
in which $\pi_{(n,0,0)}$ is given in (84), the first sum is calculated as  
\begin{align}
\sum\nolimits_{m=1}^{n-1}\pi_{(n,m,0)}&= \frac{p^2(1-\gamma)}{(1-p)\gamma}\left\{ (1-p)^{n-1} - [1+(n-1)\gamma][(1-p)(1-\gamma)]^{n-1}\right\}  \notag \\
& \quad + \frac{(p+2\gamma-2p\gamma) \gamma^2}{(p+\gamma-p\gamma)^2-p\gamma}\left\{ (1-\gamma)^{n-1} - [1+(n-1)p][(1-p)(1-\gamma)]^{n-1}\right\}  \notag \\
& \quad - \frac{p^2\gamma(1-\gamma)(p+\gamma-p\gamma)(p+\gamma-2p\gamma)}{2[(p+\gamma-p\gamma)^2-p\gamma]} n(n-1)[(1-p)(1-\gamma)]^{n-2}  \qquad  (n\geq2)
\end{align}
and when $n\geq3$, the last sum is equal to 
\begin{align}
&\sum\nolimits_{m=2}^{n-1}\sum\nolimits_{l=1}^{m-1}\pi_{(n,m,l)} \notag \\
={}&p^2(1-\gamma)^2\left\{ \frac{p}{\gamma(\gamma-p)}(1-p)^{n-2} - \frac{\gamma}{p(\gamma-p)}(1-\gamma)^{n-2} 
+ \left[\frac{1}{p} + \frac{1}{\gamma} + (n-2) \right][(1-p)(1-\gamma)]^{n-2}  \right\}   \notag \\
  {}& - \frac{p\gamma(p+\gamma-p\gamma)}{(p+\gamma-p\gamma)^2-p\gamma}\bigg( \frac{(p+\gamma-2p\gamma)(1-\gamma)}{p}\left\{ (1-\gamma)^{n-2} -[(1-p)(1-\gamma)]^{n-2} -p(n-2)[(1-p)(1-\gamma)]^{n-2} \right\}  \notag \\
  {}& \qquad - \frac{p^2(1-\gamma)^2}{2}(n-1)(n-2)[(1-p)(1-\gamma)]^{n-2}  \bigg)  \notag \\
  {}&+ \frac{\gamma^2(p+2\gamma-2p\gamma)}{(p+\gamma-p\gamma)^2-p\gamma}\left\{p(n-2)(1-\gamma)^{n-1}-2(1-p)(1-\gamma)^{n-1} + 2[(1-p)(1-\gamma)]^{n-1} +p(n-2)[(1-p)(1-\gamma)]^{n-1}  \right\} \notag \\
  {}&+ \frac{(p\gamma)^2(p+\gamma-p\gamma)}{2[(p+\gamma-p\gamma)^2-p\gamma]}(n-1)(n-2)[(1-p)(1-\gamma)]^{n-1}  \qquad (n\geq3)
\end{align}
\end{Theorem}

Observing that equation (99) is zero when we let $n=2$, which can be verified directly. Thus, we can merge the case $n=2$ into the second equation of (97). The probability expression of the AoI is relatively complex, we will prove these expressions in Appendix H. Furthermore, in order to verify that equation (97) indeed forms a proper probability distribution, in appendix H we also show that all the stationary probabilities we determined before add up to 1, although the calculations are very long.

\subsection{Stationary distribution of packet system time and packet waiting time}
Since we have known the stationary probability of every state vector $(n,m,l)$, $n>m>l\geq0$, similarly, by summing over all the tuples $(n,l)$ and $(n,m)$, the distribution of packet system time and packet waiting time can be determined as well.

In fact, we have obtained the packet system time's distribution in Lemma 5, i.e., the sums $t_n$, $n\geq0$. We rewrite the results in the following. 

Let $T$ be the random variable of the packet system time in a status updating system with Ber/Geo/1/2* queue, then its stationary distribution is given as 
\begin{equation}
\Pr\{T=m\}=\begin{cases}
\frac{(1-p)\gamma^2}{(p+\gamma-p\gamma)^2-p\gamma}   &  \qquad   (m=0)  \\
\frac{p\gamma (p+2\gamma-2p\gamma)}{(p+\gamma-p\gamma)^2-p\gamma}(1-\gamma)^m 
- \frac{p\gamma (p+\gamma-p\gamma)}{(p+\gamma-p\gamma)^2-p\gamma} [(1-p)(1-\gamma)]^m   & \qquad   (m\geq1) 
\end{cases}
\end{equation}

Now, we consider the distribution of packet waiting time $W$, $W\geq0$. First of all, 
\begin{equation}
\Pr\{W=0\}=\sum\nolimits_{\text{all the state vector $(*,*,0)$}}\pi_{(*,*,0)}=\sum\nolimits_{n=1}^{\infty}\pi_{(n,0,0)}+\sum\nolimits_{n=2}^{\infty}\sum\nolimits_{m=1}^{n-1}\pi_{(n,m,0)} = \frac{\gamma(p+\gamma-2p\gamma)}{(p+\gamma-p\gamma)^2-p\gamma}  
\end{equation}
which we have obtained in equation (216) within appendix H.

When $l\geq2$, we have following equaitons
\begin{align}
\Pr\{W=l\}&=\sum\nolimits_{n=m+1}^{\infty}\sum\nolimits_{m=l+1}^{n-1}\pi_{(n,m,l)}  \notag \\
&=\sum\nolimits_{n=m+1}^{\infty}\sum\nolimits_{m=l+1}^{n-1}\pi_{(n-1,m-1,l-1)}(1-p)(1-\gamma)   \\
&=\left( \sum\nolimits_{\tilde n=\tilde m+1}^{\infty}\sum\nolimits_{\tilde m=l}^{\tilde n-1}\pi_{(\tilde n,\tilde m,l-1)} \right) (1-p)(1-\gamma)   \\
&=\Pr\{W=l-1\}(1-p)(1-\gamma) 
\end{align}

To obtain (102), we have used the first line of (76). Within (103), do the substitutions $n-1=\tilde n$, and $m-1=\tilde m$. Applying equation (104) iteratively, in general we can write 
\begin{equation}
\Pr\{W=l\}=\Pr\{W=1\}[(1-p)(1-\gamma)]^{l-1}  \qquad   (l\geq1)
\end{equation}

In equation (190), we prove that
\begin{align}
\frac{p^2(1-\gamma)^2}{(p+\gamma-p\gamma)^2-p\gamma}&=\sum\nolimits_{l=1}^{\infty}\Pr\{W=l\} 
=\sum\nolimits_{l=1}^{\infty}\Pr\{W=1\}[(1-p)(1-\gamma)]^{l-1}  
=\frac{\Pr\{W=1\}}{p+\gamma-p\gamma}  \notag 
\end{align}
from which the first probability is determined as 
\begin{equation}
\Pr\{W=1\}=\frac{p^2(1-\gamma)^2(p+\gamma-p\gamma)}{(p+\gamma-p\gamma)^2-p\gamma} 
\end{equation}

Substituting (106) into (104), and combining with equation (101), finally we obtain that 
\begin{equation}
\Pr\{W=l\}=\begin{cases}
\frac{\gamma(p+\gamma-2p\gamma)}{(p+\gamma-p\gamma)^2-p\gamma}   & \qquad   (l=0)  \\
\frac{p^2(1-\gamma)^2 (p+\gamma-p\gamma)}{(p+\gamma-p\gamma)^2-p\gamma}[(1-p)(1-\gamma)]^{l-1}   & \qquad  (l\geq1)
\end{cases}
\end{equation}

Thus, the stationary distribution of the packet waiting time $W$ is totally obtained.

\section{Numerical Simulations}
In this Section, we provide the numerical results of the AoI-distribution for the status updating system using different packet management schemes. We will draw the distribution curves of the stationary AoI assuming that the system has various queue models, including Ber/Geo/1/1 and probabilistic Ber/Geo/1/1/$\tilde g(m)$ queue, which is given in subsection A below. In part B, AoI-distribution of system with Ber/Geo/1/2 and Ber/Geo/1/2* queues are depicted. Finally, as byproducts, probability distributions of packet system time and packet waiting time are illustrated in subsection C. Apart from this, the numerical simulations of normalized negative exponent of AoI-violation probabilities are also offered.

\subsection{AoI-distribution curves: system having Ber/Geo/1/1 and Ber/Geo/1/1/$\tilde g(m)$ queues}
The AoI analysis of size 1 system is considered in Section III of the paper. Remember that we determine only part of stationary probabilities $\pi_{(n,m)}$ in Theorem 2. Even for the special case in subsection B of Section III, the undetermined coefficients $c_1$, $c_2$, and $c_3$ are temporarily ignored, since their exact expressions are complex. 

In the following, given a pair of system parameters $(p,\gamma)$, the numerical results of stationary AoI-distribution for the system using Ber/Geo/1/1 or Ber/Geo/1/1/$\tilde g(m)$ queue are computed.

Let $p=1/5$, and $\gamma=1/3$, we first determine all the probabilities $\pi_{(n,m)}$, $n>m\geq0$. To ensure that $\tilde g(m)>0$ for all $m\geq1$, we choose $N_p=4$.

According to Theorem 3, at first the probabilities $\pi_{(n,0)}$ are obtained as 
\begin{equation}
\pi_{(n,0)}=\frac{49}{215}\left(\frac{4}{5}\right)^{n-1} - \frac{98}{645}\left(\frac{8}{15}\right)^{n-1} - \frac{14}{645}n\left(\frac{8}{15}\right)^{n-1}  \qquad  (n\geq1) 
\end{equation}

In order to completely determine the probabilities $\pi_{(n,1)}$, $n\geq3$, it requires the probabilities $\pi_{(k,1)}$, $k=3,4,5$, so that the undetermined  coefficients $c_i$, $1\leq i\leq 3$ can be obtained.

Using the second relation of equations (13), by some extra calculations, it shows that  
\begin{equation}
\pi_{(3,1)}=\frac{308}{29025}, \text{ } \pi_{(4,1)}=\frac{5096}{435375}, \text{ and }  \pi_{(5,1)}=\frac{382928}{32653125}  \notag
\end{equation}

Then, by the general expressions (162) of $c_i$, $1\leq i\leq3$, we obtain that 
\begin{equation}
c_1\approx 0.07644, \text{ } c_2\approx -0.10455, \text{ and } c_3\approx 0.03154   \notag  
\end{equation}

According to Theorem 2, the probabilities $\pi_{(n,1)}$, $n\geq3$ are determined by equation (16), where the three roots of characterization equation of the deduced difference equation is $r_1=4/5$, $r_2=2/3$, and $r_3=32/75$, respectively.

For the case $n=2$, the basic relation in (13) says 
\begin{equation}
\pi_{(2,1)}=\pi_{(1,0)}p(1-\gamma)=\frac{7}{129} \times \frac{2}{15}=\frac{14}{1935}  
\end{equation}
in which the probability $\pi_{(1,0)}$ can be obtained either from (108) by letting $n=1$, or directly calculating the general expression (148).

Summarizing above results, we write that 
\begin{equation}
\pi_{(n,1)}=\begin{cases}
\frac{14}{1935}  & \qquad  (n=2)  \\
0.07644 \left(\frac{4}{5}\right)^n - 0.10455 \left(\frac{2}{3}\right)^n + 0.03154 \left(\frac{32}{75}\right)^n  & \qquad  (n\geq3) 
\end{cases}
\end{equation}

Therefore, from the recursive relation (130), we have 
\begin{align}
\pi_{(n,m)}&=\pi_{(n-m+1,1)}\left( \frac{8}{15}\right)^{m-1}\frac{4+m}{5}  \notag \\
&=\begin{cases}
\frac{14}{9675}(4+m)\left( \frac{8}{15}\right)^{m-1}   & (n-m=1)  \\
(4+m) \left[ 0.015288  \left(\frac{4}{5}\right)^n \left(\frac{2}{3}\right)^{m-1} - 0.02091  \left(\frac{2}{3}\right)^n \left(\frac{4}{5}\right)^{m-1} + 0.006308  \left(\frac{32}{75}\right)^n \left(\frac{5}{4}\right)^{m-1}\right]   &   (n-m\geq2) 
\end{cases}
\end{align}

Now, the distribution of the AoI can be calculated. Firstly, the probabilities that the AoI equals 1 and 2 are determined. 
\begin{equation}
\Pr\{\Delta=1\}=\pi_{(1,0)}=\frac{7}{129}\approx 0.05426, \quad  \Pr\{\Delta=2\}=\pi_{(2,0)}+\pi_{(2,1)}\approx 0.08537 
\end{equation}

When $n\geq3$, it shows that  
\begin{align}
\Pr\{\Delta=n\}&=\sum\nolimits_{m=0}^{n-1}\pi_{(n,m)}  \notag \\
&=\pi_{(n,0)}+\sum\nolimits_{m=1}^{n-2}\pi_{(n,m)} + \pi_{(n,n-1)} \notag \\
&=\frac{49}{215}\left(\frac{4}{5}\right)^{n-1} - \frac{98}{645}\left(\frac{8}{15}\right)^{n-1} - \frac{14}{645}n\left(\frac{8}{15}\right)^{n-1}  \notag \\
& \quad + \sum\nolimits_{m=1}^{n-2}(4+m) \left[ 0.015288  \left(\frac{4}{5}\right)^n \left(\frac{2}{3}\right)^{m-1} - 0.02091  \left(\frac{2}{3}\right)^n \left(\frac{4}{5}\right)^{m-1} + 0.006308  \left(\frac{32}{75}\right)^n \left(\frac{5}{4}\right)^{m-1}\right] \notag \\
&\qquad  + \frac{14}{9675}(3+n)\left(\frac{8}{15}\right)^{n-2}  \notag \\
&=\frac{49}{215}\left(\frac{4}{5}\right)^{n-1} - \frac{371}{2580}\left(\frac{8}{15}\right)^{n-1} - \frac{49}{2580}n\left(\frac{8}{15}\right)^{n-1} + 0.015288 \left(\frac{4}{5}\right)^n \left[ \sum\nolimits_{m=1}^{n-2}(4+m) \left(\frac{2}{3}\right)^{m-1}\right]  \notag \\
& \quad - 0.02091 \left(\frac{2}{3}\right)^n \left[ \sum\nolimits_{m=1}^{n-2}(4+m) \left(\frac{4}{5}\right)^{m-1}\right]  
+ 0.006308 \left(\frac{32}{75}\right)^n \left[ \sum\nolimits_{m=1}^{n-2}(4+m) \left(\frac{5}{4}\right)^{m-1}\right]  \notag \\
&= \frac{49}{215}\left(\frac{4}{5}\right)^{n-1} - \frac{371}{2580}\left(\frac{8}{15}\right)^{n-1} - \frac{49}{2580}n\left(\frac{8}{15}\right)^{n-1} + 0.015288 \left(\frac{4}{5}\right)^n \left[21-15\left(\frac{2}{3}\right)^{n-2} - 3n\left(\frac{2}{3}\right)^{n-2} \right]  \notag \\
&\quad -  0.02091 \left(\frac{2}{3}\right)^n \left[45-35\left(\frac{4}{5}\right)^{n-2} - 5n\left(\frac{4}{5}\right)^{n-2} \right]  
+0.006308 \left(\frac{32}{75}\right)^n \left[4n\left(\frac{5}{4}\right)^{n-2} - 8\left(\frac{5}{4}\right)^{n-2} \right] 
\end{align}

We omit the other calculation details and directly give that 
\begin{equation}
\Pr\{\Delta=n\} \approx 0.60593 \left(\frac{4}{5}\right)^n - 0.94095 \left(\frac{2}{3}\right)^n + 0.57787 \left(\frac{8}{15}\right)^n
+ 0.0407n \left(\frac{8}{15}\right)^n  \qquad (n\geq3)
\end{equation}

Combining equations (112) and (114), finally we obtain the stationary AoI-distribution as
\begin{equation}
\Pr\{\Delta=n\}\approx \begin{cases}
0.05426    & \quad  (n=1)    \\
0.08537    & \quad  (n=2)    \\
0.60593 \left(\frac{4}{5}\right)^n - 0.94095 \left(\frac{2}{3}\right)^n + 0.57787 \left(\frac{8}{15}\right)^n
+ 0.0407n \left(\frac{8}{15}\right)^n  & \quad   (n\geq3)
\end{cases}
\end{equation}

One can check that the sum of all the probabilities determined in (115) is almost equal to 1, so that (115) is indeed a proper probability distribution.

When the queue model of the system is simplified further to basic Ber/Geo/1/1 queue, the stationary distribution of the AoI is obtained directly from Corollary 3. Let $p=1/5$ and $\gamma=1/3$, equation (31) shows that 
\begin{align}
\Pr\{\Delta=n\}&= \frac{p(1-p)\gamma ^3}{(p+\gamma -p\gamma)(\gamma -p)^2}\left[(1-p)^n-(1-\gamma)^n \right] - \frac{(p\gamma)^2}{(p+\gamma -p\gamma)(\gamma -p)}n(1-\gamma)^n  \notag \\
&=\frac{5}{7}\left[ \left(\frac{4}{5}\right)^n - \left(\frac{2}{3}\right)^n\right] - \frac{1}{14} n\left(\frac{2}{3}\right)^n 
\qquad (n \geq 1) 
\end{align}

In addition, the AoI's probability mass function is calculated as 
\begin{align}
\Pr\{\Delta \leq k\}=\sum\nolimits_{n=1}^k \Pr\{\Delta=n\} 
=1- \frac{1}{7}\left[ 20 \left(\frac{4}{5}\right)^k - 13 \left(\frac{2}{3}\right)^k - k \left(\frac{2}{3}\right)^k\right] \qquad  (k\geq1)
\end{align}

\begin{figure*}[!t]
\centering
\subfloat[AoI-distribution: Ber/Geo/1/1 queue with and without service preemption]{\includegraphics[width=3.5in]{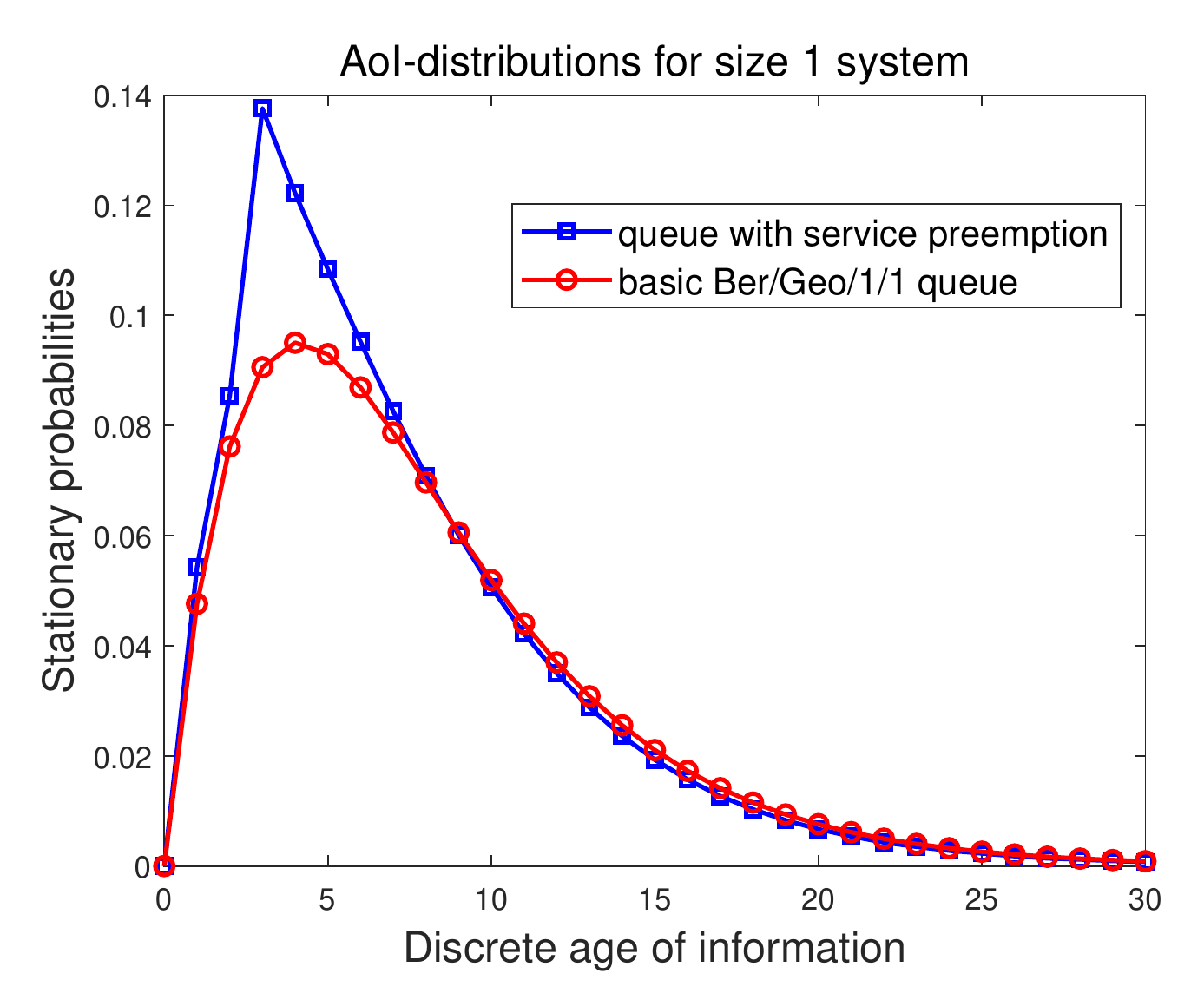}%
\label{figg0}}
\hfil
\subfloat[Cumulative probabilities of the AoI for both cases]{\includegraphics[width=3.5in]{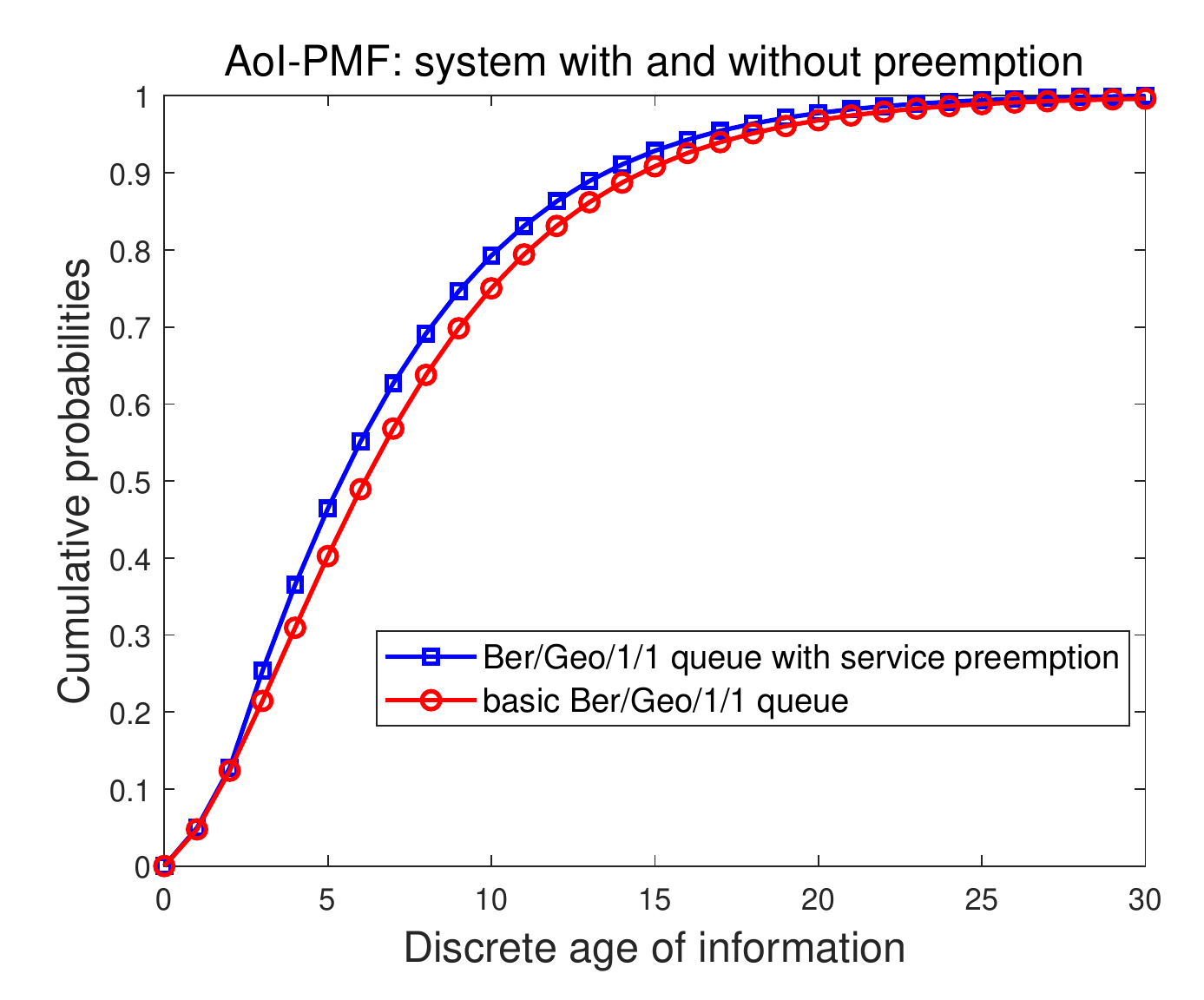}%
\label{figg1}}
\caption{The distribution and cumulative probabilities of stationary AoI for size 1 status updating system.}
\label{fig3}
\end{figure*}

In Figure \ref{fig3}, we draw the AoI-distribution and its cumulative probabilities for the system with simple Ber/Geo/1/1 queue and the queue with service preemption.

However, the numerical results do not support the expected conclusion, i.e., introducing service preemption is beneficial to enhance the timeliness of the status updating system. The reasons may include following aspects:

(i) The function $\tilde g(m)$ defined in equation (12) provides a “weak” service preemption. For example, let $m=20$, that is the packet age is 20, the preemption probability is only $\frac{1}{6}$, which is too small. The function can be modified by dropping the requirement that when $m\to\infty$, $\tilde g(m)\to1$, since in practice, the packet age may not be too larger. For instance, we can limit that $m\leq \frac{\beta}{\gamma}$, where $\beta\in \mathbb{R}$, and redefine the function $\tilde g(m)$ in the range $1\leq m\leq \beta/\gamma$. 

(ii) The parameter $\gamma$ is large. On average, the service time of each packet is approximately $\frac{1}{\gamma}$, as a result, the probability that a packet having large age (larger than 3 in our example) is small. Thus, the service preemption also occurs with a small probability, which limits the role of service preemption. 

(iii) Thirdly, even the probability is small, it is still possible that a small-age packet is replaced before its service is complete.

We believe that when all the system parameters are tuned well, service preemption scheme will help to improve the system performance, such as reducing the average value of the AoI. Discussing such better systems is also interesting, we will revisit this topic in our latter works.

\subsection{Stationary AoI-distribution: status updating system with Ber/Geo/1/2 and Ber/Geo/1/2* queues}
In equations (59) and (60), assume that the Ber/Geo/1/2 queue is used, we have obtained the probability distribution and the cumulative probabilities of the system's AoI.

Let $p=1/5$ and $\gamma=1/3$, in the following, we first compute their numerical results, then the AoI-distribution curve and AoI cumulative probabilities are depicted, along with the corresponding results for the cases that system uses Ber/Geo/1/1 queue and Ber/Geo/1/2* queue.

From (59), the stationary AoI distribution is obtained as 
\begin{equation}
\Pr\{\Delta=n\}=\frac{10}{17}\left[ \left(\frac{4}{5}\right)^n - \left(\frac{2}{3}\right)^n \right] - \frac{9}{136}n\left( \frac{2}{3} \right)^n + \frac{1}{136}n^2\left( \frac{2}{3} \right)^n   \qquad (n\geq1) 
\end{equation}

In addition, for $k\geq1$, expression (60) shows that 
\begin{equation}
\Pr\{\Delta\leq k\}=1- \frac{40}{17}\left( \frac{4}{5} \right)^k + \frac{23}{17}\left( \frac{2}{3} \right)^k + \frac{3}{68}k\left( \frac{2}{3} \right)^k - \frac{1}{68}k^2\left( \frac{2}{3} \right)^k  
\end{equation}

It is easy to obtain the AoI violation probability from expression (60), which is given as 
\begin{align}
\Pr\{\Delta > k\}&= \frac{(1-p)^3\gamma^4}{N(p,\gamma)(\gamma-p)^2}(1-p)^k - \frac{p(1-\gamma)}{N(p,\gamma)(\gamma-p)}\left[ \frac{(1-p)^2\gamma^3}{\gamma-p}+ p^2(1-\gamma) \right] (1-\gamma)^k \notag \\
&\qquad - \frac{p^2\gamma}{N(p,\gamma)}\left[ \frac{(1-p)(1-\gamma)\gamma}{\gamma-p}-\frac{2-\gamma}{2} \right] k(1-\gamma)^k+ \frac{(p\gamma)^2}{2N(p,\gamma)}k^2(1-\gamma)^k   
\end{align}

Thus, the normalized negative exponent of (120) can be determined. We have 
\begin{align}
E_{v,k,2}&=-\frac1k \log\Pr\{\Delta>k\}  \notag \\
&=-\frac1k \log \bigg\{ \frac{(1-p)^3\gamma^4}{N(p,\gamma)(\gamma-p)^2}(1-p)^k - \frac{p(1-\gamma)}{N(p,\gamma)(\gamma-p)}\left[ \frac{(1-p)^2\gamma^3}{\gamma-p}+ p^2(1-\gamma) \right] (1-\gamma)^k \notag \\
&\qquad - \frac{p^2\gamma}{N(p,\gamma)}\left[ \frac{(1-p)(1-\gamma)\gamma}{\gamma-p}-\frac{2-\gamma}{2} \right] k(1-\gamma)^k+ \frac{(p\gamma)^2}{2N(p,\gamma)}k^2(1-\gamma)^k  \bigg\}   \\
&>-\frac1k \log \bigg\{ \frac{(1-p)^3\gamma^4}{N(p,\gamma)(\gamma-p)^2}(1-p)^k  + \frac{(p\gamma)^2}{2N(p,\gamma)}k^2(1-p)^k  \bigg\}   \\
&= \log\frac{1}{1-p} -\frac1k \log \left\{ \frac{(1-p)^3\gamma^4}{N(p,\gamma)(\gamma-p)^2} + \frac{(p\gamma)^2}{2N(p,\gamma)}k^2 \right\} \notag 
\end{align}
where in (122) we simply drop the negative terms of equation (121) and notice that $1-\gamma<1-p$. The subscript “2” in $E_{v,k,2}$ represents that the system has size 2.

Now, we consider the case where the queue used in system is Ber/Geo/1/2*. The stationary AoI-distribution is determined by the complex expressions (97)-(99).

Applying the same settings, i.e., $p=1/5$ and $\gamma=1/3$, the numerical results of AoI-distribution and AoI-cumulative probabilities can be calculated explicitly. After careful calculations and repeated verifications, we obtain that  
\begin{equation}
\Pr\{\Delta=n\}=\frac{2}{5}\left( \frac{4}{5} \right)^{n-1} - \frac{131}{102}\left( \frac{2}{3} \right)^{n-1} + \frac{194}{255}\left( \frac{8}{15} \right)^{n-1} + \frac{11}{102}n\left( \frac{2}{3} \right)^{n-1} + \frac{14}{255}n\left( \frac{8}{15} \right)^{n-1}    
\qquad  (n\geq1)
\end{equation}

Observing that equation (123) gives a unified formula for the distribution of AoI, which is valid for all the cases $n\geq1$. 

According to (123), it is not hard for us to obtain the AoI violation probability as 
\begin{equation}
\Pr\{\Delta>k\}=2\left( \frac{4}{5} \right)^k - \frac{49}{17}\left( \frac{2}{3} \right)^k + \frac{32}{17}\left( \frac{8}{15} \right)^k + \frac{11}{34}k\left( \frac{2}{3} \right)^k + \frac{14}{119}k\left( \frac{8}{15} \right)^k    \qquad  (k\geq1)
\end{equation}

Therefore, the AoI-distribution function is given as 
\begin{equation}
\Pr\{\Delta\leq k\}= 1-2\left( \frac{4}{5} \right)^k + \frac{49}{17}\left( \frac{2}{3} \right)^k - \frac{32}{17}\left( \frac{8}{15} \right)^k - \frac{11}{34}k\left( \frac{2}{3} \right)^k - \frac{14}{119}k\left( \frac{8}{15} \right)^k    \qquad  (k\geq1)
\end{equation}

\begin{figure*}[!t]
\centering
\subfloat[Stationary AoI-distribution: Ber/Geo/1/2 and Ber/Geo/1/2* queues]{\includegraphics[width=3.5in]{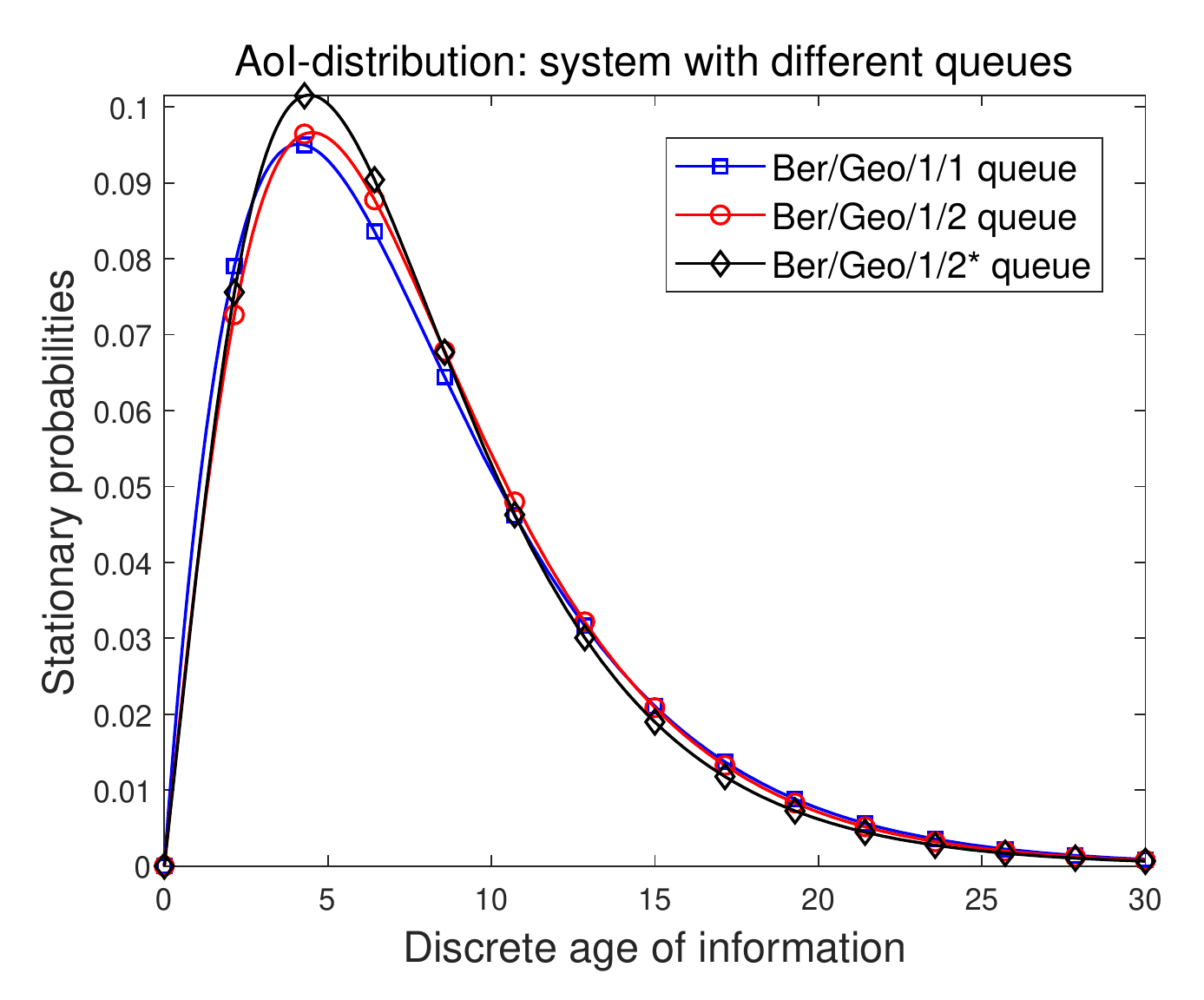}%
\label{figg2}}
\hfil
\subfloat[Cumulative probabilities of the AoI for both cases]{\includegraphics[width=3.5in]{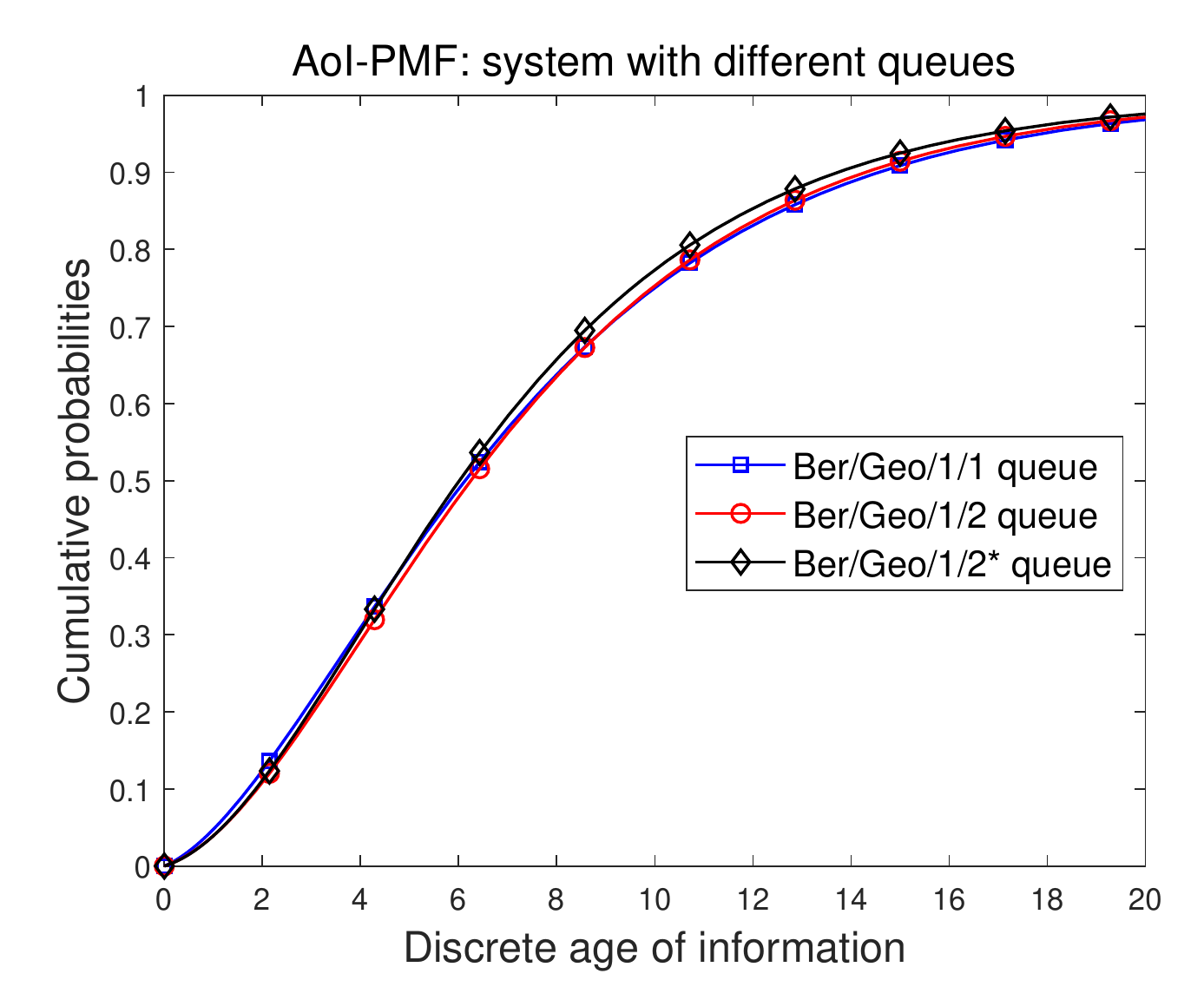}%
\label{figg3}}
\caption{Cumulative probabilities of AoI for system with different queue models.}
\label{fig4}
\end{figure*}

Provided all aboves results, now we can do the numerical simulations.

In Figure \ref{fig4}, the stationary AoI-distribution of system using different packet management policies are depicted, including system with Ber/Geo/1/1, Ber/Geo/1/2 and Ber/Geo/1/2* queues. The corresponding cumulative probabilities are also offered.

Numerical results show that the AoI-distribution curves are very similar, the differences of them are tiny. It is clear that the AoI-distribution for Ber/Geo/1/2* case has maximal peak value and smaller tail probabilities. In addition, when the age of information takes small values, we see that the blue line is above the red one, and reverses when the age of information go across the peak of the distribution curve, which means with larger probability, system using Ber/Geo/1/1 queue has small AoI, but when the AoI gets slightly larger, system having Ber/Geo/1/2 queue takes these value with larger probability. In terms of large AoI range, the tail probability curves of two cases are almost coincident, both of them are over the probability curve of Ber/Geo/1/2* queue case. This demonstrates that replacing the waiting packet can decrease the probability the AoI takes larger values.

To show the influence of different traffic load on the system's AoI distribution, in the following Figure \ref{fig5}, the distribution curves under different $\rho_d$ are depicted, where $\rho_d=p/\gamma$ is defined as the discrete traffic intensity. Let service rate $\gamma=0.6$, and set packet arrival rate as $p=\rho_d\cdot\gamma$, we change $\rho_d$ from small to large and draw the distribution curves in Figure \ref{figg_1}-Figure \ref{figg_6}. 

It shows that as $\rho_d$ is getting larger, the differences of three curves become more apparent, especially when $\rho_d$ is greater than 0.6. Similarly, the distribution curve of Ber/Geo/1/2* case has maximal peak and decreases fastest among three curves. It is more interesting to observe the trends of graphs corresponding to Ber/Geo/1/1 and Ber/Geo/1/2 queues. We see that the curves topple to the right side slightly and the peak probability gets smaller, when system size increases from 1 to 2. It is meaningful to work out how the trend is go on as system size $c$ becomes larger and determine the limiting AoI-distribution when $c$ is infinite. We shall discuss these topics in the future works.

\begin{figure*}[!t]
\centering
\subfloat[$\rho_d=0.05$]{\includegraphics[width=2.3in]{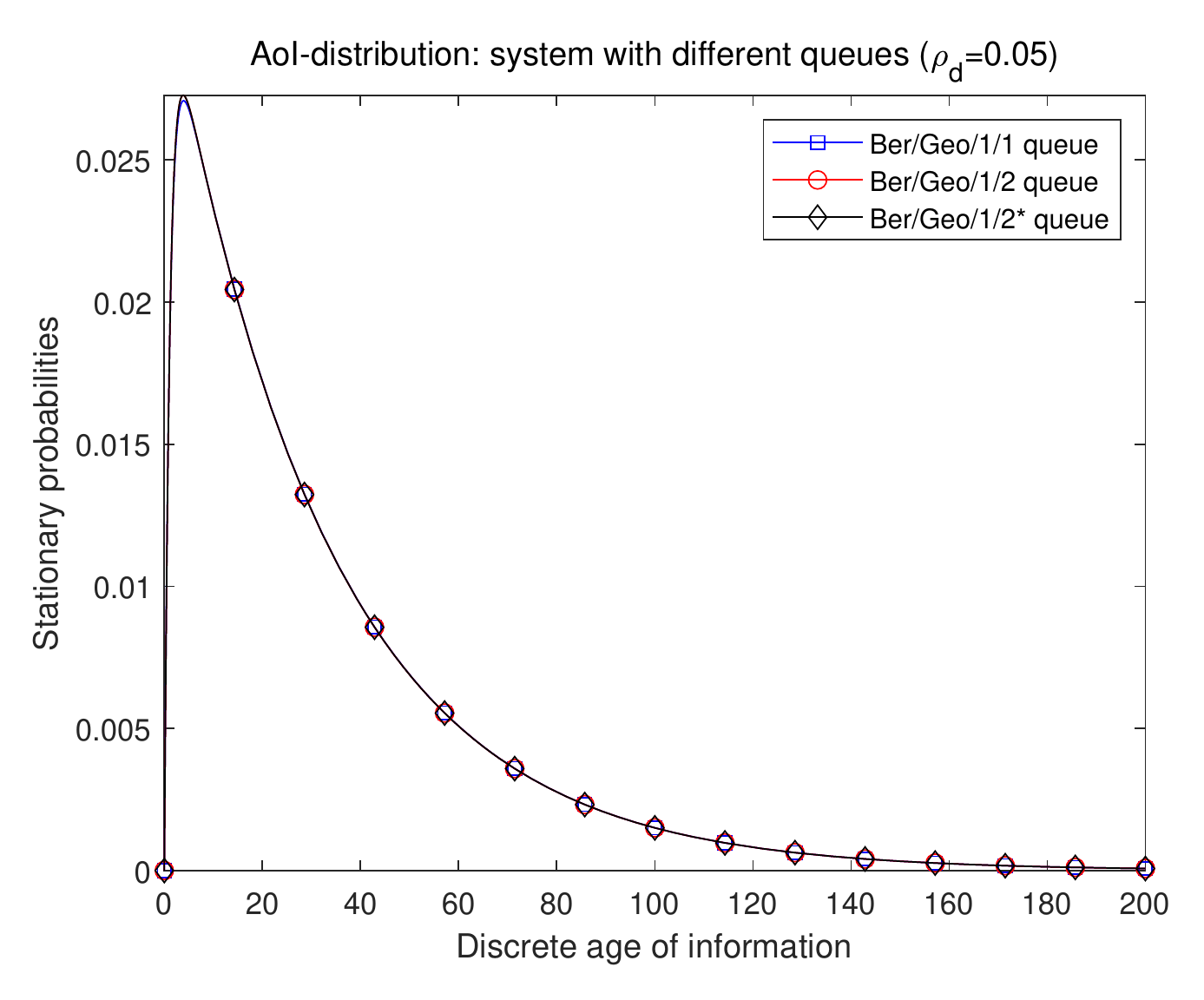}%
\label{figg_1}}
\hfil
\subfloat[$\rho_d=0.2$]{\includegraphics[width=2.3in]{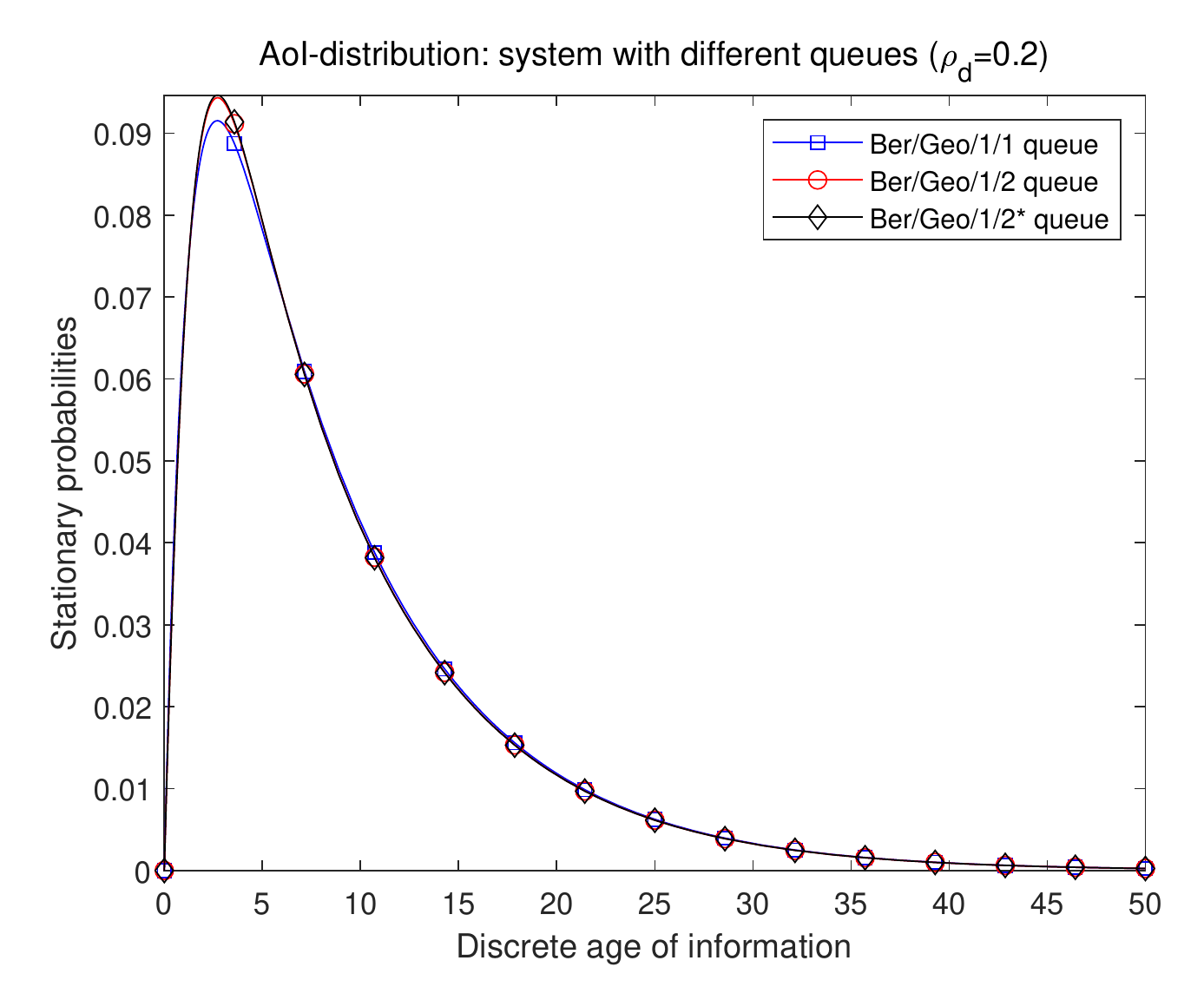}%
\label{figg_2}}
\hfil
\subfloat[$\rho_d=0.4$]{\includegraphics[width=2.3in]{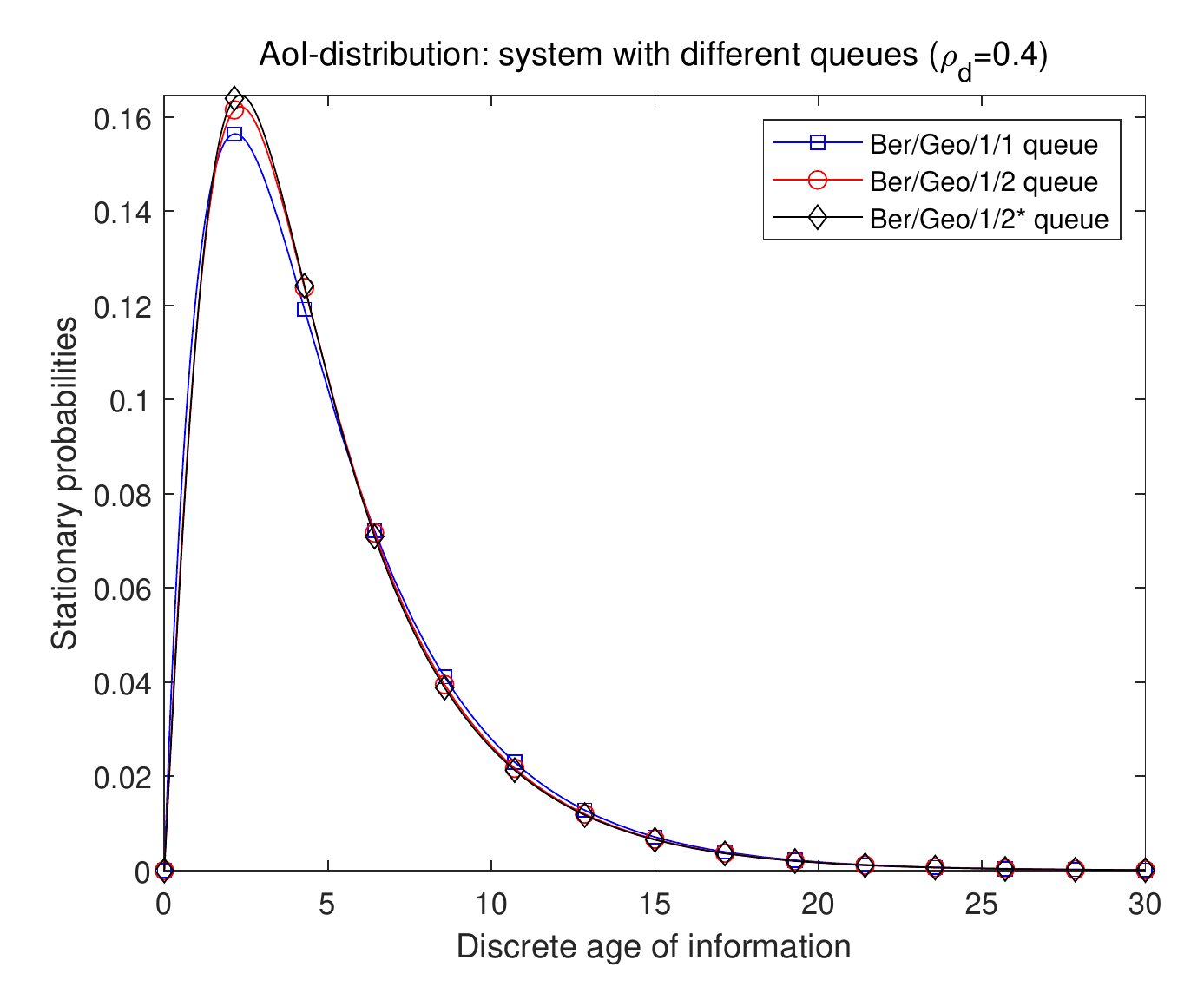}%
\label{figg_3}}
\hfil
\subfloat[$\rho_d=0.6$]{\includegraphics[width=2.3in]{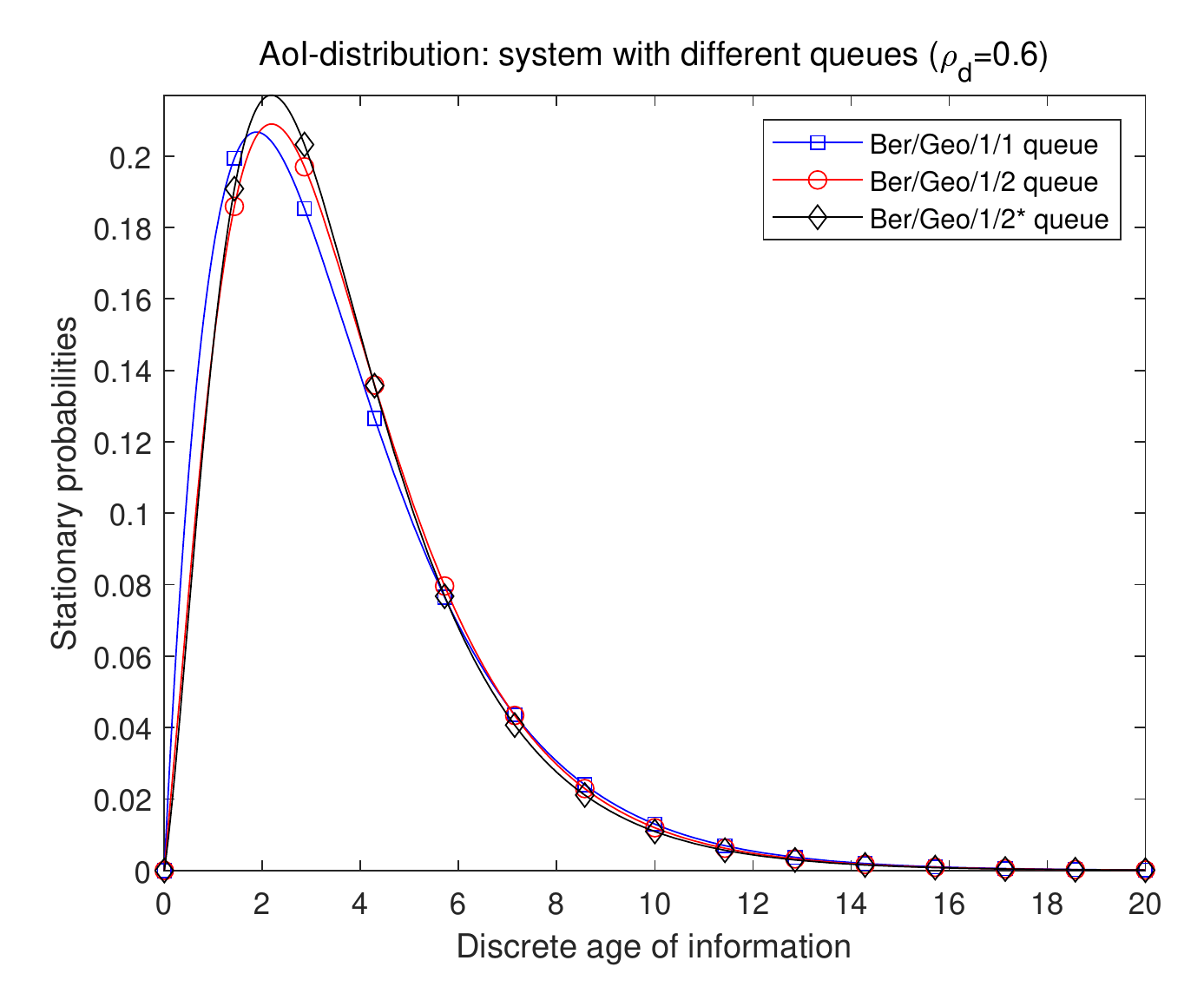}%
\label{figg_4}}
\hfil
\subfloat[$\rho_d=0.8$]{\includegraphics[width=2.3in]{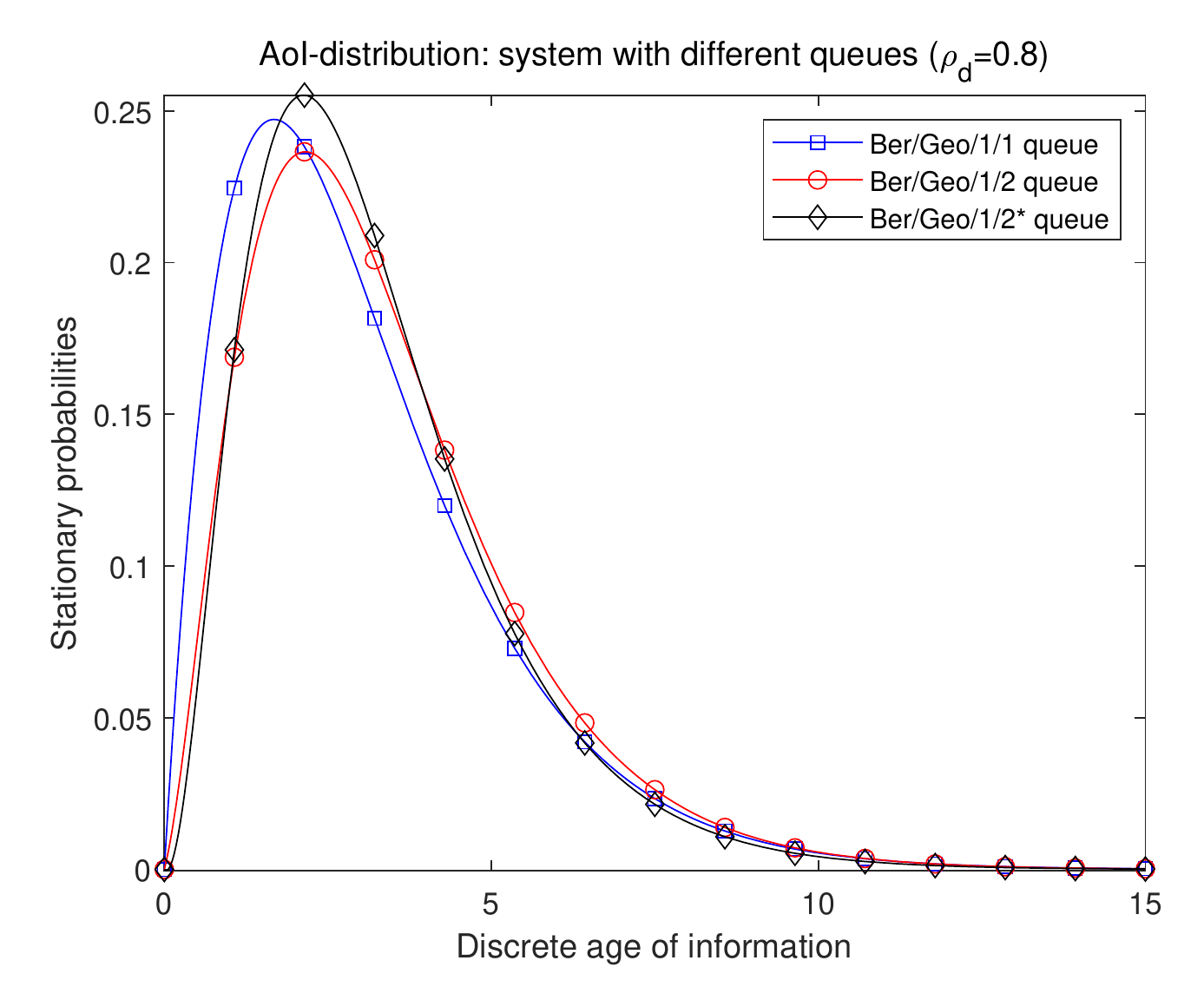}%
\label{figg_5}}
\hfil
\subfloat[$\rho_d=0.99$]{\includegraphics[width=2.3in]{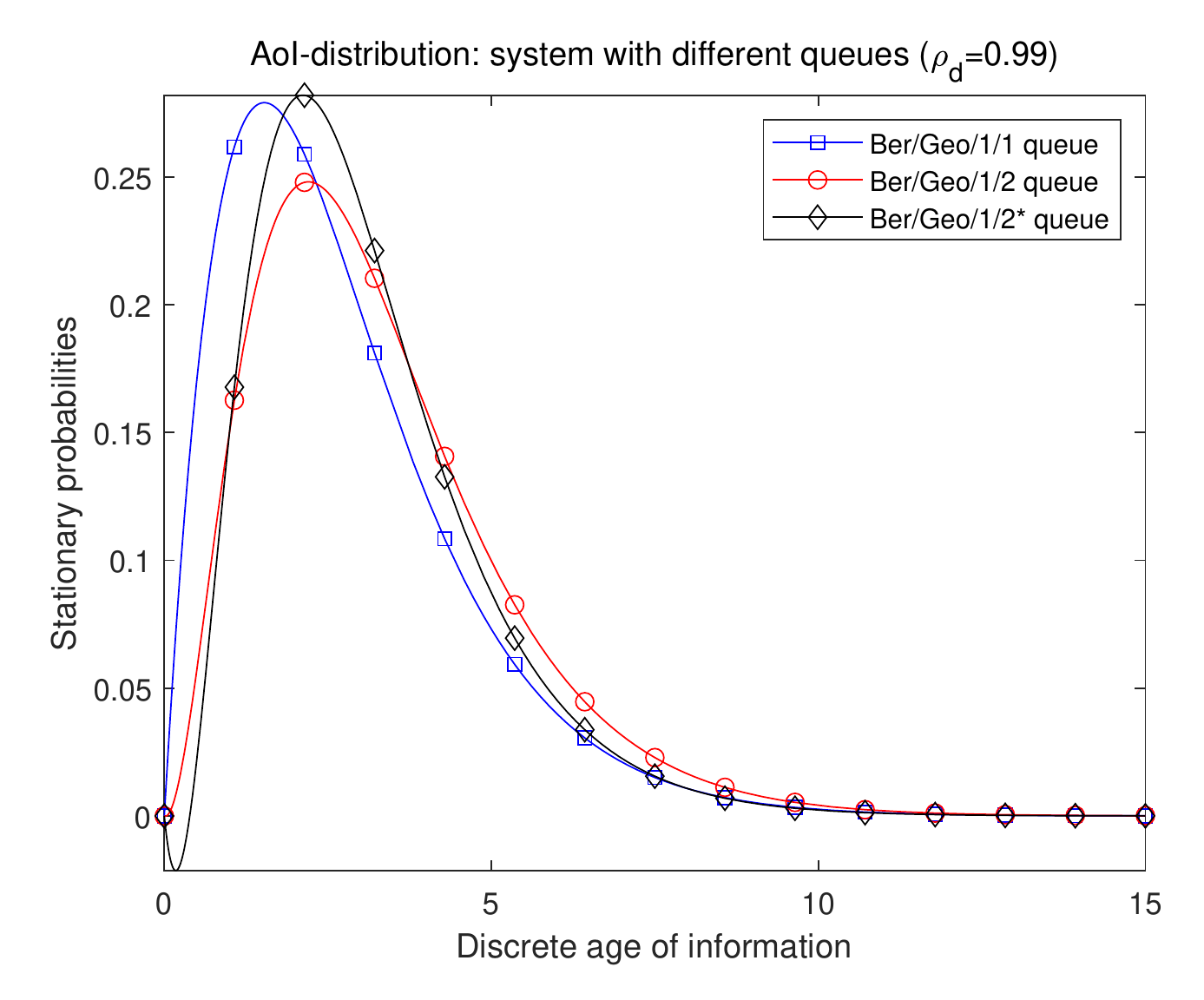}%
\label{figg_6}}
\caption{Stationary AoI distributions: status updating system with different $\rho_d$.}
\label{fig5}
\end{figure*}

\subsection{Distributions of packet system time $T$, the packet waiting time $W$ and the exponents of quality}
In this part, as byproducts, we provide the numerical simulations for the stationary distributions of packet system time $T$ and packet waiting time $W$. Notice that in fact the normalized probability distributions when $T$ or $W$ is greater than 1 are illustrated.

At first, for the system with Ber/Geo/1/2 queue, from Corollary 5 we can obtain that 
\begin{equation}
\Pr\{T=m\}=\frac{\gamma^2}{(1-\gamma)(p+\gamma-p\gamma)}[1+(m-1)p](1-\gamma)^m  \qquad  (m\geq1)
\end{equation}
and the waiting time $W$ is distributed as 
\begin{equation}
\Pr\{W=l\}=\gamma (1-\gamma)^{l-1}  \qquad  (\l\geq1)
\end{equation}
which is a geometric distribution with parameter $\gamma$.

For the case the queue model used in system is Ber/Geo/1/2*, these stationary distributions are determined in equation (100) and (107). We have 
\begin{equation}
\Pr\{T=m\}=\frac{\gamma(p+2\gamma-2p\gamma)}{p+\gamma-p\gamma}(1-\gamma)^{m-1} - \frac{\gamma}{1-\gamma}[(1-p)(1-\gamma)]^m  \qquad  (m\geq1)
\end{equation}

In addition, the distribution of $W$ is determined as 
\begin{equation}
\Pr\{W=l\}=(p+\gamma-p\gamma)[(1-p)(1-\gamma)]^{l-1} \qquad  (l\geq1)
\end{equation}

\begin{figure*}[!t]
\centering
\subfloat[Stationary distribution of packet system time]{\includegraphics[width=3.5in]{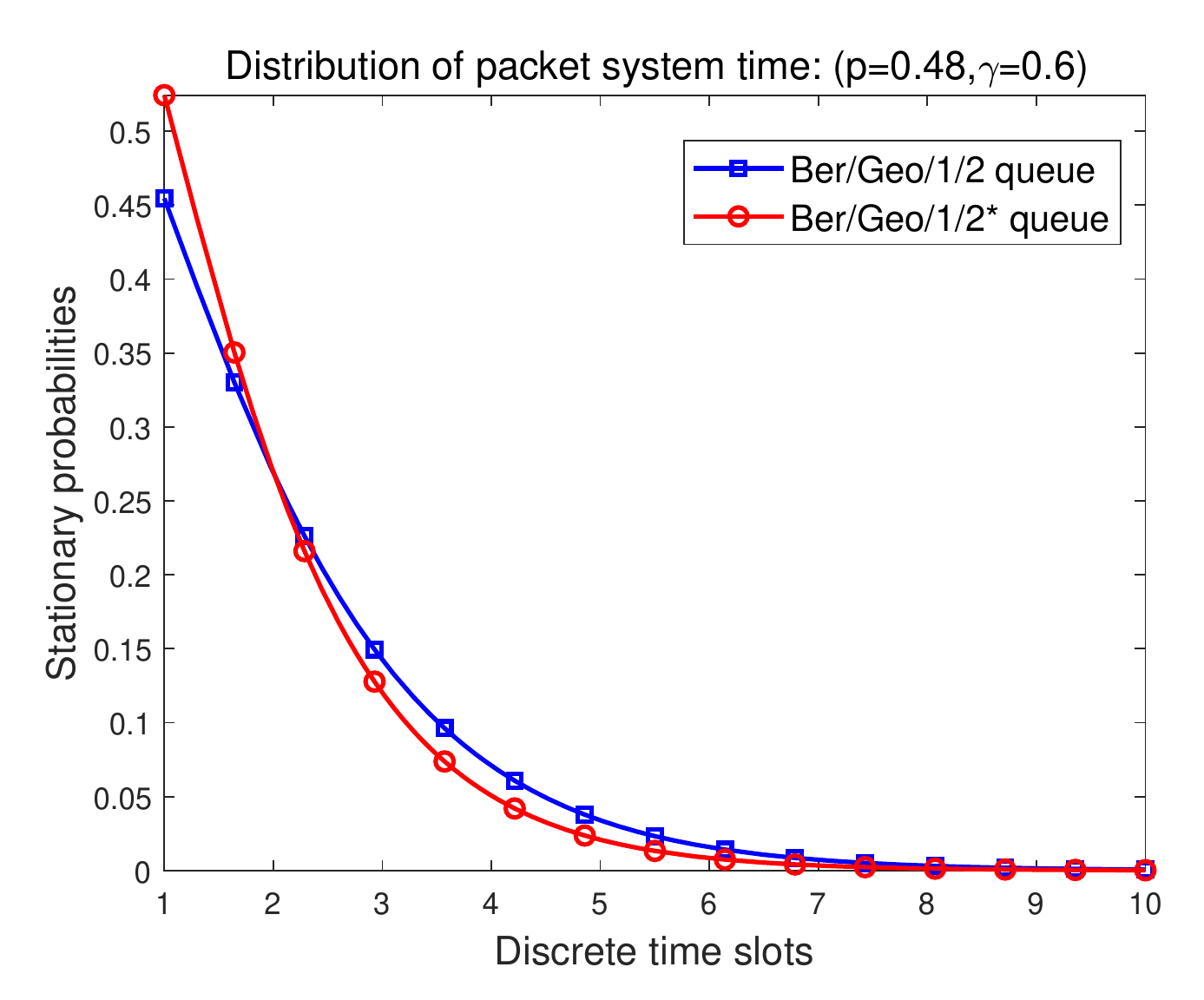}%
\label{fik1}}
\hfil
\subfloat[The distribution of packet waiting time]{\includegraphics[width=3.5in]{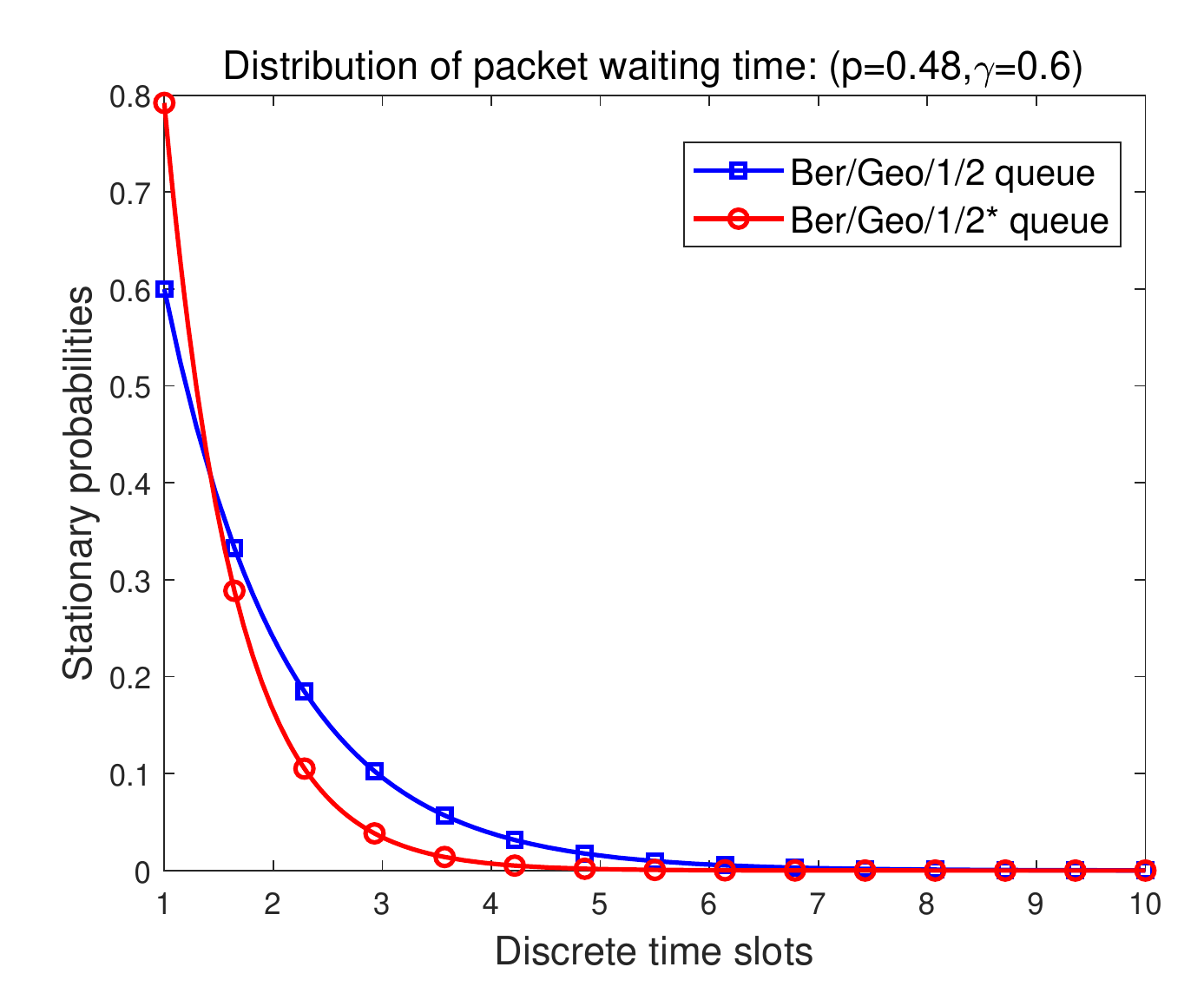}%
\label{fik2}}
\caption{The distribution of packet system and packet waiting time for size 2 status updating system.}
\label{fig6}
\end{figure*}

When the size of status updating system is 2, let $p=0.48$ and $\gamma=0.6$, we draw the distribution curves of $T$ and $W$ for two queue models in Figure \ref{fig6}.

It is understandable that system using Ber/Geo/1/2* queue has lower waiting time (if we calculate the average waiting time of both cases), since the packet in queue can be refreshed by the fresher packets. For the packet system time $T$, although the service rate is identical, however, when the packet replacement is allowable, in the long term the system time will be reduced since the waiting time is shorter. As long as a newer packet refreshes the old one in queue during the service time of the former packet, then at next time, actually a packet of lower age is served in system's server. Since packet system time is the sum of waiting time and service time, and the system equipped with Ber/Geo/1/2* queue has lower waiting time, as a result, whose packet system time is also smaller.

\begin{figure}[!t]
\centering
\includegraphics[width=4in]{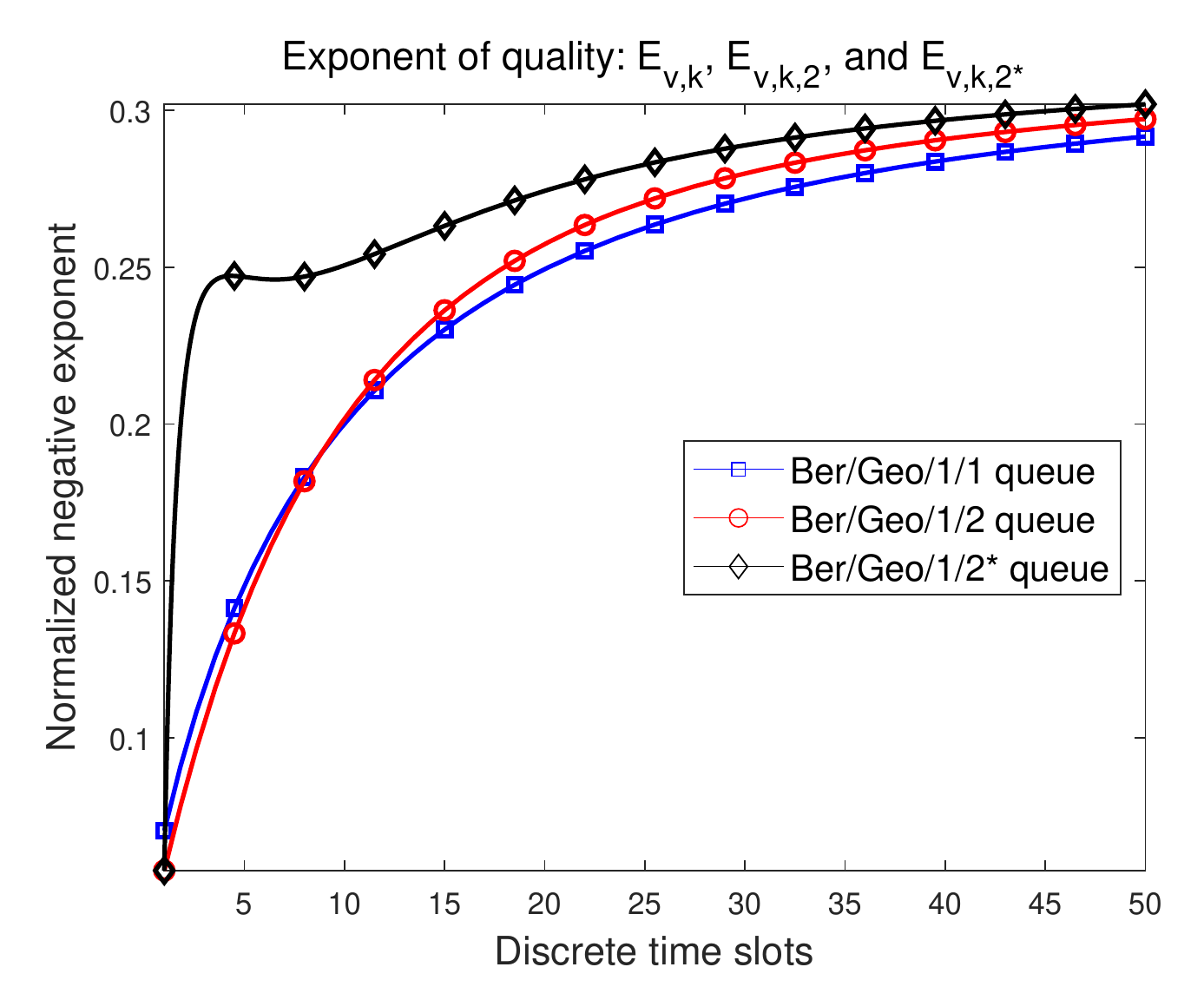}
\caption{Exponents of quality of status updating system using different queue models.}
\label{fig7}
\end{figure}

Let the status updating system uses different queue models, according to equation (33) and numerical results (119), (124), in Figure \ref{fig7} we give the graphs of three exponents of qualify, each of which corresponds to one queue. The three exponents are denoted as $E_{v,k}$, $E_{v,k,2}$, and $E_{v,k,2^*}$, respectively.

It shows that among the three cases, system with Ber/Geo/1/2* queue has the largest decreasing rate of the AoI violation probability. Compared with Ber/Geo/1/1 queue case, the exponent $E_{v,k,2}$ is greater than that of a system using Ber/Geo/1/1 queue when the AoI is large, more precisely, for the cases $\Delta\geq15$.

\section{Conclusion}
Within the past few years, the discussion about the age of information has developed to be a quite rich research theme. As a timeliness metric, the limiting time average AoI of a status updating system has been considered and determined under a variety of queue models in many articles. It is worth noting that two new topics about the AoI are causing concern recently, one is the AoI analysis for discrete time status updating system where the queue model is different from that used in a continuous time system, and the other is finding stationary distribution of the AoI when the system reaches ths steady state.

In this paper, we consider applying several packet management policies to discrete time updating system and determining the stationary AoI distribution in the destination. Notice that not the random dynamics of AoI but the entire age-state can be characterized, which contains the AoI, the age of packet that is in service, and the age of packet waiting in queue, for each case, a larger-dimensional stochastic process is constituted. We find the steady state of established AoI process by solving a group of balance equaitons which is called the stationary equations of the process. The AoI distribution is then obtained because it is one of the marginal distributions of the stationary distribution of defined AoI process before. Follow this line of thinking, we obtain the explicit expressions of the AoI distribution for the system using Ber/G/1/1, Ber/Geo/1/2, and Ber/Geo/1/2* queues. Precisely, the status updating system with Ber/G/1/1 queue is discussed in Section III. In this Section, we also assume that the packet service can be preempted by the fresher ones randomly, so that the queue model analyzed is more general. In next two Sections, size 2 status updating system is considered. We obtain the AoI distribution of the system which has Ber/Geo/1/2 or Ber/Geo/1/2* queue. In the last part of Section IV, we also discuss how the stationary AoI distribution can be determined if system uses an arbitrary Ber/Geo/1/$c$ queue.

To find the distribution of AoI, observing that we actually determine the stationary state of a larger-dimensional AoI stochastic process. For the cases the system has size 2, apart from the AoI, the distributions of both packet system time and packet waiting time are also obtained by computing the other two marginal distributions. The numerical simulations of the AoI-distribution are provided in Section VI, where we depict the AoI distribution curves under different queue models, along with system time's distribution and waiting time's distribution.

The idea and methods proposed in this paper enable us to obtain the complete description of steady state AoI of state updating system with several kinds of queue models. We hope that the AoI distribution of more systems can be derived by this methods, i.e., constituting a multi-dimensional AoI stochastic process and solving its steady state. In the further works, firstly we wonder if the stationary AoI distribution is determined when the queue model is Ber/Geo/1/$c$, $c\geq3$, and what is the limiting AoI distribution for the case $c=\infty$. On the other hand, we have not talked about the average AoI, the variance of the AoI for each case, notice that all of these can be calculated directly. It is interesting to make comparisons between these results and those deduced in continous time queue models, such as system with M/M/1/1, M/M/1/2 and M/M/1/2* queues. Also the AoI based optimal system design problems can be modified to take into consideration more constraints, for instance, the AoI variance, the violation probability, and the so-called quality of service exponent (QoE).

\appendices
\section{Proof of Theorem 2}
In this Appendix, we solve the system of equations (5) and find all the stationary probabilities $\pi_{(n,m)}$, $n>m\geq0$. For the sake of convenience, we copy these equations in the following. 
\begin{equation}
\begin{cases}
\pi_{(n,m)}=\pi_{(n-1,m-1)}[1-pg(m-1)]  \Pr\{B>m-1|B>m-2\}  &  (n>m\geq2)   \\
\pi_{(n,1)}=\pi_{(n-1,0)}p\Pr\{B>1\} + \left(\sum\nolimits_{j=1}^{n-2}\pi_{(n-1,j)}pg(j) \right) \Pr\{B>1\}  &  (n\geq3)  \\
\pi_{(2,1)}=\pi_{(1,0)}p\Pr\{B>1\}    \\
\pi_{(n,0)}=\pi_{(n-1,0)}(1-p) +  \left(\sum\nolimits_{k=n}^{\infty}\pi_{(k,n-1)}\right) [1-pg(n-1)]\Pr\{B=n-1|B>n-2\}  &   (n\geq2)  \\
\pi_{(1,0)}=\left(\sum\nolimits_{n=1}^{\infty}\pi_{(n,0)}\right)p\Pr\{B=1\} + \left(\sum\nolimits_{n=2}^{\infty}\sum\nolimits_{m=1}^{n-1}\pi_{(n,m)}pg(m)\right)\Pr\{B=1\} 
\end{cases}
\end{equation}

Firstly, when $n>m\geq2$, the first line of (130) shows that 
\begin{align}
\pi_{(n,m)} &= \pi_{(n-1,m-1)}[1-pg(m-1)] \Pr\{B>m-1|B>m-2\}  \notag \\
&= \pi_{(n-2,m-2)}[1-pg(m-2)] \Pr\{B>m-2|B>m-3\} \times [1-pg(m-1)]\Pr\{B>m-1|B>m-2\}  \notag \\
&\qquad \qquad \qquad \qquad \quad \qquad \vdots \notag \\
&=\pi_{(n-m+1,1)}[1-pg(1)]\Pr\{B>1\} \times \dots \times [1-pg(m-1)]\Pr\{B>m-1|B>m-2\}  \notag \\
&=\pi_{(n-m+1,1)}\left(\prod\nolimits_{l=1}^{m-1}[1-pg(l)]\right)\Pr\{B>m-1\}   \\
&=\begin{cases}
\pi_{(1,0)}p(1-q_1)\left(\prod\nolimits_{l=1}^{m-1}[1-pg(l)]\right)\Pr\{B>m-1\} &    (n-m=1)  \\
\bigg[\pi_{(n-m,0)}p(1-q_1)+\left( \sum\nolimits_{j=1}^{n-m-1}\pi_{(n-m,j)}pg(j)\right) (1-q_1)\bigg] \left(\prod\nolimits_{l=1}^{m-1}[1-pg(l)]\right)\Pr\{B>m-1\} &  (n-m\geq2) 
\end{cases}
\end{align}

To obtain equation (131), we have used the second and third lines of (130). In (132), denote that $\prod\nolimits_{l=1}^{m-1}[1-pg(l)]$ equals 1 when $m=1$. Therefore, equation (132) is valid for all $n>m\geq1$.

Next, we consider the probabilities $\pi_{(n,0)}$. For $n\geq2$, the fourth line of (130) says that  
\begin{align}
\pi_{(n,0)}=\pi_{(n-1,0)}(1-p) + \left(\sum\nolimits_{k=n}^{\infty}\pi_{(k,n-1)}\right) [1-pg(n-1)] \Pr\{B=n-1|B>n-2\} \notag 
\end{align}

The sum in above equation is dealt with as follows. We show that 
\begin{align}
&\sum\nolimits_{k=n}^{\infty}\pi_{(k,n-1)} \notag \\
={}& \pi_{(n,n-1)}+\sum\nolimits_{k=n+1}^{\infty}\pi_{(k,n-1)}  \notag \\
={}& \pi_{(1,0)}p(1-q_1)\left(\prod\nolimits_{l=1}^{n-2}[1-pg(l)]\right)\Pr\{B>n-2\}  \notag \\
  {}& +\sum\nolimits_{k=n+1}^{\infty} \bigg[\pi_{(k-n+1,0)}p(1-q_1) + \left( \sum\nolimits_{j=1}^{k-n}\pi_{(k-n+1,j)}pg(j)\right) (1-q_1) \bigg]  \left(\prod\nolimits_{l=1}^{n-2}[1-pg(l)]\right)\Pr\{B>n-2\}  \\
={}& p(1-q_1)\pi_{(1,0)}\left(\prod\nolimits_{l=1}^{n-2}[1-pg(l)]\right)\Pr\{B>n-2\}   \notag \\
  {}& + p(1-q_1)\left(\sum\nolimits_{k=n+1}^{\infty}\pi_{(k-n+1,0)} \right) \left(\prod\nolimits_{l=1}^{n-2}[1-pg(l)]\right)\Pr\{B>n-2\} \notag \\  
  {}& + p(1-q_1) \left( \sum\nolimits_{k=n+1}^{\infty}\sum\nolimits_{j=1}^{k-n}\pi_{(k-n+1,j)}g(j)\right)  \left( \prod\nolimits_ {l=1}^{n-2} [1-pg(l)]\right)\Pr\{B>n-2\} \notag \\
={}& p(1-q_1) \left[ \left(\sum\nolimits_{n=1}^{\infty}\pi_{(n,0)}\right) + \left( \sum\nolimits_{n=2}^{\infty}\sum\nolimits_{m=1}^{n-1}\pi_{(n,m)}g(m)\right) \right]  \left( \prod\nolimits_ {l=1}^{n-2} [1-pg(l)]\right)\Pr\{B>n-2\}   \notag \\
={}& p(1-q_1) \frac{\pi_{(1,0)}}{pq_1} \left( \prod\nolimits_ {l=1}^{n-2} [1-pg(l)]\right)\Pr\{B>n-2\}   \\
={}& \frac{\pi_{(1,0)(1-q_1)}}{q_1} \left( \prod\nolimits_ {l=1}^{n-2} [1-pg(l)]\right)\Pr\{B>n-2\}
\end{align}

In (133) we substitute the expressions (132). In order to obtain equation (134), the last line of (130) is used. Therefore, for $n\geq2$, we obtain that  
\begin{align}
\pi_{(n,0)}=\pi_{(n-1,0)}(1-p)  + \frac{\pi_{(1,0)}(1-q_1)}{q_1}\left(\prod\nolimits_ {l=1}^{n-1} [1-pg(l)]\right)q_{n-1} 
\end{align}

Applying the relation (136) repeatedly yields  
\begin{align}
\pi_{(n,0)} &= \pi_{(n-1,0)}(1-p) + \frac{\pi_{(1,0)}(1-q_1)}{q_1}\left(\prod\nolimits_ {l=1}^{n-1} [1-pg(l)]\right)q_{n-1} \notag \\
&= \left[ \pi_{(n-2,0)}(1-p) + \frac{\pi_{(1,0)}(1-q_1)}{q_1}\left(\prod\nolimits_ {l=1}^{n-2} [1-pg(l)]\right)q_{n-2} \right] (1-p) + \frac{\pi_{(1,0)}(1-q_1)}{q_1}\left(\prod\nolimits_ {l=1}^{n-1} [1-pg(l)]\right)q_{n-1} \notag \\
&= \pi_{(n-2,0)}(1-p)^2 + \frac{\pi_{(1,0)}(1-q_1)}{q_1} \bigg[ (1-p) \left(\prod\nolimits_ {l=1}^{n-2} [1-pg(l)]\right) q_{n-2}+ \left(\prod\nolimits_ {l=1}^{n-1} [1-pg(l)]\right)q_{n-1}\bigg] \notag \\ 
  {}& \qquad  \qquad  \qquad  \qquad  \quad \qquad  \qquad  \qquad \qquad  \qquad  \quad \qquad     \vdots \notag \\
&= \pi_{(1,0)}(1-p)^{n-1} + \frac{\pi_{(1,0)}(1-q_1)}{q_1} \bigg[ \sum\nolimits_{j=0}^{n-2}(1-p)^j  \left(\prod\nolimits_ {l=1}^{n-1-j} [1-pg(l)]\right)q_{n-1-j}\bigg] 
\end{align}

This gives the results (6). 

Now, we determine the first probability $\pi_{(1,0)}$. Since all the stationary probabilities add up to 1, we have the relation 
\begin{align}
1&=\sum\nolimits_{m=0}^{\infty}\sum\nolimits_{n=m+1}^{\infty}\pi_{(n,m)} \notag \\
&=\sum\nolimits_{n=1}^{\infty}\pi_{(n,0)} + \sum\nolimits_{m=1}^{\infty}\sum\nolimits_{n=m+1}^{\infty}\pi_{(n,m)} \notag \\
&=\sum\nolimits_{n=1}^{\infty}\pi_{(n,0)} + \sum\nolimits_{m=1}^{\infty}\pi_{(m+1,m)}  + \sum\nolimits_{m=1}^{\infty}\sum\nolimits_{n=m+2}^{\infty}\pi_{(n,m)} \notag \\  
&=\sum\nolimits_{n=1}^{\infty}\pi_{(n,0)} + \sum\nolimits_{m=1}^{\infty}\pi_{(1,0)}p(1-q_1) \left(\prod\nolimits_{l=1}^{m-1}[1-pg(l)]\right)\Pr\{B>m-1\}  \notag \\
& \quad + \sum\nolimits_{m=1}^{\infty}\sum\nolimits_{n=m+2}^{\infty}\bigg[\pi_{(n-m,0)}p(1-q_1) + \left( \sum\nolimits_{j=1}^{n-m-1}\pi_{(n-m,j)}pg(j) \right) (1-q_1)\bigg] \left(\prod\nolimits_{l=1}^{m-1}[1-pg(l)]\right)\Pr\{B>m-1\}  \notag \\
&=\sum\nolimits_{n=1}^{\infty}\pi_{(n,0)} + p(1-q_1)\pi_{(1,0)} \left[ \sum\nolimits_{m=1}^{\infty}\left(\prod\nolimits_{l=1}^{m-1}[1-pg(l)]\right)\Pr\{B>m-1\}\right]  \notag \\
&\quad + p(1-q_1)\left(\sum\nolimits_{n=m+2}^{\infty}\pi_{(n-m,0)}\right)  \left[ \sum\nolimits_{m=1}^{\infty}\left(\prod\nolimits_{l=1}^{m-1}[1-pg(l)]\right)\Pr\{B>m-1\}\right]  \notag \\  
& \quad +  p(1-q_1)\left(\sum\nolimits_{n=m+2}^{\infty}\sum\nolimits_{j=1}^{n-m-1}\pi_{(n-m,j)}g(j)\right) \left[ \sum\nolimits_{m=1}^{\infty}\left(\prod\nolimits_{l=1}^{m-1}[1-pg(l)]\right)\Pr\{B>m-1\}\right]  \notag \\  
&= \left(\sum\nolimits_{n=1}^{\infty}\pi_{(n,0)}\right) +  p(1-q_1) \bigg[ \left(\sum\nolimits_{n=1}^{\infty}\pi_{(n,0)}\right)  + \left(\sum\nolimits_{n=2}^{\infty}\sum\nolimits_{m=1}^{n-1}\pi_{(n,m)}g(m) \right) \bigg]  \notag \\ 
& \quad \times \left[\sum\nolimits_{m=1}^{\infty} \left(\prod\nolimits_{l=1}^{m-1}[1-pg(l)]\right)\Pr\{B>m-1\}\right] \notag \\
&=\left(\sum\nolimits_{n=1}^{\infty}\pi_{(n,0)}\right) + \frac{\pi_{(1,0)}(1-q_1)}{q_1} \left[\sum\nolimits_{m=1}^{\infty} \left(\prod\nolimits_{l=1}^{m-1}[1-pg(l)]\right)\left(\sum\nolimits_{l=m}^{\infty}q_l\right) \right] 
\end{align}

Using the equation (137), we calculate the first sum in (138). 
\begin{align}
\sum\nolimits_{n=1}^{\infty}\pi_{(n,0)} &=\pi_{(1,0)}\sum\nolimits_{n=1}^{\infty}\bigg\{(1-p)^{n-1}+\frac{1-q_1}{q_1}\bigg[ \sum\nolimits_{j=0}^{n-2}(1-p)^j \left(\prod\nolimits_{l=1}^{n-1-j}[1-pg(l)]\right)q_{n-1-j}\bigg]\bigg\}  \notag \\
&= \pi_{(1,0)}\bigg\{\frac1p+\frac{1-q_1}{q_1} \bigg[ \sum\nolimits_{n=2}^{\infty} \sum\nolimits_{j=0}^{n-2}(1-p)^j \left(\prod\nolimits_{l=1}^{n-1-j}[1-pg(l)]\right)q_{n-1-j}\bigg] \bigg\} 
\end{align}

Substituting (139) into (138), finally we obtain that  
\begin{align}
\frac{1}{\pi_{(1,0)}}&= \frac1p+\frac{1-q_1}{q_1}\bigg[ \sum\nolimits_{n=2}^{\infty} \sum\nolimits_{j=0}^{n-2}(1-p)^j  \left(\prod\nolimits_{l=1}^{n-1-j}[1-pg(l)]\right)q_{n-1-j}\bigg]  \notag \\
&\qquad \quad + \frac{1-q_1}{q_1} \left[\sum\nolimits_{m=1}^{\infty} \left(\prod\nolimits_{l=1}^{m-1}[1-pg(l)]\right)\left(\sum\nolimits_{l=m}^{\infty}q_l\right) \right] 
\end{align}

Thus, we prove the equation (7) in Theorem 1. 

For $n>m\geq2$, equation (131) shows that probabilities $\pi_{(n,m)}$ can be obtained by 
\begin{equation}
\pi_{(n,m)}=\pi_{(n-m+1,1)}\left(\prod\nolimits_{l=1}^{m-1}[1-pg(l)]\right)\Pr\{B>m-1\} \notag 
\end{equation}

Therefore, we only need to find those probabilities $\pi_{(n,1)}$, $n\geq3$. The second line in equations (130) shows that  
\begin{align}
\pi_{(n,1)}=\pi_{(n-1,0)}p(1-q_1) + \left( \sum\nolimits_{j=1}^{n-2}\pi_{(n-1,j)}pg(j) \right) (1-q_1)  
\end{align}

In the following, we rewrite (141) as a difference equation of probabilities $\pi_{(n,1)}$, $n\geq3$. The sum in above equation is dealt with as 
\begin{align}
\sum\nolimits_{j=1}^{n-2}\pi_{(n-1,j)}pg(j) &= \pi_{(n-1,1)}pg(1) + \sum\nolimits_{j=2}^{n-2}\pi_{(n-1,j)}pg(j)  \notag \\
&= \pi_{(n-1,1)}pg(1) + \sum\nolimits_{j=2}^{n-2}\pi_{(n-j,1)}\left(\prod\nolimits_{l=1}^{j-1}[1-pg(l)]\right)\Pr\{B>j-1\} pg(j) \notag \\ 
&= \sum\nolimits_{j=1}^{n-2}\pi_{(n-j,1)} \left(\prod\nolimits_{l=1}^{j-1}[1-pg(l)]\right)\Pr\{B>j-1\} pg(j) \notag 
\end{align}
if we denote $\prod\nolimits_{l=1}^{0}[1-pg(l)]$ equals 1. 

Define $\alpha_j=\left(\prod\nolimits_{l=1}^{j-1}[1-pg(l)]\right)\Pr\{B>j-1\} pg(j)$, the equation (141) is transformed to  
\begin{equation}
\pi_{(n,1)}=\pi_{(n-1,0)}p(1-q_1) +(1-q_1) \sum\nolimits_{j=1}^{n-2}\alpha_j\pi_{(n-j,1)}
\end{equation}
which forms a difference equation of probabilities $\pi_{(n,1)}$ for $n\geq3$. 

When the distribution of the packet service time $B$ and the preemption probabilities $g(m)$, $m\geq1$ are given, the probabilities $\pi_{(n,1)}$ can be derived by solving the difference equation (142). 

So far, we have obtained all the results in Theorem 2, thus the proof is complete.

\section{Proof of Theorem 3}
For the status updating system with Ber/Geo/1/1 queue, let the preemption function is defined as (12), we solve the following stationary equations of the process $AoI_{Ber/Geo/1/1/\tilde g(m)}$ and determine all the probabilities $\pi_{(n,m)}$, $n>m\geq0$.  
\begin{equation}
\begin{cases}
\pi_{(n,m)}=\pi_{(n-1,m-1)}(1-p) \frac{N_p+m}{N_p+m-1}(1-\gamma)  &   (n>m\geq2) \\
\pi_{(n,1)}=\pi_{(n-1,0)}p(1-\gamma) + \left(\sum\nolimits_{j=1}^{n-2}\pi_{(n-1,j)}\left[1-(1-p)\frac{N_p+j+1}{N_p+j} \right]  \right)(1-\gamma)  &    (n\geq3)  \\
\pi_{(2,1)}=\pi_{(1,0)}p(1-\gamma)  \\
\pi_{(n,0)}=\pi_{(n-1,0)}(1-p)  + \left(\sum\nolimits_{k=n}^{\infty}\pi_{(k,n-1)}\right) (1-p)\frac{N_p+n}{N_p+n-1}\gamma & (n\geq2) \\
\pi_{(1,0)}=\left( \sum\nolimits_{n=1}^{\infty}\pi_{(n,0)} \right)p\gamma  + \left(\sum\nolimits_{n=2}^{\infty}\sum\nolimits_{m=1}^{n-1}\pi_{(n,m)}\left[1-(1-p)\frac{N_p+m+1}{N_p+m}\right] \right)\gamma  
\end{cases}
\end{equation}

The theorem is proved directly using the general formulas (6)-(9) in Theorem 2. 

First of all, for $n\geq2$, equation (136) shows that 
\begin{align}
\pi_{(n,0)} &= \pi_{(n-1,0)}(1-p) + \frac{\pi_{(1,0)}(1-q_1)}{q_1}\left(\prod\nolimits_{l=1}^{n-1}[1-pg(l)]\right)q_{n-1} \notag \\
&=\pi_{(n-1,0)}(1-p) + \frac{\pi_{(1,0)}(1-\gamma)}{\gamma}\left(\prod\nolimits_{l=1}^{n-1}(1-p)\frac{N_p+l+1}{N_p+l}\right)(1-\gamma)^{n-2}\gamma \notag \\
&=\pi_{(n-1,0)}(1-p)  + \frac{\pi_{(1,0)}}{N_p+1}\left(N_p+n\right)[(1-p)(1-\gamma)]^{n-1} 
\end{align}

Applying (144) iteratively, we can obtain that   
\begin{align}
\pi_{(n,0)} &=\pi_{(1,0)}(1-p)^{n-1} + \frac{\pi_{(1,0)}(1-p)^{n-1}}{N_p+1}  \left(\sum\nolimits_{j=0}^{n-2}(N_p+n-j)(1-\gamma)^{n-1-j} \right)   \notag \\
&= \pi_{(1,0)}(1-p)^{n-1} + \frac{\pi_{(1,0)}(1-p)^{n-1}}{N_p+1} \left(\frac{N_p\gamma-(N_p+1)\gamma^2+1}{\gamma^2} - \frac{[1+(N_p+n)\gamma](1-\gamma)^n}{\gamma^2}  \right)   \notag \\
&= \frac{\pi_{(1,0)}}{(N_p+1)\gamma^2}\bigg\{ \left(N_p\gamma+1\right)(1-p)^{n-1} - \left[1+(N_p+n)\gamma \right](1-\gamma) [(1-p)(1-\gamma)]^{n-1}\bigg\} 
\end{align}

Let $n=1$, (145) reduces to the equation $\pi_{(1,0)}=\pi_{(1,0)}$, thus we verify that expression (145) is valid for all $n\geq1$. 

To determine the first probability $\pi_{(1,0)}$, we use equation (138), which says that  
\begin{align}
1&=\left(\sum\nolimits_{n=1}^{\infty}\pi_{(n,0)}\right) + \frac{\pi_{(1,0)}(1-\gamma)}{\gamma} \left[ \sum\nolimits_{m=1}^{\infty}\left(\prod\nolimits_{l=1}^{m-1}[1-p\tilde g(l)]\right) \left(\sum\nolimits_{l=m}^{\infty}q_l \right) \right]   \notag \\
&=\left(\sum\nolimits_{n=1}^{\infty}\pi_{(n,0)}\right) + \frac{\pi_{(1,0)}(1-\gamma)}{\gamma} \sum\nolimits_{m=1}^{\infty}\left(\prod\nolimits_{l=1}^{m-1}(1-p)\frac{N_p+l+1}{N_p+l}\right) (1-\gamma)^{m-1}  \notag \\
&=\left(\sum\nolimits_{n=1}^{\infty}\pi_{(n,0)}\right) + \frac{\pi_{(1,0)}(1-\gamma)}{(N_p+1)\gamma} \sum\nolimits_{m=1}^{\infty}\left(N_p+m\right) [(1-p)(1-\gamma)]^{m-1}  
\end{align}

Using equation (145), the sum in (146) is equal to 
\begin{align}
\sum\nolimits_{n=1}^{\infty}\pi_{(n,0)}&= \frac{\pi_{(1,0)}}{(N_p+1)\gamma^2}\bigg\{\frac{N_p\gamma+1}{p} - \frac{1-\gamma}{p+\gamma-p\gamma} - \gamma(1-\gamma)\sum\nolimits_{n=1}^{\infty}(N_p+n)[(1-p)(1-\gamma)]^{n-1}  \bigg\} \notag \\
&= \frac{\pi_{(1,0)}[N_p(p+\gamma-p\gamma)+1]}{(N_p+1)p\gamma(p+\gamma-p\gamma)} - \frac{\pi_{(1,0)}(1-\gamma)}{(N_p+1)\gamma}  \sum\nolimits_{n=1}^{\infty}(N_p+n)[(1-p)(1-\gamma)]^{n-1} 
\end{align}

Combining (146) and (147), it was observed that two sums cancel out and we have  
\begin{equation}
1=\frac{\pi_{(1,0)}[N_p(p+\gamma-p\gamma)+1]}{(N_p+1)p\gamma(p+\gamma-p\gamma)}  \notag 
\end{equation}
from which $\pi_{(1,0)}$ is figured out as 
\begin{equation}
\pi_{(1,0)}=\frac{(N_p+1)p\gamma(p+\gamma-p\gamma)}{N_p(p+\gamma-p\gamma)+1}
\end{equation}

Substituting (148) into equation (145), the probabilities $\pi_{(n,0)}$, $n\geq1$ are determined finally. It shows that  
\begin{align}
\pi_{(n,0)}&=\frac{p(p+\gamma-p\gamma)}{N_p\gamma(p+\gamma-p\gamma)+\gamma}\bigg\{(N_p\gamma+1)(1-p)^{n-1} - [1+(N_p+n)\gamma](1-\gamma)[(1-p)(1-\gamma)]^{n-1} \bigg\} \notag \\
&=\xi_1(1-p)^{n-1} - \xi_2[1+(N_p+n)\gamma][(1-p)(1-\gamma)]^{n-1}
\end{align}
where the coefficients $\xi_1$ and $\xi_2$ are defined to be 
\begin{equation}
\xi_1=\frac{(N_p\gamma+1)p(p+\gamma-p\gamma)}{N_p\gamma(p+\gamma-p\gamma)+\gamma},  
\qquad 
\xi_2=\frac{p(1-\gamma)(p+\gamma-p\gamma)}{N_p\gamma(p+\gamma-p\gamma)+\gamma}   \notag  
\end{equation}

Then, we consider the other probabilities $\pi_{(n,m)}$, $n>m\geq1$. From the equation (131), $\pi_{(n,m)}$ can be written as 
\begin{align}
\pi_{(n,m)}&=\pi_{(n-m+1,1)}\left(\prod\nolimits_{l=1}^{m-1}[1-p\tilde g(l)]\right)\Pr\{B>m-1\} \notag \\
&=\pi_{(n-m+1,1)}\left(\prod\nolimits_{l=1}^{m-1}(1-p)\frac{N_p+l+1}{N_p+l}\right)\left(\sum\nolimits_{l=m}^{\infty}q_l\right) \notag \\
&=\pi_{(n-m+1,1)}[(1-p)(1-\gamma)]^{m-1}\frac{N_p+m}{N_p+1} 
\end{align}

Thus, in order to find the probabilities $\pi_{(n,m)}$, it suffices to determine all the probabilities $\pi_{(n,1)}$, $n\geq3$. 

For the considered case, in the following we prove that the general equation (9) can be reduced to an order three difference equation with constant coefficients, such that the solution of the obtained difference equation can be solved.

For $n\geq3$, the second line of (143) shows that 
\begin{equation}
\pi_{(n,1)}=\pi_{(n-1,0)}p(1-\gamma)  + \bigg( \sum\nolimits_{j=1}^{n-2}\pi_{(n-1,j)} \left[1-(1-p)\frac{N_p+j+1}{N_p+j}\right] \bigg) (1-\gamma)  
\end{equation}

Using (150), the sum in equation (151) can be written as  
\begin{align}
&\sum\nolimits_{j=1}^{n-2}\pi_{(n-1,j)}\left[1-(1-p)\frac{N_p+j+1}{N_p+j}\right]  \notag \\
={}&\sum\nolimits_{j=1}^{n-2}\left\{\pi_{(n-j,1)}[(1-p)(1-\gamma)]^{j-1}\frac{N_p+j}{N_p+1}\right\} \left[1-(1-p)\frac{N_p+j+1}{N_p+j}\right]  \notag \\
={}&\frac{1}{N_p+1}\sum\nolimits_{j=1}^{n-2}\pi_{(n-j,1)}[(1-p)(1-\gamma)]^{j-1} [(N_p+j)-(1-p)(N_p+j+1)] \notag \\
={}&\frac{1}{N_p+1}\sum\nolimits_{j=1}^{n-2}\pi_{(n-j,1)}[p(N_p+j)-(1-p)]  [(1-p)(1-\gamma)]^{j-1}  
\end{align}

Therefore, we obtain that 
\begin{equation}
\pi_{(n,1)}=\pi_{(n-1,0)}p(1-\gamma) + \frac{1-\gamma}{N_p+1}\bigg( \sum\nolimits_{j=1}^{n-2}\pi_{(n-j,1)} [p(N_p+j)-(1-p)][(1-p)(1-\gamma)]^{j-1}  \bigg)
\end{equation}

Do once iteration of equation (153) and multiplying $(1-p)(1-\gamma)$ in both sides gives 
\begin{align}
&(1-p)(1-\gamma)\pi_{(n-1,1)}  \notag \\
={}& (1-p)(1-\gamma)\pi_{(n-2,0)}p(1-\gamma)  + \frac{1-\gamma}{N_p+1}\bigg( \sum\nolimits_{j=1}^{n-3}\pi_{(n-1-j,1)} [p(N_p+j)-(1-p)][(1-p)(1-\gamma)]^j  \bigg)
\end{align}

Let $j=\tilde j-1$, the sum in equation (154) can be rewritten as 
\begin{equation}
\sum\nolimits_{\tilde j=2}^{n-2}\pi_{(n-\tilde j,1)} [p(N_p+\tilde j-1)-(1-p)][(1-p)(1-\gamma)]^{\tilde j-1}  \bigg)  \notag  
\end{equation}

Subtracting (154) from equation (153) yields 
\begin{align}
&\pi_{(n,1)}-(1-p)(1-\gamma)\pi_{(n-1,1)} \notag \\
={}&h^{(1)}(n) + \frac{1-\gamma}{N_p+1}\bigg( \pi_{(n-1,1)}[p(N_p+1)-(1-p)]  + \sum\nolimits_{j=2}^{n-2}\pi_{(n-j,1)}p[(1-p)(1-\gamma)]^{j-1}  \bigg)   \notag 
\end{align}

Rearranging the terms, we obtain that 
\begin{equation}
\pi_{(n,1)}=h^{(1)}(n) + \frac{N_p(1-\gamma)}{N_p+1}\pi_{(n-1,1)} +\frac{p(1-\gamma)}{N_p+1} \sum\nolimits_{j=1}^{n-2}\pi_{(n-j,1)}[(1-p)(1-\gamma)]^{j-1} 
\end{equation}
where $h^{(1)}(n)$ is defined as 
\begin{align}
h^{(1)}(n)&=p(1-\gamma)\left[\pi_{(n-1,0)}-(1-p)(1-\gamma)\pi_{(n-2,0)}\right]  \notag \\
&=p(1-\gamma)\bigg\{\xi_1(1-p)^{n-2} -\xi_2[1+(N_p+n-1)\gamma][(1-p)(1-\gamma)]^{n-2} \notag \\
& \qquad - \xi_1(1-p)^{n-2}(1-\gamma)  + \xi_2[1+(N_p+n-2)\gamma][(1-p)(1-\gamma)]^{n-2} \bigg\}  \notag \\
&=p\gamma(1-\gamma)\bigg\{\xi_1(1-p)^{n-2} - \xi_2[(1-p)(1-\gamma)]^{n-2}\bigg\}  \notag 
\end{align}

Repeat the same procedures to equation (155), we show that 
\begin{align}
&\pi_{(n,1)}-(1-p)(1-\gamma)\pi_{(n-1,1)} \notag \\
={}&h^{(2)}(n) + \frac{N_p(1-\gamma)}{N_p+1} \pi_{(n-1,1)}  - \frac{N_p(1-p)(1-\gamma)^2}{N_p+1} \pi_{(n-2,1)}  + \frac{p(1-\gamma)}{N_p+1} \pi_{(n-1,1)} \notag 
\end{align}
which is equivalent to 
\begin{equation}
\pi_{(n,1)}= h^{(2)}(n) + \pi_{(n-1,1)}\left(1-\gamma+\frac{N_p(1-p)(1-\gamma)}{N_p+1}\right)  - \pi_{(n-2,1)}\frac{N_p(1-p)(1-\gamma)^2}{N_p+1}
\end{equation}

The term $h^{(2)}(n)$ is equal to 
\begin{equation}
h^{(2)}(n) =h^{(1)}(n) - (1-p)(1-\gamma)h^{(1)}(n-1) =p\gamma^2(1-\gamma)\xi_1(1-p)^{n-2} 
\end{equation}

Consider the iteration of (156) again, at this time, we compute 
\begin{align}
&\pi_{(n,1)}-(1-p)\pi_{(n-1,1)} \notag \\
={}&\pi_{(n-1,1)}\left(1-\gamma + \frac{N_p(1-p)(1-\gamma)}{N_p+1}\right)  - \pi_{(n-2,1)}\left((1-p)(1-\gamma) + \frac{N_p(1-p)^2(1-\gamma)}{N_p+1}\right)  \notag \\
  {}& \quad - \pi_{(n-2,1)} \frac{N_p(1-p)(1-\gamma)^2}{N_p+1} + \pi_{(n-3,1)} \frac{N_p[(1-p)(1-\gamma)]^2}{N_p+1} \notag 
\end{align}
which can be transformed to  
\begin{align}
&\pi_{(n,1)}- \pi_{(n-1,1)}\left[(1-p) + (1-\gamma) + \frac{N_p(1-p)(1-\gamma)}{N_p+1} \right] + \pi_{(n-2,1)}\bigg[(1-p)(1-\gamma)  \notag \\
&\qquad + \frac{N_p(1-p)^2(1-\gamma)}{N_p+1} + \frac{N_p(1-p)(1-\gamma)^2}{N_p+1} \bigg]  - \pi_{(n-3,1)}\frac{N_p[(1-p)(1-\gamma)]^2}{N_p+1}=0
\end{align}

Eventually, we obtain the constant coefficient three-order difference equation (158), whose characterization equation is 
\begin{align}
&r^3- \left[(1-p) + (1-\gamma) + \frac{N_p(1-p)(1-\gamma)}{N_p+1} \right]r^2    \notag \\
& \qquad + \bigg[(1-p)(1-\gamma) + \frac{N_p(1-p)^2(1-\gamma)}{N_p+1}+ \frac{N_p(1-p)(1-\gamma)^2}{N_p+1} \bigg]r - \frac{N_p[(1-p)(1-\gamma)]^2}{N_p+1}=0   \\
\Longleftrightarrow {}&\left[r-(1-p)\right]\left[r-(1-\gamma)\right]\left[r-\frac{N_p(1-p)(1-\gamma)}{N_p+1}\right] =0  \notag 
\end{align}

Thus, the three roots of (159) are determined as 
\begin{equation}
r_1=1-p, r_2=1-\gamma, \text{ and } r_3=\frac{N_p(1-p)(1-\gamma)}{N_p+1}   \notag 
\end{equation}

According to the difference equation theory, the solution of probabilities $\pi_{(n,1)}$, $n\geq3$ can be written as   
\begin{equation}
\pi_{(n,1)}=c_1r_1^n + c_2r_2^n + c_3r_3^n   
\end{equation}
where $c_1$, $c_2$ and $c_3$ are undetermined coefficients, and can be obtained by solving the following system of equations 
\begin{equation}
\pi_{(k,1)} = c_1(1-p)^k + c_2(1-\gamma)^k  + c_3\left(\frac{N_p(1-p)(1-\gamma)}{N_p+1}\right)^k  \quad (k=3,4,5)
\end{equation}

In form, we solve that 
\begin{equation}
\begin{cases}
c_1=\frac{r_2r_3\pi_{(3,1)} - (r_2+r_3)\pi_{(4,1)} + \pi_{(5,1)}}{(r_2-r_1)(r_3-r_1)r_1^3}     \\
c_2=\frac{r_1r_3\pi_{(3,1)} - (r_1+r_3)\pi_{(4,1)} + \pi_{(5,1)}}{(r_1-r_2)(r_3-r_2)r_2^3}     \\
c_3=\frac{r_1r_2\pi_{(3,1)} - (r_1+r_2)\pi_{(4,1)} + \pi_{(5,1)}}{(r_1-r_3)(r_2-r_3)r_3^3}   
\end{cases}
\end{equation}

The probabilities $\pi_{(k,1)}$, $k=3,4,5$ can be determined by applying the second line of (143) directly. Substituting these probabilities, eventually $c_i$, $i=1,2,3$ are represented by certain polynomials consisting of $p$, $\gamma$, and $N_p$, but the exact expressions can be very complex. Thus, we do not compute $c_i$ further, while determining their accurate values when the numerical simulations of AoI-distribution are considered.

Collecting equations (149), (150), and (160), we obtain all the stationary probabilities $\pi_{(n,m)}$ of the stochastic process $AoI_{Ber/Geo/1/1/\tilde g(m)}$. This completes the proof of Theorem 3.

\section{Proof of Lemma 2}
In this Appendix, we calculate following two sums  
\begin{align}
M_1&=\sum\nolimits_{n=1}^{\infty}\pi_{(n,0,0)} =\frac{(1-p)\gamma^2}{(p+\gamma -2p\gamma)+p^2(1-\gamma)^2} \notag \\
M_2&=\sum\nolimits_{n=2}^{\infty}\sum\nolimits_{m=2}^{n-1}\pi_{(n,m,0)} =\frac{p\gamma(1-\gamma)}{(p+\gamma -2p\gamma)+p^2(1-\gamma)^2}  \notag
\end{align}

In Lemma 1, we have obtained that for $n>m\geq 1$, the stationary probabilities $\pi_{(n,m,0)}$ are determined by
\begin{align}
\pi_{(n,m,0)}&=\pi_{(n-m,0,0)}p(1-\gamma) \left[(1-p)(1-\gamma)\right]^{m-1}  \notag  \\
&\qquad +\frac{\gamma}{1-\gamma}\left(\sum\nolimits_{k=n-m+1}^{\infty}\pi_{(k,n-m,0)} \right) \left\{ (1-\gamma)^m-[(1-p)(1-\gamma)]^m \right\}  
\end{align}

We add up all the probabilities $\{\pi_{(n,m,0)},n>m \geq 1 \}$, showing that an equation containing $M_1$ and $M_2$ can be deduced.  
\begin{align}
M_2&=\sum\nolimits_{n=2}^{\infty}\sum\nolimits_{m=1}^{n-1}\pi_{(n,m,0)}  \notag \\
&= \sum\nolimits_{n>m\geq 1}\bigg[ \pi_{(n-m,0,0)}p(1-\gamma)\left[(1-p)(1-\gamma)\right]^{m-1} +\frac{\gamma}{1-\gamma}\left(\sum\nolimits_{k=n-m+1}^{\infty}\pi_{(k,n-m,0)} \right)  \left\{(1-\gamma)^m-[(1-p)(1-\gamma)]^m \right\}  \bigg]  \notag \\
&=\sum\nolimits_{t=n-m=1}^{\infty}\sum\nolimits_{m=1}^{\infty} \bigg( \pi_{(t,0,0)}p(1-\gamma)  \left[(1-p)(1-\gamma)\right]^{m-1}+\frac{\gamma}{1-\gamma}\left(\sum\nolimits_{k=t+1}^{\infty}\pi_{(k,t,0)} \right) \left\{(1-\gamma)^m-[(1-p)(1-\gamma)]^m \right\}  \bigg)  \notag \\
&=\sum\nolimits_{t=1}^{\infty}\bigg\{ \pi_{(t,0,0)}p(1-\gamma)\sum\nolimits_{m=1}^{\infty}[(1-\gamma)(1-p)]^{m-1} +\frac{\gamma}{1-\gamma} \left(\sum\nolimits_{k=t+1}^{\infty}\pi_{(k,t,0)} \right)  \sum\nolimits_{m=1}^{\infty} \left\{(1-\gamma)^m-[(1-p)(1-\gamma)]^m \right\}   \bigg\} \notag \\
&=\sum\nolimits_{t=1}^{\infty}\bigg( \pi_{(t,0,0)}p(1-\gamma)\frac{1}{p+\gamma-p\gamma} +\frac{\gamma}{1-\gamma} \left(\sum\nolimits_{k=t+1}^{\infty}\pi_{(k,t,0)} \right) \frac{p(1-\gamma)}{\gamma(p+\gamma-p\gamma)} \bigg) \notag \\
&=\frac{p(1-\gamma)}{p+\gamma-p\gamma}\sum\nolimits_{t=1}^{\infty}\pi_{(t,0,0)}  +\frac{p}{p+\gamma-p\gamma}\sum\nolimits_{t=1}^{\infty}\sum\nolimits_{k=t+1}^{\infty}\pi_{(k,t,0)} \notag \\
&=\frac{p(1-\gamma)}{p+\gamma-p\gamma}M_1 + \frac{p}{p+\gamma-p\gamma}M_2 
\end{align}

To obtain equation (164), notice that 
\begin{equation}
\sum\nolimits_{t=1}^{\infty}\sum\nolimits_{k=t+1}^{\infty}\pi_{(k,t,0)}=\sum\nolimits_{n=2}^{\infty}\sum\nolimits_{m=1}^{n-1}\pi_{(n,m,0)}=M_2  \notag
\end{equation}

Therefore, we derive the first relation between the two sums $M_1$ and $M_2$ as 
\begin{equation}
p(1-\gamma) M_1 = (1-p)\gamma M_2
\end{equation}

Another relation can be found from the boundary condition which says that 
\begin{align}
1&=\sum\nolimits_{n=1}^{\infty} \pi_{(n,0,0)}+\sum\nolimits_{n=2}^{\infty}\sum\nolimits_{m=1}^{n-1}\pi_{(n,m,0)} +\sum\nolimits_{n=3}^{\infty}\sum\nolimits_{m=2}^{n-1}\sum\nolimits_{l=1}^{m-1}\pi_{(n,m,l)} \notag \\
&=M_1+M_2+\sum\nolimits_{n>m>l\geq 1}\pi_{(n,m,l)}
\end{align}
where in (166) we need to compute only the last sum. It shows that  
\begin{align}
&\sum\nolimits_{n>m>l\geq 1} \pi_{(n,m,l)} \notag \\
={}& \sum\nolimits_{n>m>l\geq 1} \pi_{(n-l,m-l,0)} p(1-\gamma)^{l} \notag  \\   
={}& \sum\nolimits_{n>m>l\geq 1} \bigg( \pi_{(n-m,0,0)}p(1-\gamma)[(1-p)(1-\gamma)]^{m-l-1}  \notag \\  
  {}& \qquad + \frac{\gamma}{1-\gamma} \left(\sum\nolimits_{k=n-m+1}^{\infty}\pi_{(k,n-m,0)} \right)  \left\{ (1-\gamma)^{m-l}-[(1-p)(1-\gamma)]^{m-l} \right\} \bigg)p(1-\gamma)^l    \notag \\
={}&\sum\nolimits_{n>m>l\geq 1}\bigg(\pi_{(n-m,0,0)}p^2 (1-p)^{m-l-1}(1-\gamma)^m   \notag \\
  {}&\qquad + \frac{p\gamma}{1-\gamma}\left(\sum\nolimits_{k=n-m+1}^{\infty}\pi_{(k,n-m,0)} \right)  \left\{ (1-\gamma)^{m}- (1-p)^{m-l}(1-\gamma)^m \right\} \bigg)  \notag \\
={}&\sum\nolimits_{n>m\geq 2}\bigg(\pi_{(n-m,0,0)}p^2 \frac{1-(1-p)^{m-1}}{p}(1-\gamma)^m \notag \\
  {}& \qquad +\frac{p\gamma}{1-\gamma} \left( \sum\nolimits_{k=n-m+1}^{\infty}\pi_{(k,n-m,0)} \right) \bigg[(m-1)(1-\gamma)^m - \frac{(1-p)-(1-p)^m}{p}(1-\gamma)^m \bigg]  \bigg) \notag \\
={}&\sum\nolimits_{n>m\geq 2}\bigg(\pi_{(n-m,0,0)}p(1-\gamma)\bigg\{ (1-\gamma)^{m-1} -[(1-p)(1-\gamma)]^{m-1} \bigg\}\notag \\
  {}&\qquad  +\frac{\gamma}{1-\gamma} \left( \sum\nolimits_{k=n-m+1}^{\infty}\pi_{(k,n-m,0)} \right) \bigg\{ (pm-1)(1-\gamma)^m + [(1-p)(1-\gamma)]^m \bigg\}  \bigg) \notag \\
={}&\sum\nolimits_{t=n-m=1}^{\infty}\sum\nolimits_{m=2}^{\infty} \bigg( \pi_{(t,0,0)}p(1-\gamma)\bigg\{ (1-\gamma)^{m-1} -[(1-p)(1-\gamma)]^{m-1} \bigg\} \notag \\
  {}&\qquad  +\frac{\gamma}{1-\gamma} \left( \sum\nolimits_{k=t+1}^{\infty}\pi_{(k,t,0)} \right)  \bigg\{ (pm-1)(1-\gamma)^m + [(1-p)(1-\gamma)]^m \bigg\}  \bigg)   \notag \\
={}&\sum\nolimits_{t=1}^{\infty} \bigg\{ \pi_{(t,0,0)}p(1-\gamma) \frac{p(1-\gamma)}{\gamma (p+\gamma-p\gamma)} \notag \\
  {}& \qquad +\frac{\gamma}{1-\gamma}\left(\sum\nolimits_{k=t+1}^{\infty}\pi_{(k,t,0)} \right) \bigg[ \frac{(1-\gamma)^2[p-(1-p)\gamma]}{\gamma^2} + \frac{(1-p)^2(1-\gamma)^2}{p+\gamma-p\gamma} \bigg] \bigg\} 
\end{align}

The coefficient of $\sum\nolimits_{k=t+1}^{\infty}\pi_{(k,t,0)}$ in (167) is equal to 
\begin{align}
&\frac{\gamma}{1-\gamma}\left[\frac{(1-\gamma)^2[p-(1-p)\gamma]}{\gamma^2} + \frac{(1-p)^2(1-\gamma)^2}{p+\gamma-p\gamma}\right]  \notag \\
={}&\frac{(1-\gamma)[p-(1-p)\gamma]}{\gamma} + \frac{\gamma (1-p)^2(1-\gamma)}{p+\gamma-p\gamma} \notag \\
={}&(1-\gamma)\left[\frac{p-(1-p)\gamma}{\gamma} + \frac{(1-p)^2\gamma}{p+\gamma-p\gamma}\right]  \notag \\
={}&(1-\gamma)\frac{[p-(1-p)\gamma][p+(1-p)\gamma]+(1-p)^2\gamma^2}{\gamma(p+\gamma-p\gamma)} \notag \\
={}&\frac{p^2(1-\gamma)}{\gamma(p+\gamma-p\gamma)}  \notag 
\end{align}

Therefore, we obtain that 
\begin{align}
\sum\nolimits_{n>m>l\geq1}\pi_{(n,m,l)}&= \frac{p^2 (1-\gamma)^2}{\gamma (p+\gamma -p\gamma)}\left( \sum\nolimits_{t=1}^{\infty} \pi_{(t,0,0)} \right) + \frac{p^2(1-\gamma)}{\gamma (p+\gamma -p\gamma)} \left(\sum\nolimits_{t=1}^{\infty} \sum\nolimits_{k=t+1}^{\infty}\pi_{(k,t,0)} \right) \notag \\
&=\frac{p^2 (1-\gamma)^2}{\gamma (p+\gamma -p\gamma)} M_1 +\frac{p^2(1-\gamma)}{\gamma (p+\gamma -p\gamma)} M_2
\end{align}

Substituting (168) into (166) yields that 
\begin{equation}
1=\left[1+\frac{p^2 (1-\gamma)^2}{\gamma (p+\gamma -p\gamma)} \right]  M_1 + \left[1+\frac{p^2(1-\gamma)}{\gamma (p+\gamma -p\gamma)}\right] M_2
\end{equation}
which gives another relationship between $M_1$ and $M_2$. 

Combining equations (165) and (169), we can solve that 
\begin{equation}
M_1= \frac{(1-p)\gamma^2}{(p+\gamma -2p\gamma)\gamma +p^2(1-\gamma)^2}, 
\qquad
M_2=\frac{p\gamma (1-\gamma)}{(p+\gamma -2p\gamma)\gamma +p^2(1-\gamma)^2}  \notag
\end{equation}

In addition, according to the last line of equations (39), the first probability $\pi_{(1,0,0)}$ is determined as 
\begin{equation}
\pi_{(1,0,0)}=\left(\sum\nolimits_{n=1}^{\infty}\pi_{(n,0,0)} \right)p\gamma =p\gamma M_1=\frac{p(1-p)\gamma ^3}{(p+\gamma -2p\gamma)\gamma +p^2(1-\gamma)^2} \notag
\end{equation}

So far, we complete the proof of Lemma 2.

\section{Proof of Lemma 3}
In this part, we compute the infinite sum $\sum\nolimits_{k=n}^{\infty}\pi_{(k,n-1,0)}$ and prove Lemma 3.

First of all, according to (39), probabilities $\pi_{(k,n-1,0)}$, $k\geq n\geq 3$ are determined by 
\begin{align}
\pi_{(k,n-1,0)}&=\pi_{(k-1,n-2,0)}(1-p)(1-\gamma) + \left( \sum\nolimits_{j=k}^{\infty}\pi_{(j,k-1,n-2)} \right)\gamma 
\end{align}

Summing up (170) from $k=n$ to infinity, we have 
\begin{align}
\sum\nolimits_{k=n}^{\infty}\pi_{(k,n-1,0)}&= \sum\nolimits_{k=n}^{\infty}\bigg\{\pi_{(k-1,n-2,0)}(1-p)(1-\gamma)  + \left( \sum\nolimits_{j=k}^{\infty}\pi_{(j,k-1,n-2)}\right) \gamma  \bigg\} \notag \\
&= \left(\sum\nolimits_{k=n-1}^{\infty}\pi_{(k,n-2,0)} \right) (1-p)(1-\gamma) + \left(\sum\nolimits_{k=n}^{\infty}\sum\nolimits_{j=k}^{\infty}\pi_{(j-n+2,k-n+1,0)}p(1-\gamma)^{n-2} \right) \gamma \notag \\
&= \left(\sum\nolimits_{k=n-1}^{\infty}\pi_{(k,n-2,0)} \right) (1-p)(1-\gamma) + p\gamma \left(\sum\nolimits_{n=2}^{\infty}\sum\nolimits_{m=1}^{\infty}\pi_{(n,m,0)}  \right)(1-\gamma)^{n-2} \notag \\
&= \left(\sum\nolimits_{k=n-1}^{\infty}\pi_{(k,n-2,0)} \right) (1-p)(1-\gamma) + p\gamma M_2 (1-\gamma)^{n-2}
\end{align}
where $M_2$ is defined as 
\begin{equation}
M_2=\sum\nolimits_{n=2}^{\infty}\sum\nolimits_{m=1}^{n-1}\pi_{(n,m,0)} \notag
\end{equation}
and has been obtained in Lemma 2. 

Equation (171) gives a recursive relation for the infinite sums $\sum\nolimits_{k=n}^{\infty}\pi_{(k,n-1,0)}$, $n\geq 2$, from which the general formula of the sums can be derived. We have that
\begin{align}
\sum\nolimits_{k=n}^{\infty}\pi_{(k,n-1,0)} &= \left( \sum\nolimits_{k=n-1}^{\infty}\pi_{(k,n-2,0)} \right) (1-p)(1-\gamma) +p\gamma M_2(1-\gamma)^{n-2} \notag \\
&= \bigg[ \left( \sum\nolimits_{k=n-2}^{\infty}\pi_{(k,n-3,0)} \right) (1-p)(1-\gamma)  +p\gamma M_2(1-\gamma)^{n-3} \bigg] (1-p)(1-\gamma) + p\gamma M_2 (1-\gamma)^{n-2} \notag \\
&= \left( \sum\nolimits_{k=n-2}^{\infty}\pi_{(k,n-3,0)} \right) \left[(1-p)(1-\gamma) \right]^2 + p\gamma M_2 \left\{ [(1-p)(1-\gamma)](1-\gamma)^{n-3} + (1-\gamma)^{n-2}  \right\} \notag \\
& \qquad \qquad \qquad \qquad \qquad \qquad \qquad \qquad \qquad  \vdots   \notag \\
&= \left( \sum\nolimits_{k=2}^{\infty}\pi_{(k,1,0)} \right) \left[(1-p)(1-\gamma) \right]^{n-2}  +p\gamma M_2 \left( \sum\nolimits_{j=0}^{n-3}[(1-p)(1-\gamma)]^j (1-\gamma)^{n-2-j} \right)
\end{align}

In (172), the sum $\sum\nolimits_{k=2}^{\infty}\pi_{(k,1,0)}$ is calculated as 
\begin{align}
\sum\nolimits_{k=2}^{\infty}\pi_{(k,1,0)}&= \sum\nolimits_{k=2}^{\infty}\bigg\{ \pi_{(k-1,0,0)}p(1-\gamma) + \left(\sum\nolimits_{j=k}^{\infty}\pi_{(j,k-1,0)} \right)p\gamma \bigg\}  \notag \\
&= p(1-\gamma) \left( \sum\nolimits_{k=1}^{\infty}\pi_{(k,0,0)} \right) + p\gamma \left( \sum\nolimits_{k=2}^{\infty}\sum\nolimits_{j=k}^{\infty}\pi_{(j,k-1,0)}  \right) \notag \\
&= p(1-\gamma)M_1 + p\gamma M_2
\end{align}
where $M_1$ is also defined and calculated in Lemma 2. 

Combining (172) and (173), we have that 
\begin{align}
\sum\nolimits_{k=n}^{\infty}\pi_{(k,n-1,0)}&= \left[ p(1-\gamma)M_1 +p\gamma M_2 \right]\left[(1-p)(1-\gamma)\right]^{n-2}  + p\gamma M_2 \left[ \sum\nolimits_{j=0}^{n-3}(1-p)^j (1-\gamma)^{n-2} \right]  \notag \\
&= \left[ p(1-\gamma)M_1 +p\gamma M_2 \right]\left[(1-p)(1-\gamma)\right]^{n-2} + \gamma M_2 \Big((1-\gamma)^{n-2}-[(1-p)(1-\gamma)]^{n-2} \Big)  \notag \\
&= \left[p(1-\gamma)M_1 -(1-p)\gamma M_2 \right]  \left[(1-p)(1-\gamma) \right]^{n-2} + \gamma M_2 (1-\gamma)^{n-2} \notag \\
&= \gamma M_2 (1-\gamma)^{n-2} \qquad (n\geq 2)  
\end{align}

To obtain equation (174), notice that 
\begin{equation}
p(1-\gamma)M_1=(1-p)\gamma M_2 \notag
\end{equation}
which is proved in equation (165). Thus, we complete the proof of Lemma 3.

\section{The proof of Theorem 8}
In this appendix, we determine the probability the AoI equals $n$ in steady state and the AoI cumulative probabilities, assuming that the queue model used in the status updating system is Ber/Geo/1/2.  

First of all, the sums $\sum\nolimits_{m=1}^{n-1}\pi_{(n,m,0)}$, $n\geq 2$, and $\sum\nolimits_{m=2}^{n-1}\sum\nolimits_{l=1}^{m-1}\pi_{(n,m,l)}$, $n\geq 3$ are calculated. 
\begin{align}
&\sum\nolimits_{m=1}^{n-1}\pi_{(n,m,0)} \notag \\
={}&\sum\nolimits_{m=1}^{n-1}\bigg\{ \frac{p^2\gamma^3}{N(p,\gamma)(\gamma-p)}\bigg[ (1-p)^n(1-\gamma)^m  - (1-p)^m(1-\gamma)^n \bigg] + \frac{p\gamma^3}{N(p,\gamma)}\bigg[ (1-\gamma)^{n-1} - (1-p)^m(1-\gamma)^{n-1} \bigg]  \bigg\} \notag \\
={}& \frac{p^2\gamma^3}{N(p,\gamma)(\gamma-p)}\bigg[ (1-p)^n \frac{(1-\gamma)-(1-\gamma)^n}{\gamma} -\frac{(1-p)-(1-p)^n}{p}(1-\gamma)^n \bigg]  \notag \\
  {}& \qquad \quad  + \frac{p\gamma^3}{N(p,\gamma)} \bigg[ (n-1)(1-\gamma)^{n-1}  -\frac{(1-p)-(1-p)^n}{p}(1-\gamma)^{n-1}  \bigg] \notag \\
={}& \frac{p\gamma^2}{N(p,\gamma)(\gamma-p)} \bigg\{ p(1-\gamma)(1-p)^n-(1-p)\gamma(1-\gamma)^n + (\gamma-p)[(1-p)(1-\gamma)]^n \bigg\}   \notag \\
  {}& \qquad \quad + \frac{\gamma^3}{N(p,\gamma)} \bigg\{ (np-1)(1-\gamma)^{n-1}+ (1-p)[(1-p)(1-\gamma)]^{n-1}  \bigg\}
\end{align}

The calculation of the latter sum is more complicated and tedious. We give the details below. 
\begin{align}
&\sum\nolimits_{m=2}^{n-1}\sum\nolimits_{l=1}^{m-1}\pi_{(n,m,l)} \notag \\
={}& \sum\nolimits_{m=2}^{n-1}\sum\nolimits_{l=1}^{m-1}\pi_{(n-l,m-l,0)}p(1-\gamma)^l  \notag \\
={}& \sum\nolimits_{m=2}^{n-1}\sum\nolimits_{l=1}^{m-1}\bigg[ \frac{p^2\gamma^3}{N(p,\gamma)(\gamma-p)}\bigg\{(1-p)^{n-l}(1-\gamma)^{m-l} -(1-p)^{m-l}(1-\gamma)^{n-l} \bigg\} \notag \\
  {}& \qquad \qquad \qquad \quad + \frac{p\gamma^3}{N(p,\gamma)}\bigg\{(1-\gamma)^{n-l-1}  - (1-p)^{m-l}(1-\gamma)^{n-l-1} \bigg\} \bigg] p(1-\gamma)^l    \notag \\
={}& \sum\nolimits_{m=2}^{n-1}\sum\nolimits_{l=1}^{m-1}\bigg[ \frac{(p\gamma)^3}{N(p,\gamma)(\gamma-p)}\bigg\{(1-p)^{n-l}(1-\gamma)^{m} -(1-p)^{m-l}(1-\gamma)^{n} \bigg\}  \notag \\
  {}& \qquad \qquad \qquad \quad + \frac{p^2\gamma^3}{N(p,\gamma)}\bigg\{(1-\gamma)^{n-1} - (1-p)^{m-l}(1-\gamma)^{n-1} \bigg\} \bigg]    \notag \\
={}&\sum\nolimits_{m=2}^{n-1}\bigg[\frac{(p\gamma)^3}{N(p,\gamma)(\gamma-p)}\bigg\{\frac{(1-p)^{n-m+1}-(1-p)^n}{p} (1-\gamma)^m - \frac{(1-p)-(1-p)^m}{p}(1-\gamma)^n \bigg\}  \notag \\
  {}& \qquad \qquad \qquad \quad +\frac{p^2\gamma^3}{N(p,\gamma)}\bigg\{ (m-1)(1-\gamma)^{n-1} - \frac{(1-p)-(1-p)^m}{p}(1-\gamma)^{n-1}  \bigg\} \bigg] \notag \\
={}& \sum\nolimits_{m=2}^{n-1}\bigg[ \frac{p^2\gamma^3}{N(p,\gamma)(\gamma-p)}\bigg\{ (1-p)^{n-m+1}(1-\gamma)^m -(1-p)^n(1-\gamma)^m - (1-p)(1-\gamma)^n +(1-p)^m(1-\gamma)^n \bigg\}  \notag \\
  {}& \qquad \qquad \qquad \quad  + \frac{p\gamma^3}{N(p,\gamma)}\bigg\{ (m-1)p(1-\gamma)^{n-1} -(1-p)(1-\gamma)^{n-1}+(1-p)^m(1-\gamma)^{n-1} \bigg\}  \bigg]  \notag \\
={}& \frac{p^2\gamma^3}{N(p,\gamma)(\gamma-p)}\bigg\{ (1-p)^{n+1}\sum\nolimits_{m=2}^{n-1}\left(\frac{1-\gamma}{1-p}\right)^m - (1-p)^n\frac{(1-\gamma)^2-(1-\gamma)^n}{\gamma} \notag \\
  {}& \qquad \qquad \qquad \qquad \qquad  -(n-2)(1-p)(1-\gamma)^n  +\frac{(1-p)^2-(1-p)^n}{p}(1-\gamma)^n \bigg\} \notag \\
  {}& \qquad \qquad \qquad \quad + \frac{p\gamma^3}{N(p,\gamma)} \bigg\{\left[ \sum\nolimits_{m=2}^{n-1}(mp-1) \right](1-\gamma)^{n-1} + \frac{(1-p)^2-(1-p)^n}{p}(1-\gamma)^{n-1}  \bigg\} 
\end{align}

The two sums in (176) are computed as follows. Firstly, we have 
\begin{align}
(1-p)^{n+1}\sum\nolimits_{m=2}^{n-1}\left(\frac{1-\gamma}{1-p}\right)^m = (1-p)^{n+1}\frac{\left(\frac{1-\gamma}{1-p}\right)^2-\left(\frac{1-\gamma}{1-p}\right)^n}{1-\frac{1-\gamma}{1-p}}=\frac{(1-\gamma)^2(1-p)^n-(1-p)^2(1-\gamma)^n}{\gamma-p} \notag
\end{align}

In addition, 
\begin{align}
\left[\sum\nolimits_{m=2}^{n-1}(mp-1)\right](1-\gamma)^{n-1}= \left[\frac{(n+1)(n-2)}{2}p-(n-2)\right](1-\gamma)^{n-1}=\left[\frac{p}{2}n^2-\left(\frac{p}{2}+1\right)n+(2-p) \right](1-\gamma)^{n-1}  \notag 
\end{align}

Substituting above results into (176) gives that  
\begin{align}
&\sum\nolimits_{m=2}^{n-1}\sum\nolimits_{l=1}^{m-1}\pi_{(n,m,l)} \notag \\
={}&\frac{p^2\gamma^3}{N(p,\gamma)(\gamma-p)}\bigg\{\frac{p(1-\gamma)^2}{\gamma(\gamma-p)}(1-p)^n  +\frac{(\gamma-2p)(1-p)^2}{p(\gamma-p)}(1-\gamma)^n   - \frac{\gamma-p}{p\gamma}[(1-p)(1-\gamma)]^n -(n-2)(1-p)(1-\gamma)^n  \bigg\}  \notag \\
  {}& \qquad \quad  +\frac{p\gamma^3}{N(p,\gamma)}\bigg\{ \left[\frac p2 n^2 -\left( \frac p2 +1\right)n+(2-p) \right] (1-\gamma)^{n-1}-\frac{1-p}{p}[(1-p)(1-\gamma)]^{n-1} \bigg\} \notag \\
={}&\frac{p^3\gamma^2(1-\gamma)^2}{N(p,\gamma)(\gamma-p)^2}(1-p)^n + \frac{p\gamma^3(\gamma-2p)(1-p)^2}{N(p,\gamma)(\gamma-p)^2}(1-\gamma)^n -\frac{\gamma^2(p+\gamma-p\gamma)}{N(p,\gamma)(1-\gamma)}[(1-p)(1-\gamma)]^n   \notag \\
  {}& \qquad \quad  - \frac{p^2\gamma^3(1-p)}{N(p,\gamma)(\gamma-p)}(n-2)(1-\gamma)^n  +\frac{\gamma^3[p^2n^2-(p^2+2p)n+2]}{2N(p,\gamma)}(1-\gamma)^{n-1} 
\end{align}

Now, the stationary AoI distribution can be determined. For $n\geq3$, we only need to add up all the stationary probabilities having identical first age-component. That is 
\begin{align}
\Pr\{\Delta=n\}&=\pi_{(n,0,0)} + \sum\nolimits_{m=1}^{n-1}\pi_{(n,m,0)} + \sum\nolimits_{m=2}^{n-1}\sum\nolimits_{l=1}^{m-1}\pi_{(n,m,l)}  \notag \\
&=\frac{p(1-p)^2\gamma^4}{N(p,\gamma)(\gamma-p)^2}[(1-p)^n-(1-\gamma)^n] - \frac{p^2\gamma^3}{N(p,\gamma)}\left[ \frac{1}{2(1-\gamma)}+\frac{1-p}{\gamma-p}\right]n(1-\gamma)^n + \frac{p^2\gamma^3}{2N(p,\gamma)}n^2(1-\gamma)^{n-1}
\end{align}

The calculations to obtain expression (178) are direct, so that we omit the details. Instead, we will check that (178) indeed forms a proper probability distribution by proving that the sum of all the probabilities is equal to 1.

Notice that equation (175) is zero when $n=1$ and the second sum (177) equals zero for both $n=1$ and $n=2$. Thus, actually the probability expression (178) is valid for all $n\geq 1$.

Notice that 
\begin{align}
\sum\nolimits_{n=1}^{\infty}\Pr\{\Delta=n\}&= \sum\nolimits_{n=1}^{\infty} \bigg\{ \frac{p(1-p)^2\gamma^4}{N(p,\gamma)(\gamma-p)^2}[(1-p)^n-(1-\gamma)^n]  \notag \\
&\qquad \qquad  - \frac{p^2\gamma^3}{N(p,\gamma)}\left[ \frac{1}{2(1-\gamma)}+\frac{1-p}{\gamma-p}\right]n(1-\gamma)^n + \frac{p^2\gamma^3}{2N(p,\gamma)}n^2(1-\gamma)^{n-1} \bigg\}   \notag \\
&=\frac{p(1-p)^2\gamma^4}{N(p,\gamma)(\gamma-p)^2} \left[ \frac{1-p}{p} - \frac{1-\gamma}{\gamma} \right]  - \frac{p^2\gamma^3}{N(p,\gamma)}\left[ \frac{1}{2(1-\gamma)}+\frac{1-p}{\gamma-p}\right] \frac{1-\gamma}{\gamma^2} + \frac{p^2\gamma^3}{2N(p,\gamma)} \frac{2-\gamma}{\gamma^3}   \\
&=\frac{(1-p)^2\gamma^3}{N(p,\gamma)(\gamma-p)} - \frac{p^2\gamma}{2N(p,\gamma)} - \frac{p^2\gamma(1-p)(1-\gamma)}{N(p,\gamma)(\gamma-p)} + \frac{p^2(2-\gamma)}{2N(p,\gamma)}  \notag \\
&=\frac{(1-p)\gamma}{N(p,\gamma)(\gamma-p)}\Big\{ (1-p)\gamma^2 - p^2(1-\gamma) \Big\} + \frac{p^2(1-\gamma)}{N(p,\gamma)}
\notag \\
&=\frac{(1-p)\gamma(p+\gamma-p\gamma) + p^2(1-\gamma)}{N(p,\gamma)}
\end{align}

In equation (179), we compute the sum 
\begin{align}
\sum\nolimits_{n=1}^{\infty}n^2(1-\gamma)^{n-1}
=\sum\nolimits_{n=1}^{\infty}n(n-1)(1-\gamma)^{n-1} + \sum\nolimits_{n=1}^{\infty}n(1-\gamma)^{n-1} 
=(1-\gamma)\frac{2}{\gamma^3} + \frac{1}{\gamma^2}=\frac{2-\gamma}{\gamma^3}  \notag 
\end{align}
and (180) is obtained since $(1-p)\gamma^2 - p^2(1-\gamma)=(\gamma-p)(p+\gamma-p\gamma)$.

Further, the numerator of (180) can be rewritten as 
\begin{align}
&(1-p)\gamma(p+\gamma-p\gamma) + p^2(1-\gamma)  \notag \\
={}& \gamma(p+\gamma-p\gamma) -p\gamma(p+\gamma-p\gamma) + p^2(1-\gamma) \notag \\
={}& \gamma(p+\gamma-2p\gamma) + p\gamma^2 -p\gamma(p+\gamma-p\gamma) + p^2(1-\gamma) \notag \\
={}& \gamma(p+\gamma-2p\gamma) -p^2\gamma(1-\gamma) + p^2(1-\gamma) \notag \\
={}& \gamma(p+\gamma-2p\gamma) + p^2(1-\gamma)^2  \notag \\
={}& N(p,\gamma)
\end{align}
which proves that the sum of all the probabilities is 1. 

Now, we compute the AoI cumulative probabilities. Let $k\geq1$, it shows that 
\begin{align}
&\Pr\{\Delta\leq k\} = \sum\nolimits_{n=1}^k \Pr\{\Delta=n\}  \notag \\
={}& \sum\nolimits_{n=1}^k \bigg\{ \frac{p(1-p)^2\gamma^4}{N(p,\gamma)(\gamma-p)^2}[(1-p)^n-(1-\gamma)^n]  - \frac{p^2\gamma^3}{N(p,\gamma)}\left[ \frac{1}{2(1-\gamma)}+\frac{1-p}{\gamma-p}\right]n(1-\gamma)^n + \frac{p^2\gamma^3}{2N(p,\gamma)}n^2(1-\gamma)^{n-1} \bigg\}  \notag 
\end{align}

Notice that we have following sums 
\begin{equation}
\sum\nolimits_{n=1}^k n(1-\gamma)^{n-1} = \frac{1-(1-\gamma)^k-k\gamma(1-\gamma)^k}{\gamma^2}   \notag 
\end{equation}
and
\begin{equation}
\sum\nolimits_{n=1}^k n^2(1-\gamma)^{n-1}= \sum\nolimits_{n=1}^k n(n-1)(1-\gamma)^{n-1} 
+ \sum\nolimits_{n=1}^k n(1-\gamma)^{n-1}   \notag
\end{equation}
in which
\begin{align}
\sum\nolimits_{n=1}^k n(n-1)(1-\gamma)^{n-1}&=(1-\gamma)\left[ \sum\nolimits_{n=0}^{k}x^n \right]''_{x=1-\gamma} \notag \\
&= (1-\gamma) \bigg\{ \frac{2}{\gamma^3} - \frac{\gamma^2 k^2(1-\gamma)^{k-1} + (2\gamma-\gamma^2)k(1-\gamma)^{k-1} + 2(1-\gamma)^k}{\gamma^3}\bigg\} \notag\\
&= \frac{2(1-\gamma) - \gamma^2 k^2(1-\gamma)^k - (2\gamma-\gamma^2)k(1-\gamma)^k - 2(1-\gamma)^{k+1}}{\gamma^3} \notag 
\end{align} 

Therefore, the probability that AoI is greater than $k$ is equal to 
\begin{align}
\Pr\{\Delta\leq k\}&= \frac{p(1-p)^2\gamma^4}{N(p,\gamma)(\gamma-p)^2}\bigg\{\frac{(1-p)-(1-p)^{k+1}}{p}- \frac{(1-\gamma)-(1-\gamma)^{k+1}}{\gamma} \bigg\}  \notag \\
& \quad - \frac{p^2\gamma^3}{N(p,\gamma)}\left[\frac{1}{2(1-\gamma)} + \frac{1-p}{\gamma-p}\right](1-\gamma)\frac{1-(1-\gamma)^k - k\gamma(1-\gamma)^k}{\gamma^2}  \notag \\
& \qquad + \frac{p^2\gamma^3}{2N(p,\gamma)}\bigg\{ \frac{2(1-\gamma) - \gamma^2 k^2(1-\gamma)^k - (2\gamma-\gamma^2)k(1-\gamma)^k - 2(1-\gamma)^{k+1}}{\gamma^3}+ \frac{1-(1-\gamma)^k - k\gamma(1-\gamma)^k}{\gamma^2} \bigg\}  \notag  \\
&=1- \frac{(1-p)^3\gamma^4}{N(p,\gamma)(\gamma-p)^2}(1-p)^k + \frac{p(1-\gamma)}{N(p,\gamma)(\gamma-p)}\left[ \frac{(1-p)^2\gamma^3}{\gamma-p}+ p^2(1-\gamma) \right] (1-\gamma)^k \notag \\
&\qquad + \frac{p^2\gamma}{N(p,\gamma)}\left[ \frac{(1-p)(1-\gamma)\gamma}{\gamma-p}-\frac{2-\gamma}{2} \right] k(1-\gamma)^k - \frac{(p\gamma)^2}{2N(p,\gamma)}k^2(1-\gamma)^k   \qquad  (k\geq1)
\end{align}

We omit the calculations to obtain the final result (182), and finish the proof of Theorem 8.

\section{Proof of Theorem 11}
Here, we prove that the probabilities $\pi_{(n,0,0)}$ are equal to 
\begin{equation}
\pi_{(n,0,0)}=p(1-p)^{n-1}-p(1-\gamma) \left[1 + \frac{p\gamma(p+\gamma-p\gamma)}{(p+\gamma-p\gamma)^2-p\gamma} n\right] [(1-p)(1-\gamma)]^{n-1} \quad  (n\geq1) 
\end{equation}

Equations (76) shows that 
\begin{equation}
\pi_{(n,0,0)}=\pi_{(n-1,0,0)}(1-p) + \left( \sum\nolimits_{k=n}^{\infty}\pi_{(k,n-1,0)} \right) (1-p)\gamma  \qquad (n\geq2)
\end{equation}

Using formula (78), the sum in (184) is dealt with as 
\begin{align}
& \sum\nolimits_{k=n}^{\infty}\pi_{(k,n-1,0)}   \notag \\
={}& \sum\nolimits_{k=n}^{\infty}\left\{ \pi_{(k-n+1,0,0)}p(1-\gamma)[(1-p)(1-\gamma)]^{n-2} + p\gamma (n-1)t_{k-n+1} [(1-p)(1-\gamma)]^{n-2} \right\} \notag \\
={}& \left( \sum\nolimits_{k=1}^{\infty}\pi_{(k,0,0)}\right)p(1-\gamma)[(1-p)(1-\gamma)]^{n-2} + p\gamma (n-1) \left( \sum\nolimits_{k=1}^{\infty}t_k \right) [(1-p)(1-\gamma)]^{n-2}  
\end{align}

Define $t_0=\sum\nolimits_{n=1}^{\infty}\pi_{(n,0,0)}$, then we have the equation 
\begin{equation}
t_0 + \sum\nolimits_{k=1}^{\infty}t_k = 1
\end{equation}
since $t_0$ and $t_k$, $k\geq1$ constitutes the probability distribution of the second age-component in state vector $(n,m,l)$. Applying also the last line in (76), which says that $\sum\nolimits_{k=1}^{\infty}\pi_{(k,0,0)}=\frac{\pi_{(1,0,0)}}{p\gamma}$, we obtain that 
\begin{align}
& \sum\nolimits_{k=n}^{\infty}\pi_{(k,n-1,0)}   \notag \\
={}& \frac{\pi_{(1,0,0)}}{p\gamma} p(1-\gamma)[(1-p)(1-\gamma)]^{n-2} + p\gamma (n-1) \left( 1- \frac{\pi_{(1,0,0)}}{p\gamma} \right) [(1-p)(1-\gamma)]^{n-2} \notag \\
={}& \frac{\pi_{(1,0,0)(1-\gamma)}}{\gamma} [(1-p)(1-\gamma)]^{n-2} + \left( p\gamma- \pi_{(1,0,0)} \right) (n-1) [(1-p)(1-\gamma)]^{n-2} 
\end{align}

Therefore, 
\begin{equation}
\pi_{(n,0,0)}=\pi_{(n-1,0,0)}(1-p) + \pi_{(1,0,0)}[(1-p)(1-\gamma)]^{n-1} + (1-p)\gamma \left( p\gamma - \pi_{(1,0,0)} \right) (n-1)[(1-p)(1-\gamma)]^{n-2}
\end{equation}

Using equation (188) repeatedly yields that 
\begin{align}
\pi_{(n,0,0)}&=\pi_{(1,0,0)}(1-p)^{n-1} + \pi_{(1,0,0)}\left( \sum\nolimits_{j=0}^{n-2}(1-p)^j [(1-p)(1-\gamma)]^{n-1-j} \right) \notag \\
&\qquad + (1-p)\gamma \left(p\gamma-\pi_{(1,0,0)}\right) \left( \sum\nolimits_{j=0}^{n-2}(1-p)^j (n-1-j)[(1-p)(1-\gamma)]^{n-2-j} \right)  \notag \\
&=\pi_{(1,0,0)}(1-p)^{n-1} + \pi_{(1,0,0)} \frac{1-\gamma}{\gamma} \left\{ (1-p)^{n-1} - [(1-p)(1-\gamma)]^{n-1}\right\} \notag \\
&\qquad + (1-p)\gamma \left(p\gamma-\pi_{(1,0,0)}\right) \frac{(1-p)^{n-1}-[1+(n-1)\gamma][(1-p)(1-\gamma)]^{n-1}}{(1-p)\gamma^2} \notag \\
&=p(1-p)^{n-1} - p(1-\gamma)[(1-p)(1-\gamma)]^{n-1} - \left( p\gamma - \pi_{(1,0,0)}\right) n[(1-p)(1-\gamma)]^{n-1}
\end{align}

Let $n=1$ in (189), it gives the obvious equation $\pi_{(1,0,0)}=\pi_{(1,0,0)}$. Thus, expression is appliable for all $n\geq1$.

Using the last line of (76) again, it shows that 
\begin{equation}
\frac{\pi_{(1,0,0)}}{p\gamma}=\sum\nolimits_{n=1}^{\infty} \left\{ p(1-p)^{n-1} - p(1-\gamma)[(1-p)(1-\gamma)]^{n-1} - \left( p\gamma - \pi_{(1,0,0)}\right) n[(1-p)(1-\gamma)]^{n-1} \right\} \notag 
\end{equation}
from which we derive that 
\begin{equation}
\pi_{(1,0,0)}=\frac{p(1-p)\gamma^3}{(p+\gamma-p\gamma)^2-p\gamma}
\end{equation}

Notice that the denominator of $\pi_{(1,0,0)}$ is always positive since we have 
\begin{equation}
(p+\gamma-p\gamma)^2-p\gamma > (p+\gamma-p\gamma)^2-\gamma^2 = \left[(p+\gamma-p\gamma)+\gamma \right] p(1-\gamma)>0  \notag 
\end{equation}

Substitute (190) into (189), eventually we prove that the probability $\pi_{(n,0,0)}$ is equal to 
\begin{align}
\pi_{(n,0,0)}&=p(1-p)^{n-1} - p(1-\gamma)[(1-p)(1-\gamma)]^{n-1} - \frac{p^2\gamma(1-\gamma)(p+\gamma-p\gamma)}{(p+\gamma-p\gamma)^2-p\gamma} n[(1-p)(1-\gamma)]^{n-1}  \notag \\
&=p(1-p)^{n-1} - p(1-\gamma)\left[1 + \frac{p\gamma(p+\gamma-p\gamma)}{(p+\gamma-p\gamma)^2-p\gamma}n\right][(1-p)(1-\gamma)]^{n-1}
\end{align}

This completes the proof of Theorem 11.

\section{Proof of Lemma 5}
In this appendix, we prove Lemma 5 which determines the following sums 
\begin{equation}
t_n = \sum\nolimits_{k=n+1}^{\infty} \sum\nolimits_{j=0}^{n-1}\pi_{(k,n,j)} \qquad (n\geq1)  \notag 
\end{equation}
and $t_0=\sum\nolimits_{k=1}^{\infty}\pi_{(k,0,0)}$.

We have that 
\begin{align}
t_n &=\sum\nolimits_{k=n+1}^{\infty} \sum\nolimits_{j=0}^{n-1}\pi_{(k,n,j)} \notag \\
&= \sum\nolimits_{k=n+1}^{\infty}\pi_{(k,n,0)} + \sum\nolimits_{k=n+1}^{\infty} \sum\nolimits_{j=1}^{n-1}\pi_{(k,n,j)} \notag \\
&=\sum\nolimits_{k=n+1}^{\infty}\pi_{(k,n,0)} + \sum\nolimits_{k=n+1}^{\infty} \sum\nolimits_{j=1}^{n-1}\left( \sum\nolimits_{y=0}^{n-j-1}\pi_{(k-j,n-j,y)} \right) p(1-\gamma)[(1-p)(1-\gamma)]^{j-1}   \\
&=\sum\nolimits_{k=n+1}^{\infty}\pi_{(k,n,0)} + p(1-\gamma) \sum\nolimits_{j=1}^{n-1} \left( \sum\nolimits_{k=n-j+1}^{\infty} \sum\nolimits_{y=0}^{n-j-1}\pi_{(k,n-j,y)} \right) [(1-p)(1-\gamma)]^{j-1} \notag   \\
&=\sum\nolimits_{k=n+1}^{\infty}\pi_{(k,n,0)} + p(1-\gamma) \sum\nolimits_{j=1}^{n-1}t_{n-j}[(1-p)(1-\gamma)]^{j-1} 
\end{align}
where in equation (192), the expression (79) is substituted. Applying (78), the first sum of (193) is calculated as  
\begin{align}
&\sum\nolimits_{k=n+1}^{\infty}\pi_{(k,n,0)} \notag \\
={}&\sum\nolimits_{k=n+1}^{\infty} \left\{ \pi_{(k-n,0,0)}p(1-\gamma)[(1-p)(1-\gamma)]^{n-1} + p\gamma n t_{k-n}[(1-p)(1-\gamma)]^{n-1} \right\} \notag \\
={}& \left( \sum\nolimits_{k=1}^{\infty}\pi_{(k,0,0)}\right) p(1-\gamma)[(1-p)(1-\gamma)]^{n-1} + p\gamma \left( \sum\nolimits_{k=1}^{\infty}t_k\right) n [(1-p)(1-\gamma)]^{n-1}  \notag \\
={}& \frac{\pi_{(1,0,0)}(1-\gamma)}{\gamma}[(1-p)(1-\gamma)]^{n-1} + \left(p\gamma-\pi_{(1,0,0)}\right)n[(1-p)(1-\gamma)]^{n-1}
\end{align}
in which we have used the relations
\begin{equation}
\sum\nolimits_{k=1}^{\infty}\pi_{(k,0,0)}=\frac{\pi_{(1,0,0)}}{p\gamma} 
\text{ and } \sum\nolimits_{k=1}^{\infty}t_k = 1-t_0=1-\sum\nolimits_{k=1}^{\infty}\pi_{(k,0,0)}   \notag 
\end{equation}

First of all, we can derive that  
\begin{equation}
t_0=\frac{\pi_{(1,0,0)}}{p\gamma}= \frac{(1-p)\gamma^2}{(p+\gamma-p\gamma)^2-p\gamma}
\end{equation}
since $\pi_{(1,0,0)}$ is known in (190).

For $n\geq2$, we have that
\begin{equation}
t_n=p(1-\gamma) \sum\nolimits_{j=1}^{n-1}t_{n-j}[(1-p)(1-\gamma)]^{j-1} +  \frac{\pi_{(1,0,0)}(1-\gamma)}{\gamma}[(1-p)(1-\gamma)]^{n-1} + \left(p\gamma-\pi_{(1,0,0)}\right)n[(1-p)(1-\gamma)]^{n-1}
\end{equation}

According to (196), similarly we can write the expression of $t_{n-1}$. Observing that the following difference 
\begin{equation}
t_n - (1-p)(1-\gamma)t_{n-1} = p(1-\gamma)t_{n-1} + \left( p\gamma- \pi_{(1,0,0)}\right)[(1-p)(1-\gamma)]^{n-1} \notag 
\end{equation}
gives the equation 
\begin{equation}
t_n = t_{n-1}(1-\gamma) + \left( p\gamma- \pi_{(1,0,0)}\right)[(1-p)(1-\gamma)]^{n-1}
\end{equation}

Using (197) repeatedly, we have  
\begin{align}
t_n&=t_1(1-\gamma)^{n-1} +  \left( p\gamma- \pi_{(1,0,0)}\right) \left( \sum\nolimits_{j=0}^{n-2}(1-\gamma)^j [(1-p)(1-\gamma)]^{n-1-j} \right)  \notag \\
&=t_1(1-\gamma)^{n-1} +  \left( p\gamma- \pi_{(1,0,0)}\right) \frac{1-p}{p}\left\{ (1-\gamma)^{n-1} - [(1-p)(1-\gamma)]^{n-1}\right\} \notag \\
&=t_1(1-\gamma)^{n-1} + \frac{p\gamma(1-p)(1-\gamma)(p+\gamma-p\gamma)}{(p+\gamma-p\gamma)^2-p\gamma} \left\{ (1-\gamma)^{n-1} - [(1-p)(1-\gamma)]^{n-1}\right\} 
\end{align}

Let $n=1$, equation (198) reduces to $t_1=t_1$, thus (198) is valid for $n\geq1$.

Since 
\begin{align}
\sum\nolimits_{n=1}^{\infty}t_n &= \sum\nolimits_{n=1}^{\infty} \left[ t_1(1-\gamma)^{n-1} + \frac{p\gamma(1-p)(1-\gamma)(p+\gamma-p\gamma)}{(p+\gamma-p\gamma)^2-p\gamma} \left\{ (1-\gamma)^{n-1} - [(1-p)(1-\gamma)]^{n-1}\right\} \right] \notag \\
&=\frac{t_1}{\gamma} +  \frac{p\gamma(1-p)(1-\gamma)(p+\gamma-p\gamma)}{(p+\gamma-p\gamma)^2-p\gamma} \left(\frac{1}{\gamma} - \frac{1}{p+\gamma-p\gamma} \right)  \notag \\
&=\frac{t_1}{\gamma} + \frac{p^2(1-p)(1-\gamma)^2}{(p+\gamma-p\gamma)^2-p\gamma}  \\
&= 1-t_0 \notag \\
&= 1- \sum\nolimits_{k=1}^{\infty}\pi_{(k,0,0)}  \notag \\
&= 1- \frac{\pi_{(1,0,0)}}{p\gamma}  \notag \\
&= \frac{p(1-\gamma)(p+\gamma-p\gamma)}{(p+\gamma-p\gamma)^2-p\gamma} 
\end{align}

From equations (199) and (200), we can solve that 
\begin{equation}
t_1=\frac{p\gamma(1-\gamma) \left[ (p+\gamma-p\gamma) - p(1-p)(1-\gamma)\right]}{(p+\gamma-p\gamma)^2-p\gamma}
\end{equation}

Substituting (201) into equaiton (198) obtains 
\begin{equation}
t_n=\frac{p\gamma(1-\gamma)(p+2\gamma-2p\gamma)}{(p+\gamma-p\gamma)^2-p\gamma}(1-\gamma)^{n-1} 
- \frac{p\gamma(1-p)(1-\gamma)(p+\gamma-p\gamma)}{(p+\gamma-p\gamma)^2-p\gamma}[(1-p)(1-\gamma)]^{n-1} \qquad (n\geq1)
\end{equation}

Combining results (195) and (202), we finally determine the sums $t_n$, $n\geq0$. This completes the proof of Lemma 5.

\section{Proof of Theorem 14}
In this appendix, we calculate the probability expression of the AoI and determine the stationary AoI-distribution for the system with Ber/Geo/1/2* queue.

First of all, when $n=1$ we have 
\begin{equation}
\Pr\{\Delta=1\}= \pi_{(1,0,0)}=\frac{p(1-p)\gamma^3}{(p+\gamma-p\gamma)^2-p\gamma}
\end{equation}

In general, the probability that the AoI takes value $n$ is obtained by collecting all the stationary probabilities having $n$ as first age-component. Therefore, it shows that 
\begin{equation}
\Pr\{\Delta=n\}=\pi_{(n,0,0)}+\sum\nolimits_{m=1}^{n-1}\pi_{(n,m,0)} + \sum\nolimits_{m=2}^{n-1}\sum\nolimits_{l=1}^{m-1}\pi_{(n,m,l)}  
\end{equation}

Substituting equation (86), we compute the first sum in (204) as follows. 
\begin{align}
\sum\nolimits_{m=1}^{n-1}\pi_{(n,m,0)}&=\sum\nolimits_{m=1}^{n-1}\bigg\{ p^2(1-p)^{n-2}[(1-\gamma)^m-(1-\gamma)^n] + \frac{(p\gamma)^2(p+2\gamma-2p\gamma)}{(p+\gamma-p\gamma)^2-p\gamma}m(1-p)^{m-1}(1-\gamma)^{n-1}  \notag \\
& \qquad - \frac{p^2\gamma(1-\gamma)(p+\gamma-p\gamma)}{(p+\gamma-p\gamma)^2-p\gamma}[p(1-\gamma)n+(\gamma-p)m][(1-p)(1-\gamma)]^{n-2} \bigg\}  \notag \\
&=p^2(1-p)^{n-2}\left\{\frac{(1-\gamma)-(1-\gamma)^n}{\gamma}-(n-1)(1-\gamma)^n \right\} \notag \\  
&\qquad + \frac{(p\gamma)^2(p+2\gamma-2p\gamma)}{(p+\gamma-p\gamma)^2-p\gamma}\left\{ \frac{1-(1-p)^{n-1}-p(n-1)(1-p)^{n-1}}{p^2} \right\} (1-\gamma)^{n-1} \notag \\
&\qquad -  \frac{p^2\gamma(1-\gamma)(p+\gamma-p\gamma)}{(p+\gamma-p\gamma)^2-p\gamma}\left[ p(1-\gamma)n(n-1) + (\gamma-p) \frac{n(n-1)}{2} \right] [(1-p)(1-\gamma)]^{n-2}   \notag \\
&=\frac{p^2(1-\gamma)}{(1-p)\gamma}\left\{(1-p)^{n-1}-[1+(n-1)\gamma][(1-p)(1-\gamma)]^{n-1} \right\} \notag \\
&\qquad +\frac{(p+2\gamma-2p\gamma)\gamma^2}{(p+\gamma-p\gamma)^2-p\gamma}\left\{(1-\gamma)^{n-1}-[1+(n-1)p][(1-p)(1-\gamma)]^{n-1} \right\}  \notag \\
&\qquad - \frac{p^2\gamma(1-\gamma)(p+\gamma-p\gamma)(p+\gamma-2p\gamma)}{2[(p+\gamma-p\gamma)^2-p\gamma]}n(n-1)[(1-p)(1-\gamma)]^{n-2} 
\end{align}
which gives equation (98).

In order to verify that the sum of all the stationary probabilities is 1, here we continue to calculate the sum of (205) over the range $n\geq2$. During the following calculations, the formulas 
\begin{equation}
\sum\nolimits_{n=1}^{\infty}nx^{n-1}=\frac{1}{(1-x)^2}, \qquad  \sum\nolimits_{n=2}^{\infty}n(n-1)x^{n-2}=\frac{2}{(1-x)^3} \qquad (0<x<1)
\end{equation}
will be used multiple times. We show that 
\begin{align}
&\sum\nolimits_{n=2}^{\infty}\sum\nolimits_{m=1}^{n-1}\pi_{(n,m,0)} \notag \\
={}& \frac{p^2(1-\gamma)}{(1-p)\gamma}\left( \frac{1-p}{p} - \frac{(1-p)(1-\gamma)}{p+\gamma-p\gamma} - \frac{\gamma(1-p)(1-\gamma)}{(p+\gamma-p\gamma)^2} \right)  \notag \\
  {}& +\frac{(p+2\gamma-2p\gamma)\gamma^2}{(p+\gamma-p\gamma)^2-p\gamma}\left(\frac{1-\gamma}{\gamma}-\frac{(1-p)(1-\gamma)}{p+\gamma-p\gamma}-\frac{p(1-p)(1-\gamma)}{(p+\gamma-p\gamma)^2} \right) \notag \\
  {}& -\frac{p^2\gamma(1-\gamma)(p+\gamma-p\gamma)(p+\gamma-2p\gamma)}{2[(p+\gamma-p\gamma)^2-p\gamma]}\frac{2}{(p+\gamma-p\gamma)^3}  \notag \\
={}& \frac{p^2(1-\gamma)}{(1-p)\gamma}\frac{(1-p)\gamma^2}{p(p+\gamma-p\gamma)^2} + \frac{(p+2\gamma-2p\gamma)\gamma^2}{(p+\gamma-p\gamma)^2-p\gamma}\frac{p^2(1-\gamma)}{\gamma(p+\gamma-p\gamma)^2} - \frac{p^2\gamma(1-\gamma)(p+\gamma-2p\gamma)}{[(p+\gamma-p\gamma)^2-p\gamma](p+\gamma-p\gamma)^2} \notag \\
={}& \frac{p\gamma(1-\gamma)}{(p+\gamma-p\gamma)^2} + \frac{p^2\gamma(1-\gamma)(p+2\gamma-2p\gamma)}{[(p+\gamma-p\gamma)^2-p\gamma](p+\gamma-p\gamma)^2} -  \frac{p^2\gamma(1-\gamma)(p+\gamma-2p\gamma)}{[(p+\gamma-p\gamma)^2-p\gamma](p+\gamma-p\gamma)^2} \notag \\
={}&  \frac{p\gamma(1-\gamma)}{(p+\gamma-p\gamma)^2} + \frac{p^2\gamma^2(1-\gamma)}{[(p+\gamma-p\gamma)^2-p\gamma](p+\gamma-p\gamma)^2} \notag \\
={}& \frac{p\gamma(1-\gamma)}{(p+\gamma-p\gamma)^2} \left\{1+\frac{p\gamma}{(p+\gamma-p\gamma)^2-p\gamma}\right\} \notag \\
={}&\frac{p\gamma(1-\gamma)}{(p+\gamma-p\gamma)^2-p\gamma}
\end{align}

Therefore, we obtain the partial sum of the stationary probabilities. Next, the latter sum in (204) is computed. Using probability expression (96), we have   
\begin{align}
&\sum\nolimits_{m=2}^{n-1}\sum\nolimits_{l=1}^{m-1}\pi_{(n,m,l)} \notag \\
={}&\sum\nolimits_{m=2}^{n-1}\sum\nolimits_{l=1}^{m-1} \bigg\{ p^3(1-p)^{n-m+l-2}(1-\gamma)^m \left[ 1 - (1-\gamma)^{n-m} \right]  \notag \\
&\quad - \frac{p^2\gamma(p+\gamma-p\gamma)}{(p+\gamma-p\gamma)^2-p\gamma}\left[p^2(1-\gamma)(n-m) + (1-p)\gamma\right] (1-p)^{n-m+l-2}(1-\gamma)^{n-1}  \notag \\
& \quad  +  \frac{(p\gamma)^2(p+2\gamma-2p\gamma)}{(p+\gamma-p\gamma)^2-p\gamma}(1-p)^{l-1}(1-\gamma)^{n-1}\left[1 - (1-p)^{m-l}\right] +\frac{(p\gamma)^2(p+\gamma-p\gamma)}{(p+\gamma-p\gamma)^2-p\gamma}[(1-p)(1-\gamma)]^{n-1}  \bigg\} \notag\\
={}&\sum\nolimits_{m=2}^{n-1}\bigg( p^2 \left\{(1-p)^{n-m-1} -(1-p)^{n-2}\right\} \left\{ (1-\gamma)^m -(1-\gamma)^n\right\}\notag \\
  {}& \quad -\frac{p\gamma(p+\gamma-p\gamma)}{(p+\gamma-p\gamma)^2-p\gamma}\left[p^2(1-\gamma)(n-m)+(1-p)\gamma \right] \left\{ (1-p)^{n-m-1}(1-\gamma)^{n-1} - (1-p)^{n-2}(1-\gamma)^{n-1}\right\}  \notag \\
  {}& \quad + \frac{p\gamma^2(p+2\gamma-2p\gamma)}{(p+\gamma-p\gamma)^2-p\gamma}\left\{(1-\gamma)^{n-1}-(1-p)^{m-1}(1-\gamma)^{n-1}-p(m-1)(1-p)^{m-1}(1-\gamma)^{n-1} \right\} \notag \\
 {}& \quad + \frac{(p\gamma)^2(p+\gamma-p\gamma)}{(p+\gamma-p\gamma)^2-p\gamma}(m-1)[(1-p)(1-\gamma)]^{n-1}  \bigg)
\end{align}
in which
\begin{align}
S_1&=\sum\nolimits_{m=2}^{n-1}p^2 \left\{(1-p)^{n-m-1} -(1-p)^{n-2}\right\} \left\{ (1-\gamma)^m -(1-\gamma)^n\right\}  \notag \\
&=p^2\sum\nolimits_{m=2}^{n-1}\left\{(1-p)^{n-m-1}(1-\gamma)^m -(1-p)^{n-m-1}(1-\gamma)^n -(1-p)^{n-2}(1-\gamma)^m +(1-p)^{n-2}(1-\gamma)^n \right\} \notag \\
&=p^2\bigg(\frac{(1-\gamma)^2}{\gamma-p}\left[(1-p)^{n-2}-(1-\gamma)^{n-2}\right] -\frac{(1-\gamma)^2}{p}\left[(1-\gamma)^{n-2}-[(1-p)(1-\gamma)]^{n-2}\right]  \notag \\
& \qquad -\frac{(1-\gamma)^2}{\gamma}\left[(1-p)^{n-2}-[(1-p)(1-\gamma)]^{n-2}\right] +(1-\gamma)^2(n-2)[(1-p)(1-\gamma)]^2 \bigg)  \notag \\
&=p^2(1-\gamma)^2\left\{\frac{p}{\gamma(\gamma-p)}(1-p)^{n-2} - \frac{\gamma}{p(\gamma-p)}(1-\gamma)^{n-2}+\left[\frac{1}{p}+\frac{1}{\gamma}+(n-2) \right] [(1-p)(1-\gamma)]^{n-2} \right\} 
\end{align}
and
\begin{align}
S_2&=\sum\nolimits_{m=2}^{n-1}\frac{p\gamma(p+\gamma-p\gamma)}{(p+\gamma-p\gamma)^2-p\gamma}\left[p^2(1-\gamma)(n-m)+(1-p)\gamma \right] \left\{ (1-p)^{n-m-1}(1-\gamma)^{n-1} - (1-p)^{n-2}(1-\gamma)^{n-1}\right\}  \notag \\
&=\frac{p\gamma(p+\gamma-p\gamma)}{(p+\gamma-p\gamma)^2-p\gamma}\left( \sum\nolimits_{m=2}^{n-1}\left[p^2(1-\gamma)(n-m)+(1-p)\gamma \right] (1-p)^{n-m-1} \right) (1-\gamma)^{n-1}  \notag \\
&\qquad - \frac{p\gamma(p+\gamma-p\gamma)}{(p+\gamma-p\gamma)^2-p\gamma}\left\{p^2(1-\gamma)^2 \frac{(n-1)(n-2)}{2}[(1-p)(1-\gamma)]^{n-2}+\gamma (n-2)[(1-p)(1-\gamma)]^{n-1}\right\}
\end{align}

Let $n-m=j$, it shows that 
\begin{align}
&\sum\nolimits_{m=2}^{n-1}\left[p^2(1-\gamma)(n-m)+(1-p)\gamma \right] (1-p)^{n-m-1} \notag \\
={}&\sum\nolimits_{j=1}^{n-2}\left[p^2(1-\gamma)j+(1-p)\gamma \right] (1-p)^{j-1} \notag \\
={}&p^2(1-\gamma)\frac{1-(1-p)^{n-2}-p(n-2)(1-p)^{n-2}}{p^2}+\gamma \frac{(1-p)-(1-p)^{n-1}}{p} \notag \\
={}&\frac{p(1-\gamma)-p(1-\gamma)(1-p)^{n-2}-p^2(1-\gamma)(n-2)(1-p)^{n-2}+(1-p)\gamma-(1-p)\gamma(1-p)^{n-2}}{p} \notag \\
={}&\frac{p+\gamma-2p\gamma}{p} - \frac{p+\gamma-2p\gamma}{p}(1-p)^{n-2} - p(1-\gamma)(n-2)(1-p)^{n-2}
\end{align}

Therefore, we obtain that 
\begin{align}
S_2&=\frac{p\gamma(p+\gamma-p\gamma)}{(p+\gamma-p\gamma)^2-p\gamma} \bigg\{ \frac{(p+\gamma-2p\gamma)(1-\gamma)}{p}(1-\gamma)^{n-2}  \notag \\
&\qquad  - \frac{(p+\gamma-2p\gamma)(1-\gamma)}{p}[(1-p)(1-\gamma)]^{n-2}- p(1-\gamma)^2 (n-2)[(1-p)(1-\gamma)]^{n-2} \bigg\} \notag \\
&\qquad - \frac{p\gamma(p+\gamma-p\gamma)}{(p+\gamma-p\gamma)^2-p\gamma}\left\{p^2(1-\gamma)^2 \frac{(n-1)(n-2)}{2}[(1-p)(1-\gamma)]^{n-2}+\gamma (n-2)[(1-p)(1-\gamma)]^{n-1}\right\} \notag \\
&=\frac{p\gamma(p+\gamma-p\gamma)}{(p+\gamma-p\gamma)^2-p\gamma}\bigg( \frac{(p+\gamma-2p\gamma)(1-\gamma)}{p}\bigg\{ (1-\gamma)^{n-2} - [(1-p)(1-\gamma)]^{n-2}  \notag \\
&\qquad - p(n-2)[(1-p)(1-p\gamma)]^{n-2} \bigg\} - \frac{p^2(1-\gamma)^2}{2}(n-1)(n-2) [(1-p)(1-\gamma)]^{n-2} \bigg)
\end{align}

At last, the rest part of equation (208) is dealt with together. 
\begin{align}
S_3&=\frac{p\gamma^2(p+2\gamma-2p\gamma)}{(p+\gamma-p\gamma)^2-p\gamma} \sum\nolimits_{m=2}^{n-1} \bigg\{(1-\gamma)^{n-1}-(1-p)^{m-1}(1-\gamma)^{n-1}-p(m-1)(1-p)^{m-1}(1-\gamma)^{n-1} \bigg\}  \notag \\
& \qquad + \frac{(p\gamma)^2(p+\gamma-p\gamma)}{(p+\gamma-p\gamma)^2-p\gamma}\frac{(n-1)(n-2)}{2}[(1-p)(1-\gamma)]^{n-1} \notag \\
&=\frac{p\gamma^2(p+2\gamma-2p\gamma)}{(p+\gamma-p\gamma)^2-p\gamma} \bigg\{(n-2)(1-\gamma)^{n-1}- \frac{(1-p)(1-\gamma)^{n-1}- [(1-p)(1-\gamma)]^{n-1}}{p}  \notag \\
& \qquad - \frac{(1-p)(1-\gamma)^{n-1}-[(1-p)(1-\gamma)]^{n-1}}{p} + (n-2)[(1-p)(1-\gamma)]^{n-2} \bigg\} \notag \\
& \qquad + \frac{(p\gamma)^2(p+\gamma-p\gamma)}{(p+\gamma-p\gamma)^2-p\gamma}\frac{(n-1)(n-2)}{2}[(1-p)(1-\gamma)]^{n-1} \notag \\
&=\frac{\gamma^2(p+2\gamma-2p\gamma)}{(p+\gamma-p\gamma)^2-p\gamma}\bigg\{p(n-2)(1-\gamma)^{n-1}-2(1-p)(1-\gamma)^{n-1}+2[(1-p)(1-\gamma)]^{n-1}+p(n-2)[(1-p)(1-\gamma)]^{n-1} \bigg\}  \notag \\
& \qquad + \frac{(p\gamma)^2(p+\gamma-p\gamma)}{(p+\gamma-p\gamma)^2-p\gamma}\frac{(n-1)(n-2)}{2}[(1-p)(1-\gamma)]^{n-1}
\end{align}

Collecting results (209), (212), and (213), eventually we have that 
\begin{align}
&\sum\nolimits_{m=2}^{n-1}\sum\nolimits_{l=1}^{m-1}\pi_{(n,m,l)}=S_1 - S_2+S_3  \notag \\
={}& p^2(1-\gamma)^2\left\{\frac{p}{\gamma(\gamma-p)}(1-p)^{n-2} - \frac{\gamma}{p(\gamma-p)}(1-\gamma)^{n-2}+\left[\frac{1}{p}+\frac{1}{\gamma}+(n-2) \right] [(1-p)(1-\gamma)]^{n-2} \right\}   \notag \\
  {}& - \frac{p\gamma(p+\gamma-p\gamma)}{(p+\gamma-p\gamma)^2-p\gamma}\bigg( \frac{(p+\gamma-2p\gamma)(1-\gamma)}{p}\bigg\{ (1-\gamma)^{n-2} - [(1-p)(1-\gamma)]^{n-2}  \notag \\
  {}& \qquad - p(n-2)[(1-p)(1-p\gamma)]^{n-2} \bigg\} - \frac{p^2(1-\gamma)^2}{2}(n-1)(n-2) [(1-p)(1-\gamma)]^{n-2} \bigg)  \notag \\
  {}& + \frac{\gamma^2(p+2\gamma-2p\gamma)}{(p+\gamma-p\gamma)^2-p\gamma}\bigg\{p(n-2)(1-\gamma)^{n-1}-2(1-p)(1-\gamma)^{n-1}+2[(1-p)(1-\gamma)]^{n-1}+p(n-2)[(1-p)(1-\gamma)]^{n-1} \bigg\}  \notag \\
  {}& \qquad + \frac{(p\gamma)^2(p+\gamma-p\gamma)}{2[(p+\gamma-p\gamma)^2-p\gamma]}(n-1)(n-2)[(1-p)(1-\gamma)]^{n-1} 
\end{align}
which is exactly the probability expression (99).

It can be directly verified that (214) is zero when we let $n=2$. Thus, equation (204) is actually valid for $n\geq2$. Finally, combining (203), (84), (205) and (214), we obtain the distribution of AoI and complete the proof of Theorem 14.

To ensure that equation (97) is indeed a proper probability distribution, we have to check  
\begin{equation}
\sum\nolimits_{n=1}^{\infty}\Pr\{\Delta=n\}
=\sum\nolimits_{n=1}^{\infty}\pi_{(n,0,0)} + \sum\nolimits_{n=2}^{\infty}\sum\nolimits_{m=1}^{n-1}\pi_{(n,m,0)} 
  + \sum\nolimits_{n=3}^{\infty}\sum\nolimits_{m=2}^{n-1}\sum\nolimits_{l=1}^{m-1}\pi_{(n,m,l)} =1     \notag 
\end{equation}
where
\begin{equation}
\sum\nolimits_{n=1}^{\infty}\pi_{(n,0,0)}=\frac{\pi_{(1,0,0)}}{p\gamma}=\frac{(1-p)\gamma^2}{(p+\gamma-p\gamma)^2-p\gamma} \notag 
\end{equation}
and we have obtained in (207) that 
\begin{equation}
\sum\nolimits_{n=2}^{\infty}\sum\nolimits_{m=1}^{n-1}\pi_{(n,m,0)}=\frac{p\gamma(1-\gamma)}{(p+\gamma-p\gamma)^2-p\gamma} 
\end{equation}

Therefore, to prove the claim, it suffices to show that 
\begin{align}
\sum\nolimits_{n=3}^{\infty}\sum\nolimits_{m=2}^{n-1}\sum\nolimits_{l=1}^{m-1}\pi_{(n,m,l)} &=\sum\nolimits_{n=3}^{\infty}\left(S_1-S_2+S_3\right)  \notag \\
&= 1- \frac{(1-p)\gamma^2}{(p+\gamma-p\gamma)^2-p\gamma}  - \frac{p\gamma(1-\gamma)}{(p+\gamma-p\gamma)^2-p\gamma} \notag \\
&=1- \frac{\gamma(p+\gamma-2p\gamma)}{(p+\gamma-p\gamma)^2-p\gamma} 
\end{align}

Since
\begin{align}
\gamma(p+\gamma-2p\gamma)&=[\gamma+(p-p\gamma)-(p-p\gamma)][(p+\gamma-p\gamma)-p\gamma]  \notag \\
&=[(p+\gamma-p\gamma)-p(1-\gamma)][(p+\gamma-p\gamma)-p\gamma]  \notag \\ 
&=(p+\gamma-p\gamma)^2 - p(p+\gamma-p\gamma) + p^2\gamma(1-\gamma)  \notag \\
&=(p+\gamma-p\gamma)^2 - p\gamma - p^2(1-\gamma) + p^2\gamma(1-\gamma)  \notag \\ 
&=(p+\gamma-p\gamma)^2 - p\gamma - p^2(1-\gamma)^2
\end{align}
such that we have to verify that 
\begin{equation}
\sum\nolimits_{n=3}^{\infty}\sum\nolimits_{m=2}^{n-1}\sum\nolimits_{l=1}^{m-1}\pi_{(n,m,l)} 
=\sum\nolimits_{n=3}^{\infty}\left(S_1-S_2+S_3\right) = \frac{p^2(1-\gamma)^2}{(p+\gamma-p\gamma)^2-p\gamma}
\end{equation}

Firstly, 
\begin{align}
\sum\nolimits_{n=3}^{\infty}S_1&=\sum\nolimits_{n=3}^{\infty} p^2(1-\gamma)^2\left\{\frac{p}{\gamma(\gamma-p)}(1-p)^{n-2} - \frac{\gamma}{p(\gamma-p)}(1-\gamma)^{n-2}+\left[\frac{1}{p}+\frac{1}{\gamma}+(n-2) \right] [(1-p)(1-\gamma)]^{n-2} \right\}   \notag \\
&=p^2(1-\gamma)^2\left\{\frac{p}{\gamma(\gamma-p)}\frac{1-p}{p} - \frac{\gamma}{p(\gamma-p)}\frac{1-\gamma}{\gamma} +\left(\frac{1}{p}+\frac{1}{\gamma}\right) \frac{(1-p)(1-\gamma)}{p+\gamma-p\gamma} + \frac{(1-p)(1-\gamma)}{(p+\gamma-p\gamma)^2} \right\}   \notag \\
&=p^2(1-\gamma)^2\left\{\frac{1-p}{\gamma(\gamma-p)} -\frac{1-\gamma}{p(\gamma-p)} +\frac{(p+\gamma)(1-p)(1-\gamma)}{p\gamma(p+\gamma-p\gamma)} +\frac{(1-p)(1-\gamma)}{(p+\gamma-p\gamma)^2} \right\} \notag \\
&=p^2(1-\gamma)^2\left\{\frac{p+\gamma-1}{p\gamma} +\frac{(p+\gamma)(1-p)(1-\gamma)}{p\gamma(p+\gamma-p\gamma)} +\frac{(1-p)(1-\gamma)}{(p+\gamma-p\gamma)^2} \right\} \notag \\
&=p^2(1-\gamma)^2\left\{ \frac{(p+\gamma-1)(p+\gamma-p\gamma)+(p+\gamma)(1-p)(1-\gamma)}{p\gamma(p+\gamma-p\gamma)}  +\frac{(1-p)(1-\gamma)}{(p+\gamma-p\gamma)^2} \right\} 
\end{align}
where
\begin{align}
&(p+\gamma-1)(p+\gamma-p\gamma)+(p+\gamma)(1-p)(1-\gamma)  \notag\\
={}&(p+\gamma)(p+\gamma-p\gamma) - (p+\gamma-p\gamma) + (p+\gamma)(1-p)(1-\gamma) \notag \\
={}&(p+\gamma)  - (p+\gamma-p\gamma)   \\
={}&p\gamma  \notag 
\end{align}

To obtain equation (220), notice that $p+\gamma-p\gamma+(1-p)(1-\gamma)=1$. Thus, 
\begin{equation}
\sum\nolimits_{n=3}^{\infty}S_1=p^2(1-\gamma)^2\left\{\frac{1}{p+\gamma-p\gamma} + \frac{(1-p)(1-\gamma)}{(p+\gamma-p\gamma)^2} \right\} =\frac{p^2(1-\gamma)^2}{(p+\gamma-p\gamma)^2}
\end{equation}

Next, the second part of the entire sum (218) is handled. 
\begin{align}
&\sum\nolimits_{n=3}^{\infty}S_2  \notag  \\
={}&\frac{p\gamma(p+\gamma-p\gamma)}{(p+\gamma-p\gamma)^2-p\gamma} \sum\nolimits_{n=3}^{\infty} \bigg( \frac{(p+\gamma-2p\gamma)(1-\gamma)}{p}\bigg\{ (1-\gamma)^{n-2} -[(1-p)(1-\gamma)]^{n-2}-p(n-2)[(1-p)(1-p\gamma)]^{n-2} \bigg\}\notag \\
  {}&\qquad  - \frac{p^2(1-\gamma)^2}{2}(n-1)(n-2) [(1-p)(1-\gamma)]^{n-2} \bigg)   \notag \\
={}&\frac{p\gamma(p+\gamma-p\gamma)}{(p+\gamma-p\gamma)^2-p\gamma} \bigg( \frac{(p+\gamma-2p\gamma)(1-\gamma)}{p}
\bigg\{ \frac{1-\gamma}{\gamma} - \frac{(1-p)(1-\gamma)}{p+\gamma-p\gamma}-\frac{p(1-p)(1-\gamma)}{(p+\gamma-p\gamma)^2}\bigg\}  \notag \\
  {}&- \frac{p^2(1-p)(1-\gamma)^3}{2}\frac{2}{(p+\gamma-p\gamma)^3}  \bigg)  
\end{align}
in which
\begin{align}
&\frac{(p+\gamma-2p\gamma)(1-\gamma)}{p}
\bigg\{ \frac{1-\gamma}{\gamma} - \frac{(1-p)(1-\gamma)}{p+\gamma-p\gamma}-\frac{p(1-p)(1-\gamma)}{(p+\gamma-p\gamma)^2}\bigg\}  \notag \\
={}&\frac{(p+\gamma-2p\gamma)(1-\gamma)}{p}
\bigg\{ (1-\gamma)\left[ \frac{1}{\gamma} - \frac{1-p}{p+\gamma-p\gamma}\right] - \frac{p(1-p)(1-\gamma)}{(p+\gamma-p\gamma)^2}\bigg\}  \notag \\
={}&\frac{(p+\gamma-2p\gamma)(1-\gamma)}{p}
\bigg\{  \frac{p(1-\gamma)}{\gamma(p+\gamma-p\gamma)}  - \frac{p(1-p)(1-\gamma)}{(p+\gamma-p\gamma)^2}\bigg\}  \notag \\
={}&\frac{(p+\gamma-2p\gamma)(1-\gamma)}{p}
 \frac{p(1-\gamma)}{(p+\gamma-p\gamma)} \left[ \frac{1}{\gamma} - \frac{1-p}{p+\gamma-p\gamma} \right] \notag \\
={}& \frac{p(1-\gamma)^2(p+\gamma-2p\gamma)}{\gamma(p+\gamma-p\gamma)^2}
\end{align}

Therefore, we can write 
\begin{align}
\sum\nolimits_{n=3}^{\infty}S_2&= \frac{p\gamma(p+\gamma-p\gamma)}{(p+\gamma-p\gamma)^2-p\gamma} \left( \frac{p(1-\gamma)^2(p+\gamma-2p\gamma)}{\gamma(p+\gamma-p\gamma)^2} - \frac{p^2(1-p)(1-\gamma)^3}{(p+\gamma-p\gamma)^3}  \right)   \notag \\
&=\frac{p\gamma(p+\gamma-p\gamma)}{(p+\gamma-p\gamma)^2-p\gamma} \frac{p(1-\gamma)^2}{(p+\gamma-p\gamma)^2}
\left[ \frac{p+\gamma-2p\gamma}{\gamma}- \frac{p(1-p)(1-\gamma)}{p+\gamma-p\gamma} \right] \notag \\
&=\frac{p^2\gamma(1-\gamma)^2}{[(p+\gamma-p\gamma)^2-p\gamma](p+\gamma-p\gamma)}\frac{(p+\gamma-p\gamma)^2-p\gamma}{\gamma(p+\gamma-p\gamma)} \notag \\
&=\frac{p^2(1-\gamma)^2}{(p+\gamma-p\gamma)^2} 
\end{align}

The last part is calculated as follows. 
\begin{align}
\sum\nolimits_{n=3}^{\infty}S_3 &=\frac{\gamma^2(p+2\gamma-2p\gamma)}{(p+\gamma-p\gamma)^2-p\gamma} \sum\nolimits_{n=3}^{\infty} \bigg\{ p(n-2)(1-\gamma)^{n-1} - 2(1-p)(1-\gamma)^{n-1} + 2[(1-p)(1-\gamma)]^{n-1}   \notag \\
& \qquad  + p(n-2)[(1-p)(1-\gamma)]^{n-1} \bigg\} + \frac{(p\gamma)^2(p+\gamma-p\gamma)}{2[(p+\gamma-p\gamma)^2-p\gamma]} \sum\nolimits_{n=3}^{\infty} (n-1)(n-2)[(1-p)(1-\gamma)]^{n-1} \notag \\
&=\frac{\gamma^2(p+2\gamma-2p\gamma)}{(p+\gamma-p\gamma)^2-p\gamma}\left\{\frac{p(1-\gamma)^2}{\gamma^2}-\frac{2(1-p)(1-\gamma)^2}{\gamma} + \frac{2(1-p)^2(1-\gamma)^2}{p+\gamma-p\gamma}+ \frac{p(1-p)^2(1-\gamma)^2}{(p+\gamma-p\gamma)^2} \right\}  \notag \\
&\qquad + \frac{(p\gamma)^2(p+\gamma-p\gamma)}{2[(p+\gamma-p\gamma)^2-p\gamma]}\frac{2(1-p)^2(1-\gamma)^2}{(p+\gamma-p\gamma)^3}
\end{align}
where
\begin{align}
&\frac{p(1-\gamma)^2}{\gamma^2}-\frac{2(1-p)(1-\gamma)^2}{\gamma} + \frac{2(1-p)^2(1-\gamma)^2}{p+\gamma-p\gamma}+ \frac{p(1-p)^2(1-\gamma)^2}{(p+\gamma-p\gamma)^2}  \notag \\
={}&\frac{p(1-\gamma)^2}{\gamma^2}- 2(1-p)(1-\gamma)^2 \left[ \frac{1}{\gamma} - \frac{1-p}{p+\gamma-p\gamma}\right] + \frac{p(1-p)^2(1-\gamma)^2}{(p+\gamma-p\gamma)^2}  \notag \\
={}&\frac{p(1-\gamma)^2}{\gamma^2} - \frac{2p(1-p)(1-\gamma)^2}{\gamma(p+\gamma-p\gamma)} + \frac{p(1-p)^2(1-\gamma)^2}{(p+\gamma-p\gamma)^2}  \notag \\
={}& \frac{p(1-\gamma)^2}{\gamma^2} - \frac{p(1-p)(1-\gamma)^2}{p+\gamma-p\gamma} \left[ \frac{2}{\gamma} - \frac{1-p}{p+\gamma-p\gamma}\right]  \notag \\
={}&\frac{p(1-\gamma)^2}{\gamma^2} - \frac{p(1-p)(1-\gamma)^2}{p+\gamma-p\gamma}\frac{2p+\gamma-p\gamma}{\gamma(p+\gamma-p\gamma)}  \notag \\
={}&\frac{p(1-\gamma)^2}{\gamma}\left[ \frac{1}{\gamma} - \frac{(1-p)(2p+\gamma-p\gamma)}{(p+\gamma-p\gamma)^2} \right]
\notag \\
={}&\frac{p(1-\gamma)^2}{\gamma} \frac{(p+\gamma-p\gamma)^2-(1-p)\gamma(2p+\gamma-p\gamma)}{\gamma(p+\gamma-p\gamma)^2}  \notag \\
={}&\frac{p(1-\gamma)^2}{\gamma} \frac{(p+\gamma-p\gamma)^2 - [(p+\gamma-p\gamma)-p][(p+\gamma-p\gamma)+p]}{\gamma(p+\gamma-p\gamma)^2} \notag \\
={}&\frac{p(1-\gamma)^2}{\gamma} \frac{(p+\gamma-p\gamma)^2 - [(p+\gamma-p\gamma)^2-p^2]}{\gamma(p+\gamma-p\gamma)^2} \notag \\
={}&\frac{p^3(1-\gamma)^2}{\gamma^2(p+\gamma-p\gamma)^2}
\end{align}
such that we obtain that  
\begin{align}
\sum\nolimits_{n=3}^{\infty}S_3 &=\frac{\gamma^2(p+2\gamma-2p\gamma)}{(p+\gamma-p\gamma)^2-p\gamma} \frac{p^3(1-\gamma)^2}{\gamma^2(p+\gamma-p\gamma)^2} + \frac{(p\gamma)^2(1-p)^2(1-\gamma)^2}{[(p+\gamma-p\gamma)^2-p\gamma](p+\gamma-p\gamma)^2} \notag \\
&= \frac{p^3(1-\gamma)^2(p+2\gamma-2p\gamma)}{[(p+\gamma-p\gamma)^2-p\gamma](p+\gamma-p\gamma)^2}  + \frac{(p\gamma)^2(1-p)^2(1-\gamma)^2}{[(p+\gamma-p\gamma)^2-p\gamma](p+\gamma-p\gamma)^2} \notag \\
&=\frac{p^2(1-\gamma)^2}{[(p+\gamma-p\gamma)^2-p\gamma](p+\gamma-p\gamma)^2}\bigg\{ p(p+2\gamma-2p\gamma) + (1-p)^2\gamma^2 \bigg\} \notag \\
&=\frac{p^2(1-\gamma)^2}{[(p+\gamma-p\gamma)^2-p\gamma](p+\gamma-p\gamma)^2}\bigg\{ [(p+\gamma-p\gamma)-(1-p)\gamma][(p+\gamma-p\gamma) + (1-p)\gamma] + (1-p)^2\gamma^2 \bigg\} \notag \\
&=\frac{p^2(1-\gamma)^2}{[(p+\gamma-p\gamma)^2-p\gamma](p+\gamma-p\gamma)^2}\bigg\{ (p+\gamma-p\gamma)^2 
- (1-p)^2\gamma^2  + (1-p)^2\gamma^2 \bigg\} \notag \\
&=\frac{p^2(1-\gamma)^2}{[(p+\gamma-p\gamma)^2-p\gamma](p+\gamma-p\gamma)^2} (p+\gamma-p\gamma)^2  \notag \\
&=\frac{p^2(1-\gamma)^2}{(p+\gamma-p\gamma)^2-p\gamma}
\end{align}

Combining equations (221), (224), and (227), it was observed that 
\begin{equation}
\sum\nolimits_{n=3}^{\infty}\left( S_1-S_2+S_3\right)=\frac{p^2(1-\gamma)^2}{(p+\gamma-p\gamma)^2-p\gamma}
\end{equation}
which satisfies (218), such that all the stationary probabilities indeed add up to 1. This completes this appendix.

\ifCLASSOPTIONcaptionsoff
  \newpage
\fi



\bibliographystyle{IEEEtran}
\bibliography{paperlist.bib}
\end{document}